%% file: sigma24-057.tex
\pdfoutput=1
\RequirePackage{ifpdf}
\ifpdf % We~are running pdfTeX in pdf mode
\documentclass[pdftex]{sigma}
\else
\documentclass{sigma}
\fi

\usepackage{extarrows}
\usepackage{tikz}

\numberwithin{equation}{section}

\newtheorem{Theorem}{Theorem}[section]
\newtheorem*{Theorem*}{Theorem}

\newtheorem{Conjecture}[Theorem]{Conjecture}
 { \theoremstyle{definition}
\newtheorem{Definition}[Theorem]{Definition}
\newtheorem{Assumption}[Theorem]{Assumption}

\newtheorem{Remark}[Theorem]{Remark} }

\begin{document}

\allowdisplaybreaks

\renewcommand{\thefootnote}{}

\newcommand{\arXivNumber}{1910.13639}

\renewcommand{\PaperNumber}{057}

\FirstPageHeading

\ShortArticleName{Smooth Solutions of the tt* Equation: A Numerical Aided Case Study}

\ArticleName{Smooth Solutions of the tt* Equation: \\ A Numerical Aided Case Study\footnote{This paper is a~contribution to the Special Issue on Evolution Equations, Exactly Solvable Models and Random Matrices in honor of Alexander Its' 70th birthday. The~full collection is available at \href{https://www.emis.de/journals/SIGMA/Its.html}{https://www.emis.de/journals/SIGMA/Its.html}}}

\Author{Yuqi LI}

\AuthorNameForHeading{Y.~Li}

\Address{School of Mathematical Sciences, Key Laboratory of MEA $\&$ Shanghai Key Laboratory of PMMP,\\ East China Normal University, Shanghai 200241, P.R.~China}
\Email{\href{mailto:yqli@sei.ecnu.edu.cn}{yqli@sei.ecnu.edu.cn}}

\ArticleDates{Received December 31, 2023, in final form June 13, 2024; Published online June 29, 2024}

\Abstract{An important special class of the tt* equations are the tt*-Toda equations. Guest et al.\ have given comprehensive studies on the tt*-Toda equations in a series of papers. The fine asymptotics for a large class of solutions of a special tt*-Toda equation, the case 4a in their classification, have been obtained in the paper [\textit{Comm. Math. Phys.} \textbf{374} (2020), 923--973] in the series. Most of these formulas are obtained with elaborate reasoning and the calculations involved are lengthy. There are concerns about these formulas if they have not been verified by other methods. The first part of this paper is devoted to the numerical verification of these fine asymptotics. In fact, the numerical studies can do more and should do more. A natural question is whether we can find more such beautiful formulas in the tt* equation via numerical study. The second part of this paper is devoted to the numerical study of the fine asymptotics of the solutions in an enlarged class defined from the Stoke data side. All the fine asymptotics of the solutions in the enlarged class are found by the numerical study. The success of the numerical study is largely due to the truncation structures of the tt* equation.}


\Classification{33E17; 34E05; 35Q51; 65L05}

\renewcommand{\thefootnote}{\arabic{footnote}}
\setcounter{footnote}{0}

\section{Introduction}
The tt* equations were introduced by Cecotti and Vafa when they studied the fusion of topological $N=2$ supersymmetric quantum field theory with its conjugate, the anti-topological one~\cite{CV-1}.
	They also appeared in the extraction of exact results for supersymmetric $\sigma$ models \cite{CV-2}
	and in the classification of the $N=2$ supersymmetric theories \cite{CV-3}.
	Dubrovin gave the zero-curvature representation of the tt* equations and studied their geometrical aspects \cite{Dub}.
	An important special class of the tt* equations are the tt*-Toda equations, which are the reduction of the two-dimensional ($n+1$)-periodic Toda lattice with opposite sign
	\[	%	\label{Toda-N}
		 2 (w_i)_{z\bar z} =-{\rm e}^{2 (w_{i+1}-w_i)}+{\rm e}^{2 (w_i-w_{i-1})},\qquad
			w_{i+n+1}=w_i ,
	\]
	where $\bar z$ denotes the complex conjugate of $z \in \mathbb{C}$ and $w_i=w_i(z,\bar z) \in \mathbb{R}$,
	constrained by both the~$l$-anti-symmetry constraint
	\[%\label{l-anti}
			w_0+w_{l-1}=0,\quad w_1+w_{l-2}=0, \quad \dots, \quad
			w_l+w_n=0,\quad w_{l+1}+w_{n-1}=0, \quad \dots,
	\]
	where the fixed $l \in \{0,1,\dots, n\}$,
	and the radial constraint
	\begin{gather}		\label{Radical}
		w_i(z,\bar z)=w_i(|z|), \qquad i \in \{0,1,\dots, n\}.
	\end{gather}
	The $l=0$ case of tt*-Toda equations is called the $A_n$ type.
	They were first derived by Cecotti and Vafa when they deformed the superpotentials with the $A_n$-minimal model of the Landau--Ginzburg approach \cite{CV-1}.
	The existence of global solutions for any $n$ for these $A_n$ type tt*-Toda equations can be established by the Higgs bundle method \cite{Mochi-1, Mochi-2}.
	Recently, the existence and uniqueness of these solutions were proved using the Riemann--Hilbert method \cite{GIL-4}.
	
	Almost all concrete example studies of the tt* equations were reduced to the third Painlev\'{e} equation before the work of Guest and Lin \cite{GL-1},
	where they initiated the direct study of a generalized tt* equation with two unknowns
	\begin{gather}\label{TT}
			u_{z \bar z}={\rm e}^{a u}-{\rm e}^{v-u}, \qquad
			v_{z \bar z}={\rm e}^{v-u}-{\rm e}^{-b v},
	\end{gather}
	where $a,b>0$,
	subject to the boundary condition
	\begin{gather}
			u(z)\xlongrightarrow{ |z| \rightarrow \infty} 0, \qquad	u(z) \xlongrightarrow{ z \rightarrow 0} (\gamma+o(1)) \log |z|,\nonumber\\
v(z)\xlongrightarrow{ |z| \rightarrow \infty} 0, \qquad v(z) \xlongrightarrow{z \rightarrow 0} (\delta+o(1)) \ln |z|.
		\label{BC}
	\end{gather}
	The tt*-Toda equations with two dependent variables are the cases $a,b \in \{1,2 \}$, exhausted in~\cite{GL-1}.
	
	In \cite{GIL-1}, Guest, Its and Lin proved the following property for equation \eqref{TT} with boundary condition \eqref{BC}.
	\begin{Theorem}[\cite{GIL-1}]\label{thm-GIL-1}
		For $a,b>0$ and any $(\gamma,\delta)$ in the triangular region $\gamma \ge -\frac{2}{a}$, $\delta \le \frac{2}{b}$, ${\gamma-\delta \le 2}$,
		the system \eqref{TT} has a unique smooth solution that satisfies the boundary condition~\eqref{BC}.
		Furthermore, the unique solution is real and radially-invariant.
	\end{Theorem}
\input{Figure1.tikz}
	Theorem \ref{thm-GIL-1} establishes a map from the point $(\gamma, \delta)$ in the triangular region in Figure \ref{fig-1} to the smooth solution of equation~\eqref{TT}.
	Thus, it characterizes a two-parameter family of smooth real solutions of the tt* equation in $\mathbb{C}^*$.
	Note that a result similar to Theorem \ref{thm-GIL-1} had been obtained by Guest and Lin in \cite{GL-1},
	where they required $\gamma, \delta>0$.
	But the difference is crucial since Theorem \ref{thm-GIL-1} characterizes
	{\it all} smooth radial solution of equation~\eqref{TT} \cite{GIL-2}.
	By the Riemann--Hilbert approach, Guest et al.\ obtained all connection formulae for the tt* cases, i.e., ${a,b \in \{1, 2\}}$~\cite{GIL-2}.
	The complete picture of the monodromy data, holomorphic data, and asymptotic data were finally obtained in~\cite{GIL-3}.
	
	The case $a=b=2$ of \eqref{TT}, which is the case 4a in their classification of the tt*-Toda equations, was studied more thoroughly.
	In \cite{GIL-3}, the fine asymptotics(see below for the exact definition)
	are all given for the class of solutions defined in Theorem \ref{thm-GIL-1}.
	In this case, $w_0=\frac{1}{2}u$ and $ w_1=\frac{1}{2}v$ were used as the proper independent variables.
	Then \eqref{TT} becomes
	\begin{gather}
			2 (w_0)_{z \bar z}={\rm e}^{4 w_0}-{\rm e}^{2 w_1-2 w_0}, \qquad
			2 (w_1)_{z \bar z}={\rm e}^{2 w_1-2w_0}-{\rm e}^{-4 w_1}.
		\label{TT-1}
	\end{gather}
	According to the radical constraint \eqref{Radical},
	system \eqref{TT-1} is written into an ordinary differential equation (ODE) with variable $r=|z|$
	\begin{gather}
			\frac{1}{2} w_0''+\frac{1}{2 r} w_0' ={\rm e}^{4 w_0}-{\rm e}^{2 w_1-2 w_0},\qquad
			\frac{1}{2} w_1''+\frac{1}{2 r} w_1' ={\rm e}^{2 w_1-2w_0}-{\rm e}^{-4 w_1},\label{TT-2}
		\end{gather}
	where the prime denotes $\frac{\rm d}{{\rm d}r}$.
	Near $r=0$, by \eqref{BC}, $w_0$ and $w_1$ have properties
	\begin{gather}
			2 w_0(r) \xlongrightarrow{r \rightarrow 0} (\gamma_0 +o(1) ) \ln r,\qquad
			2 w_1(r) \xlongrightarrow{r \rightarrow 0} (\gamma_1+o(1) ) \ln r.\label{OAsymp}
	\end{gather}
	Near $r=\infty$, the asymptotics of $w_0$ and $w_1$ are expressed
	by the Stokes data $s_1^\mathbb{R}$ and $s_2^\mathbb{R}$ \cite{GIL-2}:
	\begin{gather}
			w_0(r)+w_1(r)\xlongrightarrow{r \rightarrow \infty}-s_1^\mathbb{R} 2^{-\frac{3}{4}} ( \pi r)^{-\frac{1}{2}} {\rm e}^{-2 \sqrt{2}r},\nonumber\\
			w_0(r)-w_1(r)\xlongrightarrow{r \rightarrow \infty}s_2^\mathbb{R} 2^{-\frac{3}{2}} ( \pi r)^{-\frac{1}{2}} {\rm e}^{-4 r}.\label{InfAsymp}
	\end{gather}
	The map from $(\gamma_0, \gamma_1)$ to $\big(s_1^\mathbb{R}, s_2^\mathbb{R}\big)$
	is the connection formula \cite{GIL-2}
	\begin{gather}
			s_1^\mathbb{R}=-2 \cos \left( \frac{\pi}{4} (\gamma_0+1) \right)-2 \cos \left( \frac{\pi}{4}(\gamma_1+3) \right),\nonumber \\
			s_2^\mathbb{R}=-2 -4 \cos \left( \frac{\pi}{4} (\gamma_0+1) \right) \cos \left( \frac{\pi}{4}(\gamma_1+3) \right).
 \label{ConnectFormula}
	\end{gather}
	
	The $r=\infty$ asymptotics \eqref{InfAsymp} is able to
	uniquely fix the solution of \eqref{TT-2}.
	This is an initial value problem from $r=\infty$.
	However, the rough asymptotics \eqref{OAsymp} itself is not enough to fix the solution.
	To fix the solution, it must be accompanied by the rough asymptotics at $r=\infty$:
	$w_0(r) \xlongrightarrow{r \rightarrow \infty}0$, $w_1(r) \xlongrightarrow{r \rightarrow \infty} 0$.
	But this becomes a boundary value problem.
	To get an initial value problem from $r=0$,
	one should start with a more detailed asymptotics near $r=0$.
	In fact, it would be very appropriate to start with the fine asymptotics at $r=0$.

	\begin{Definition}
		An asymptotics is said to be a fine asymptotics of a system of differential equations
		if it satisfies the system's truncation equation with respect to the asymptotics.
	\end{Definition}
	
	Practically, one can obtain the fine asymptotics from a rough one by the following way:
	first truncate and simplify the differential equation system according to the rough asymptotics,
	then solve the truncated system,
	and then fix the parameters of the solution by comparing it with the rough asymptotics.
	
	As an example, let us find out the fine asymptotics of \eqref{TT-2} at $r=\infty$ that coincides with asymptotics \eqref{InfAsymp}.
	The truncation equation for the solutions of \eqref{TT-2} with respect to the asymptotics $w_0(r) \rightarrow 0$ and $w_1(r) \rightarrow 0$
	is
	\begin{gather}
			\frac{1}{2} w_0''+\frac{1}{2 r} w_0' =6 w_0 -2 w_1, \qquad
			\frac{1}{2} w_1''+\frac{1}{2 r} w_1' =6 w_1 -2 w_0.
		\label{TT-2-trunc}
	\end{gather}
	The exact solution of \eqref{TT-2-trunc} that coincides with asymptotics \eqref{InfAsymp} is
	\begin{gather}
			w_0(r)+w_1(r)=-\frac{\sqrt{2} }{\pi} s_1^\mathbb{R} K_0\big(2 \sqrt{2} r\big),\qquad
			w_0(r)-w_1(r)=\frac{1 }{\pi} s_2^\mathbb{R} K_0(4 r),
		\label{inf-asym-sol}
	\end{gather}
	where $K_0$ denotes the Bessel $K_0$ function.
	So \eqref{inf-asym-sol} is the fine asymptotics for the solutions with asymptotics $w_0(r) \rightarrow 0$ and $w_1(r) \rightarrow 0$ at $r=\infty$,
	whereas asymptotics \eqref{InfAsymp} should not be taken as a fine asymptotics since it is not an exact solution of \eqref{TT-2-trunc}.
	
	In \cite{GIL-3}, all fine asymptotics of \eqref{TT-2} at $r=0$ for the solutions described by Theorem \ref{thm-GIL-1} have been obtained.
	These fine asymptotics contain seven cases.
	For convenience, we list them in Section \ref{fs-theo}.
	Therefore, the fine asymptotics at $r=\infty$ and $r=0$ are all known for the solutions described by Theorem \ref{thm-GIL-1},
	i.e., the situations at $r=\infty$ and $r=0$ become symmetric.
	However, these fine asymptotics at $r=0$ are complicated, especially that of the vertex case.
	An intuitive explanation is still lacking.
	Moreover, nothing is known for the general case outside of the triangle in Figure \ref{fig-1}.
	This is our motivation to start the numerical study.
	The first part of this paper verifies these fine asymptotics numerically up to $100$ digits for all the seven cases at $r=0$.
	
	Fine asymptotics are subject to the class of the solutions.
	If the solution class is enlarged, new fine asymptotics will appear.
	We will enlarge the solution class from the Stoke data side in the following way.
	The connection formula \eqref{ConnectFormula} maps the $(\gamma_0, \gamma_1)$ region
	to the $\big(s_1^\mathbb{R}, s_2^\mathbb{R}\big)$ region.
	Coming down to equation \eqref{TT-2}, the region map can be represented by Figure \ref{fig-2}.
	\input{Figure2.tikz}
	Any solution represented by a point $\big(s_1^\mathbb{R}, s_2^\mathbb{R}\big)$
	in the curved triangle (including the edges and the vertexes) in Figure \ref{fig-2}
	must have asymptotic \eqref{OAsymp-Fancy} near $r=0$,
	where $(\gamma_0, \gamma_1)$ is determined by $\big(s_1^\mathbb{R}, s_2^\mathbb{R}\big)$ by the connection formula \eqref{ConnectFormula}.
	So the class of solutions described by Theorem \ref{thm-GIL-1} are parameterized by the points in the curved triangle (including the edges and the vertexes).
	We enlarge the class of solutions to the ones parameterized by the points on the whole real $\big(s_1^\mathbb{R}, s_2^\mathbb{R}\big)$ plane.
	Based on our numerical results, we will generalize the range and the explanation of the connection formula
	and obtain all the fine asymptotics of the enlarged class of solutions at $r=0$.
	Of course, the solution class can also be generalized from the side of $r=0$.
	However, the problem is much harder to solve.
	
	The paper is organized as follows.
	In Section \ref{fs-theo}, we list all the seven fine asymptotics of \eqref{TT-2} at $r=0$ obtained in \cite{GIL-3}.
	In Section \ref{sec-verify}, we numerically verify these seven fine asymptotics.
	In Section \ref{CONJ}, we study the cases where $\big(s_1^\mathbb{R}, s_2^\mathbb{R}\big)$ is outside the curved triangle and obtain our main result.
	In Section \ref{DFG}, we present a numerical study from the $r=0$ side.
	In Section \ref{sec6}, we give the conclusion and discussions.
	This paper can be seen as a complement to \cite{GIL-1,GIL-2,GIL-3}.
	
	\section[Fine asymptotics of (1.5) at r=0 of the class of solutions defined by Theorem 1.1]{Fine asymptotics of (\ref{TT-2}) at $\boldsymbol{r=0}$ of the class \\
of solutions defined by Theorem \ref{thm-GIL-1}}\label{fs-theo}
	The fine asymptotics of \eqref{TT-2} at $r=0$ of the class of solutions defined by Theorem \ref{thm-GIL-1} have all been obtained in \cite{GIL-3}.
	For convenience, we list them all here.
	We will use the following notations.
	\begin{itemize}\itemsep=0pt
		\item{$\Gamma$: $\Gamma(z)$ is the usual Gamma function defined by $\Gamma(z)=\int_{0}^{+\infty} t^{z-1}{\rm e}^{-t} {\rm d}t$ for $\operatorname{Re}(z)>0$.}
		\item{$\psi$: $\psi(z)=\frac{\rm d}{{\rm d}z} \ln (\Gamma(z))=\frac{\Gamma'(z)}{\Gamma(z)}$.}
		\item{$s$: $s=\ln (r)$ is used as an easy independent variable near $r=0$.}
		\item{$\gamma_{{\rm Eu}}$: $\gamma_{{\rm Eu}}$ is the Euler's constant $\gamma$, whose numerical values is approximately $0.5772156649$.}
		\item{$\zeta$: $\zeta(z)$ is the Riemann zeta function.}
	\end{itemize}
	The seven fine asymptotics of \eqref{TT-2} at $r=0$ obtained in \cite{GIL-3} are the following.
	\begin{itemize}\itemsep=0pt
		\item{General case:}
\begin{gather}\label{OAsymp-Fancy}
					2 w_0(r) \xlongrightarrow{r \rightarrow 0} \gamma_0 \ln r +\rho_0, \qquad
					2 w_1(r) \xlongrightarrow{r \rightarrow 0} \gamma_1 \ln r+\rho_1,
				\end{gather}
			where
			\begin{gather}
					\rho_0= -\ln \left( 2^{2 \gamma_0}
					\frac{\Gamma\big(\frac{1+\gamma_0}{4}\big) \Gamma\big(\frac{4+\gamma_0+\gamma_1}{8}\big) \Gamma\big(\frac{6+\gamma_0-\gamma_1}{8}\big)}
					{ \Gamma\big(\frac{3-\gamma_0}{4}\big)\Gamma\big(\frac{4-\gamma_0-\gamma_1}{8}\big) \Gamma\big(\frac{2-\gamma_0+\gamma_1}{8}\big)} \right),\nonumber \\
					\rho_1=-\ln \left( 2^{2 \gamma_1}
					\frac{\Gamma\big(\frac{3+\gamma_1}{4}\big) \Gamma\big(\frac{4+\gamma_0+\gamma_1}{8}\big) \Gamma\big(\frac{2-\gamma_0+\gamma_1}{8}\big)}
					{ \Gamma\big(\frac{1-\gamma_1}{4}\big)\Gamma\big(\frac{4-\gamma_0-\gamma_1}{8}\big) \Gamma\big(\frac{6+\gamma_0-\gamma_1}{8}\big)} \right). \label{rhos-DEF}
				\end{gather}
		\item{E1 case:} \begin{gather}
					2 w_0(r) \xlongrightarrow{r \rightarrow 0} \gamma_0 \ln r +a_{E1}, \qquad
					2 w_1(r) \xlongrightarrow{r \rightarrow 0} \ln r+\ln(-2 s+b_{E1}), \label{Asymp-E1}
				\end{gather}
			where
			\begin{gather*}
				a_{E1}=-\ln \left( 2^{2 \gamma_0}
				\frac{\Gamma\big(\frac{\gamma_0+1}{4}\big) \big( \Gamma\big(\frac{\gamma_0+5}{8}\big) \big)^2}
				{ \Gamma\big(\frac{3-\gamma_0}{4}\big) \big(\Gamma\big(\frac{3-\gamma_0}{8}\big) \big)^2 } \right) ,\\
				b_{E1}=\frac{1}{2} \psi\left(\frac{3-\gamma_0}{8}\right)+\frac{1}{2} \psi\left(\frac{5+\gamma_0}{8}\right) -\gamma_{{\rm Eu}} +4 \ln 2.
		\end{gather*}
		\item{E2 case:}\[%\label{Asymp-E2}
					2 w_0(r) \xlongrightarrow{r \rightarrow 0} -\ln (r) -\ln \left(-2 s+ a_{E2} \right),\qquad
					2 w_1(r) \xlongrightarrow{r \rightarrow 0} \gamma_1 \ln (r)+b_{E2},
			\]
			where
			\begin{gather*}
				a_{E2}=\frac{1}{2} \psi\left(\frac{3+\gamma_1}{8}\right)+\frac{1}{2} \psi\left(\frac{5-\gamma_1}{8}\right) -\gamma_{{\rm Eu}} +4 \ln 2 ,\\
				b_{E2}=-\ln \left( 2^{2 \gamma_1}
				\frac{\Gamma\big(\frac{\gamma_1+3}{4}\big) \big( \Gamma\big(\frac{\gamma_1+3}{8}\big) \big)^2}
				{ \Gamma\big(\frac{1-\gamma_1}{4}\big) \big(\Gamma\big(\frac{5-\gamma_1}{8}\big) \big)^2 } \right).
		\end{gather*}
		\item{E3 case:}\begin{gather}
					2 w_0(r)+ 2 w_1(r) \xlongrightarrow{r \rightarrow 0} 2(\gamma_0-1)\ln (r) +a_{E3},\nonumber\\
					2 w_1(r)- 2 w_0(r) \xlongrightarrow{r \rightarrow 0} -2 \ln (r)-\ln \left( 4 (s+b_{E3})^2 \right),\label{Asymp-E3}
				\end{gather}
			where
			\begin{gather*}
				a_{E3}=4(1-\gamma_0) \ln 2-4 \ln \left( \Gamma \left( \frac{1+\gamma_0}{4}\right) \right)+
				4 \ln \left( \Gamma \left( \frac{3-\gamma_0}{4}\right) \right), \\
				b_{E3}=-\frac{1}{4}\psi\left(\frac{3-\gamma_0}{4}\right)
				-\frac{1}{4}\psi\left(\frac{\gamma_0-3}{4}\right) +\frac{1}{3-\gamma_0}+\frac{\gamma_{{\rm Eu}}}{2}
				-2 \ln(2).
		\end{gather*}
		\item{V1 case:} \begin{gather}
					2 w_0(r) \xlongrightarrow{r \rightarrow 0} 3 \ln (r)+\ln ( P_3), \qquad
					2 w_0(r)+ 2 w_1(r) \xlongrightarrow{r \rightarrow 0} 4 \ln(r)+ \ln ( P_4),\label{Asymp-V1}
				\end{gather}
			where
				\begin{gather}
					P_3= -\frac{4}{3} (s-\ln 4)^3-4 \gamma_{{\rm Eu}} (s-\ln 4)^2-4 \gamma_{{\rm Eu}}^2 (s-\ln 4)-\frac{1}{24} \zeta(3)-\frac{4}{3} \gamma_{{\rm Eu}}^3,
					\label{P3}\\
					P_4= \frac{4}{3} (s-\ln 4)^4+ \frac{16 }{3}\gamma_{{\rm Eu}} (s-\ln 4)^3+8 \gamma_{{\rm Eu}}^2 (s-\ln 4)^2
					\nonumber\\
\hphantom{P_4=}{} +\left(\frac{16\gamma_{{\rm Eu}}^3}{3} -\frac{\zeta(3)}{12}\right) (s-\ln 4)-\frac{ \gamma_{{\rm Eu}} \zeta(3)}{12}+\frac{4\gamma_{{\rm Eu}}^4 }{3}.
					\label{P4}
\end{gather}
		\item{V2 case:}
\[
				2 w_0(r)=-2 w_1(r) \xlongrightarrow{r \rightarrow 0} -\ln (r) -\ln(-2 s -2 \gamma_{{\rm Eu}}+2 \ln 2) .
\]
		\item{V3 case:} \[%\label{Asymp-V3}
					2 w_1(r) \xlongrightarrow{r \rightarrow 0} -3 \ln (r)-\ln ( P_3), \qquad
					2 w_0(r)+ 2 w_1(r) \xlongrightarrow{r \rightarrow 0} -4 \ln(r)- \ln ( P_4),
			\]
			where $P_3$ and $P_4$ are defined by \eqref{P3} and \eqref{P4}.
	\end{itemize}

	\section[Verifying numerically the fine asymptotics of (1.5) at r=0 of the class of solutions defined by Theorem 1.1]{Verifying numerically the fine asymptotics of (\ref{TT-2})\\ at $\boldsymbol{r=0}$ of the class of solutions defined by Theorem \ref{thm-GIL-1}}\label{sec-verify}
	It is well known that equation \eqref{TT-2} has symmetry $w_0\rightarrow -w_1$, $w_1\rightarrow -w_0$,
	i.e., if $(w_0(r), w_1(r))=(f(r), g(r))$ is a solution of the tt* equation,
	then $(w_0(r), w_1(r))=(-g(r), -f(r))$ is also a~solution of the tt* equation.
	Therefore, if the solution $(w_0(r), w_1(r))=(f(r), g(r))$ has data
	$(\gamma_0,\gamma_1)=(\mu_0,\mu_1)$ and $\big(s_1^\mathbb{R}, s_2^\mathbb{R}\big)=(\nu_1,\nu_2)$,
	then the solution $(w_0(r), w_1(r))=(-g(r),\allowbreak -f(r))$ will have data
	$(\gamma_0,\gamma_1)=(-\mu_1,-\mu_0)$ and $\big(s_1^\mathbb{R}, s_2^\mathbb{R}\big)=(-\nu_1,\nu_2)$ by \eqref{OAsymp} and \eqref{InfAsymp}.
	From this symmetry, the fine asymptotics of the E2 case and the V3 case can be obtained from those of the E1 case and the V1 case respectively.
	Furthermore, as has been mentioned in \cite{GIL-3}, the~V2 case is just the sinh-Gordon,
	for which the asymptotic is already well known.
	So, we will only verify four cases: the general, E1, E3 and V1.
	Instead of verifying these asymptotics near $r=0$ directly from the initial value problem defined by ODE \eqref{TT-2} and the $r=\infty$ rough asymptotics~\eqref{InfAsymp},
	which is difficult to reach a satisfactory accuracy,
	we will start from the initial value problem defined by the ODE \eqref{TT-2} and the $r=\infty$ fine asymptotics \eqref{inf-asym-sol}
	to verify the fine asymptotics near $r=0$.\looseness=-1
	
	\subsection[Preliminary for the numerical experiments: an approximation proper for calculations near r=infty]{Preliminary for the numerical experiments:\\ an approximation proper for calculations near $\boldsymbol{r=\infty}$}

	Consider the solutions of \eqref{TT-2} with asymptotics
	$w_0(r)\xlongrightarrow{r \rightarrow \infty} 0$ and $w_1(r)\xlongrightarrow{r \rightarrow \infty} 0$.
	Near $r=\infty$ the primary asymptotics of the solutions is given by the fine asymptotics \eqref{inf-asym-sol}.
	
	Let $w_p=w_0+w_1$, $w_m=w_0-w_1$.
	Then, the equations for $w_p$ and $w_m$ are
	\begin{gather}
			\left( \frac{1}{2} \frac{{\rm d}^2}{{\rm d}r^2}+\frac{1}{2 r} \frac{\rm d}{{\rm d} r} \right) w_p={\rm e}^{2 w_p+2 w_m}-{\rm e}^{2 w_m-2 w_p}
			=2 {\rm e}^{2 w_m} \sinh \left( 2 w_p \right), \nonumber\\
			\left( \frac{1}{2} \frac{{\rm d}^2}{{\rm d}r^2}+\frac{1}{2 r} \frac{\rm d}{{\rm d} r} \right) w_m
			={\rm e}^{2 w_p+2 w_m}+{\rm e}^{2 w_m-2w_p}-2 {\rm e}^{-2 w_m}\nonumber\\[-1mm]
			\phantom{\left( \frac{1}{2} \frac{{\rm d}^2}{{\rm d}r^2}+\frac{1}{2 r} \frac{\rm d}{{\rm d} r} \right) w_m}=4 {\rm e}^{2 w_m} \sinh^2(w_p)+4 \sinh (2 w_m).	\label{wpwm}
		\end{gather}

	Note that \eqref{wpwm} is written in a form that better preserves the significant digits in the numerical integration near $r=\infty$.
	The errors in the approximation of $(w_p, w_m)$ by the fine asymptotics~\eqref{inf-asym-sol} are caused by the nonlinear terms in the expansion of \eqref{wpwm}.
	In general, the most significant correction to $w_p$ is proportional to \smash{$ w_p^{(0)} w_m^{(0)}$} \big(see \eqref{wpwm0} for the definition of \smash{$w_p^{(0)}$} and~\smash{$w_m^{(0)}$}\big), i.e.,
	\smash{$w_p =c_p K_0\big(2 \sqrt{2} r\big)+O\big(r^{-1} {\rm e}^{-(2 \sqrt{2}+4)r}\big)$}.
	Meanwhile, the most significant correction to $w_m$ is proportional to the square of \smash{$w_p^{(0)}$}, i.e.,
	\smash{$ w_m =c_m K_0(4 r)+O\big(r^{-1} {\rm e}^{-4 \sqrt{2} r}\big)$}.
	
	These results are sufficient for the rough numerical investigations for smooth solutions of the~tt* equation.
	They are called rough simply because they can be refined.
	For high precision numerical integration of \eqref{wpwm} from the $r=\infty$ side, the relative error will not enlarge too much when $r$ is still large.
	For $w_m(r)$, the relative error is about \smash{$O\big(r^{-\frac{1}{2}} {\rm e}^{-4 (\sqrt{2}-1) r}\big)$}.
	If we give the initial values by the fine asymptotics \eqref{inf-asym-sol} with $r=45$,
	the relative error of the initial values are of order $10^{-33}$, which is not so satisfactory.
	If we want to reach a relative error of order~$10^{-100}$ by this way,
	$r=138$ is needed to give the initial values.
	We will see, after considering the most significant contribution of the nonlinear terms,
	the starting $r$ can be greatly reduced.
	
	Suppose
		 \begin{gather*}
			w_p(r)=w_p^{(0)}(r)+w_p^{(1)}(r)+w_p^{(2)}(r)+\cdots,\qquad
			w_m(r)=w_m^{(0)}(r)+w_m^{(1)}(r)+w_m^{(2)}(r)+\cdots,
		\end{gather*}
	where
	\begin{gather}
			w_p^{(0)}(r) = -\frac{\sqrt{2}}{\pi} s_1^\mathbb{R} K_0\big(2 \sqrt{2} r\big),\qquad
			w_m^{(0)}(r) = \frac{1}{\pi} s_2^\mathbb{R} K_0(4 r).
		 \label{wpwm0}
	\end{gather}
	
	Then $w_p^{(1)}$ and $w_m^{(1)}$ satisfy
	\[
			\left( \frac{1}{2} \frac{{\rm d}^2}{{\rm d}r^2}+\frac{1}{2 r} \frac{\rm d}{{\rm d} r} \right) w_p^{(1)}-4 w_p^{(1)}=8 w_p^{(0)} w_m^{(0)},\qquad
			\left( \frac{1}{2} \frac{{\rm d}^2}{{\rm d}r^2}+\frac{1}{2 r} \frac{\rm d}{{\rm d} r} \right) w_m^{(1)}-8 w_m^{(1)}
			=4 \big(w_p^{(0)} \big)^2
	\]
	with $w_p^{(1)}(\infty)=0$ and $w_m^{(1)}(\infty)=0$.
	
	The solution of $w_p^{(1)}$ and $w_m^{(1)}$ is
	 \begin{gather*}
			w_p^{(1)}=2 I_0\big(2 \sqrt{2} r\big) \int_\infty^r K_0(2 \sqrt{2}r) \big(8 w_p^{(0)}(r) w_m^{(0)}(r) \big) r {\rm d}r\\
			\phantom{w_p^{(1)}=}{}-2 K_0\big(2 \sqrt{2} r\big) \int_\infty^r I_0\big(2 \sqrt{2} r\big) \big(8 w_p^{(0)}(r) w_m^{(0)}(r) \big) r {\rm d}r,\\
			w_m^{(1)}=2 I_0(4 r) \int_\infty^r K_0(4 r) \big(4 \big( w_p^{(0)}(r) \big)^2 \big) r {\rm d}r
			-2 K_0(4 r) \int_\infty^r I_0(4 r) \big( 4\big( w_p^{(0)}(r) \big)^2 \big) r {\rm d}r,
		\end{gather*}
	where $I_0$ is the Bessel $I_0$ function.

	Then
	\begin{gather}
			w_p(r)=w_p^{(0)}(r)+w_p^{(1)}(r)+O\big(r^{-\frac{3}{2}} {\rm e}^{-6 \sqrt{2} r} \big),\nonumber\\
			w_m(r)=w_m^{(0)}(r)+w_m^{(1)}(r)+O\big(r^{-\frac{3}{2}} {\rm e}^{-(4+ 4\sqrt{2}) r} \big).
		 \label{iniApprox}
	\end{gather}
	The relative errors are both of order $ r^{-1} {\rm e}^{-4 \sqrt{2} r}$.
	To acquire a relative error of order $10^{-100}$, it is enough to start the numerical integration from $r=45$.
	Higher-order nonlinear terms should not be considered,
	otherwise we will run into high-dimensional integrations that are time-consuming to compute to high accuracy, for example, an accuracy of $10^{-100}$.
	
	The truncation of \eqref{iniApprox} will be used to give initial values for the numerical integration of~\eqref{wpwm} near $r=\infty$ for all of the following cases.
	
	\subsection{The general case: in the triangular \label{GeneralCase}}
	This subsection is devoted to the verification of \eqref{OAsymp-Fancy}.
	
	To be specific, we fix $(\gamma_0, \gamma_1)=\big(1, \frac{1}{3} \big)$.
	Then, $\big(s_1^\mathbb{R}, s_2^\mathbb{R}\big)=\big(\sqrt{3},-2 \big)$ by \eqref{ConnectFormula}.
	\eqref{iniApprox} means that we can start our numerical integration from $r=45$
	for moderate $\big(s_1^\mathbb{R},s_2^\mathbb{R}\big)$ to get a relative error of order less than $10^{-100}$.
	Recall that in Section \ref{fs-theo}, we have mentioned
	\begin{eqnarray}
		s=\ln r \label{s-DEF}
	\end{eqnarray}
	is a proper independent variable near $r=0$.
	Therefore, the numerical integration is naturally divided into two parts:
	on $r \in [r_m, 45]$ and on $s \in [s_f, s_m=\ln r_m]$.
	For convenience, we always choose $r_m=1$.
	$s_f$ varies with $\big(s_1^\mathbb{R},s_2^\mathbb{R}\big)$
	and will be determined after we solve the associated truncation of \eqref{TT-2} for the fine asymptotics.
	
	\subsubsection[Numerical integration from r=45 to r=1]{Numerical integration from $\boldsymbol{r=45}$ to $\boldsymbol{r=1}$}
	By the truncation of \eqref{iniApprox}, the initial values for the numerical integration of \eqref{wpwm}
	are calculated up to more than $100$ digits
	\begin{gather}
			w_p(45)=-4.5763465910740842210810671823515633075572030760030\ldots \times 10^{-57},\nonumber \\
			w_p'(45)=1.2994612025622450236510718743064448909150132699101\ldots \times 10^{-56},\nonumber\\
			w_m(45)=-3.9902150828859022626192436154419670328254784177405\ldots \times 10^{-80},\nonumber\\
			w_m'(45)=1.6005134816454403480052616718328017176197600655449\ldots \times 10^{-79}.
\label{ini-General}
	\end{gather}
	To save space, we list only the first $50$ digits in \eqref{ini-General}.
	It is not surprising that
	$w_m(45)$ in \eqref{ini-General} coincides with \smash{$w_m^{(0)}(45)=-\frac{\sqrt{6}}{\pi} K_0\big(90 \sqrt{2}\big)$} for the first $33$ digits
	and that $w_p(45)$ in \eqref{ini-General} coincides with \smash{$w_p^{(0)}(45)=-\frac{2}{\pi} K_0(180)$} for all the listed $50$ digits.
	Formula \eqref{iniApprox} gives only the order of the error, not the actual value.
	We obtain the errors of \eqref{ini-General} by comparing the initial values~\eqref{ini-General} with a more accurate numerical solution starting from $r=55$.
	Table~\ref{tab1} shows both the absolute error and the relative error of the initial values at $r=45$.
	
	\begin{table}[ht]\renewcommand{\arraystretch}{1.2}
			\centering
\caption{Errors of the initial values for the general case with $(\gamma_0,\gamma_1)=\big(1,\frac{1}{3}\big)$.}\label{tab1}
\vspace{1mm}

		\begin{tabular}{ c|c c c c}
			\hline \hline
			$r=45$& $w_p$ & $w_p'$ & $w_m$ &$w_m'$\\
			\hline
		 Absolute error  &$1.98 \times 10^{-170}$ &$1.68 \times 10^{-169}$ &$2.43 \times 10^{-193}$&$2.36\times 10^{-192}$ \\
	  Relative error  &$4.32\times 10^{-114}$ &$1.30 \times 10^{-113}$ &$6.09 \times 10^{-114}$&$1.47\times 10^{-113}$
		\end{tabular}
	\end{table}
	
In this paper, we use the Gauss--Legendre method,
	which is an implicit Runge--Kutta method suitable for high-precision numerical integration,
	to numerically integrate ODEs.
	Integrating~\eqref{wpwm} numerically from $r=45$ to $r=1$
	by a $100$-stage Gauss--Legendre method with step size $\frac{1}{100}$,
	we obtain the numerical values of $w_p$, $w_p'$, $w_m$ and $w_m'$ at $r=1$:
	\begin{gather}
			w_p(1)=-3.2972969594742103001480456261339460432792854660454\ldots\times 10^{-2},\nonumber\\
			w_p'(1)=1.0829838290019404254859616425541702465151021916881\ldots\times 10^{-1},\nonumber\\
			w_m(1)=-6.6648017026562016812805168052539563362254856278250\ldots\times 10^{-3},\nonumber\\
			w_m'(1)=2.8961723214345113722967491163879906375020596216242\ldots\times 10^{-2}.
		 \label{values-general-1}
	\end{gather}
	Note that \eqref{values-general-1} only lists the first $50$ digits of the numerical solution.
	Numerical experiments show that the errors caused by the numerical integration are all negligible.
	This is easy to understand because the precision order of the numerical integration, which is twice the stage number, is large and the step size is small.
	
	Comparing \eqref{values-general-1} with the more accurate solution starting from $r=55$,
	we obtain the errors of \eqref{values-general-1} as Table \ref{tab2}.
	\begin{table}[ht]\renewcommand{\arraystretch}{1.2}	\centering
		\caption{Errors of the numerical solution at $r=1$ for the general case with $(\gamma_0,\gamma_1)=\big(1,\frac{1}{3}\big)$.}\label{tab2}
\vspace{1mm}

		\begin{tabular}{ c|c c c c}
			\hline \hline
			$r=1$ & $w_p$ & $w_p'$ & $w_m$ &$w_m'$\\
			\hline
Absolute error &$2.85 \times 10^{-115}$ &$9.31 \times 10^{-115}$ &$6.64 \times 10^{-116}$&$2.82\times 10^{-115}$ \\
Relative error &$8.63\times 10^{-114}$ &$8.60 \times 10^{-114}$ &$9.97 \times 10^{-114}$&$9.75\times 10^{-114}$
		\end{tabular}
	\end{table}
	
	\subsubsection[Near r=0]{Near $\boldsymbol{r=0}$}
	Inspired by the form of \eqref{OAsymp-Fancy}, we use independent variable $s$
	and dependent variables
	\begin{gather}
			\tilde w_0 =2 w_0-\gamma_0 s, \qquad
			\tilde w_1 =2 w_1-\gamma_1 s.
	 \label{transform-general}
	\end{gather}
	Please recall that $s=\ln (r)$ is defined by \eqref{s-DEF}.
	From the numeric point of view, the advantage of using $s$ rather than $r$
	is that it can avoid the frequent adjustment of the step size when we solve \eqref{TT-2} numerically near $r=0$.
	
	The equations for $\tilde w_0$ and $\tilde w_1$ are
	\begin{gather}
			\frac{1}{4} \frac{ {\rm d}^2\tilde w_0}{{\rm d}s^2}= {\rm e}^{2 \tilde w_0 +2 (\gamma_0+1) s}-{\rm e}^{\tilde w_1- \tilde w_0+(\gamma_1-\gamma_0+2)s},\nonumber \\
			\frac{1}{4} \frac{ {\rm d}^2\tilde w_1}{{\rm d}s^2}={\rm e}^{\tilde w_1- \tilde w_0+(\gamma_1-\gamma_0+2)s}-{\rm e}^{-2 \tilde w_1+2 (1-\gamma_1) s}.
	\label{tu0tu1}
	\end{gather}
	
	We expect $\tilde w_0 \xlongrightarrow{s \rightarrow -\infty} \rho_0$
	and $\tilde w_1 \xlongrightarrow{s \rightarrow -\infty} \rho_1$.
	In the triangular, $\gamma_0>-1$, $\gamma_1<1$, $\gamma_1>\gamma_0-2$.
	So, all terms in the right of \eqref{tu0tu1} can be ignored at first.
	Thus,
	\begin{gather}
			\frac{1}{4} \frac{ {\rm d}^2\tilde w_0^{(0)}}{{\rm d}s^2}= 0 ,\qquad
			\frac{1}{4} \frac{ {\rm d}^2\tilde w_1^{(0)}}{{\rm d}s^2}=0
 \label{tu0tu1-truc}
	\end{gather}
	is the associated truncation of \eqref{tu0tu1} for the fine asymptotics of the general case.
	
	The initial values of $\tilde w_0$, $\frac{ {\rm d}\tilde w_0}{{\rm d}s}$, $\tilde w_1$ and $\frac{ {\rm d}\tilde w_1}{{\rm d}s}$ at $s=0$
	can be derived from $w_p$, $w_p'$, $w_m$ and $w_m'$ at $r=1$:
	\begin{gather}
			\tilde w_0|_{s=0}= w_p|_{r=1}+w_m|_{r=1}, \qquad
			\frac{ {\rm d} \tilde w_0}{{\rm d}s}|_{s=0}= w_p'|_{r=1}+w_m'|_{r=1} -\gamma_0, \nonumber\\
			\tilde w_1|_{s=0}= w_p|_{r=1}-w_m|_{r=1}, \qquad
			\frac{ {\rm d}\tilde w_1}{{\rm d}s}|_{s=0}= w_p'|_{r=1}-w_m'|_{r=1} -\gamma_1.
 \label{ini-s0-General}
	\end{gather}
	
	In the truncation of equation \eqref{tu0tu1} to \eqref{tu0tu1-truc},
	the ignored terms are of order $O\big({\rm e}^{2 (\gamma_0+1) s}\big)$, order $O\big({\rm e}^{(\gamma_1-\gamma_0+2) s}\big)$
	and order \smash{$O\big({\rm e}^{2 (1-\gamma_1)s}\big)$}.
	Now, we have fixed $(\gamma_0, \gamma_1)=\big(1, \frac{1}{3}\big)$.
	Thus, $\big(\tilde w_0, \tilde w_1\big)$ will approach $(\rho_0,\rho_1)|_{\gamma_0=1,\gamma_1=\frac{1}{3}}$
	with a distance of order \smash{$O\big({\rm e}^{\frac{4}{3} s}\big)$}, where
		\begin{gather*}
			\rho_0|_{\gamma_0=1,\gamma_1=\frac{1}{3}} = 0.89156581440748831917188012305422345475702308262231\dots,\\
			\rho_1|_{\gamma_0=1,\gamma_1=\frac{1}{3}} = 0.22017225140694662756648980530049931068839656816740\dots
		\end{gather*}
	by \eqref{rhos-DEF}.
	So, when ${\rm e}^{\frac{4}{3} s} \approx 10^{-100}$, i.e., $s \approx -172.7$,
	$\big(\tilde w_0, \tilde w_1\big)$ will be indistinguishable from $(\rho_0,\rho_1)|_{\gamma_0=1,\gamma_1=\frac{1}{3}}$ within our precision tolerance.
	Therefore, it is enough to integrate \eqref{tu0tu1} numerically from $s=0$ to $s_f=-175$.

	\begin{table}[ht]\renewcommand{\arraystretch}{1.2}
		\centering
\caption{Errors of the numerical solution at $s=-175$ for the general case with $(\gamma_0,\gamma_1)=\big(1,\frac{1}{3}\big)$.}\label{tab3}

\vspace{1mm}

		\begin{tabular}{ c|c c c c}
			\hline \hline
			$s=-175$ & $\tilde w_0$ & $\frac{{\rm d}\tilde w_0}{{\rm d}s}$ & $\tilde w_1$ &$\frac{{\rm d} \tilde w_1}{{\rm d}s}$\\
			\hline
			Absolute error &$1.33 \times 10^{-111}$ &$7.66 \times 10^{-114}$ &$6.54 \times 10^{-112}$&$3.76\times 10^{-114}$\\
			Relative error &$1.50 \times 10^{-111}$ &$1.08 \times 10^{-12}$ &$2.97 \times 10^{-111}$&$2.04\times 10^{-12}$
		\end{tabular}
	\end{table}
	
	Table \ref{tab3} shows that the numerical solution is as accurate as we expected.
	The relative error of $\frac{{\rm d}\tilde w_0}{{\rm d}s}$ or $\frac{{\rm d}\tilde w_1}{{\rm d}s}$ in Table \ref{tab3} seems to be large.
	But this is really nothing since it is only another demonstration of the fact that$\frac{{\rm d}\tilde w_0}{{\rm d}s}$ and $\frac{{\rm d}\tilde w_1}{{\rm d}s}$ are small.
	
	Table \ref{tab4} shows how good the asymptotic solution \eqref{OAsymp-Fancy} is.
	\begin{table}[ht]\renewcommand{\arraystretch}{1.2}\centering
	\caption{Approximate derivation from the asymptotic solution for the general case with $(\gamma_0,\gamma_1)=\big(1,\frac{1}{3}\big)$.}	\label{tab4}

\vspace{1mm}

		\begin{tabular}{@{\,}c | c c c c c c c@{\,}}
			\hline \hline
			$s$ & $-25$ & $-50$ & $-75$ &$-100$ &$-125$& $-150$&$-175$\\
			\hline
			$\ln(\rho_0-\tilde w_0)$ &$-33.1938$ &$-66.5271$&$-99.8605$&$-133.194$&$-166.527$&$-199.860$&$-233.194 $\\
			$\ln(\rho_1-\tilde w_1)$ &$-34.5412$ &$-67.8745$ &$-101.208$&$-134.541$&$-167.875$&$-201.208$&$-234.541$
		\end{tabular}
	\end{table}
	
	Table \ref{tab4} not only numerically verifies the asymptotics of the general case for $(\gamma_0,\gamma_1)=\big(1,\frac{1}{3}\big)$,
	but also confirms our estimate that $\big(\tilde w_0, \tilde w_1\big)$ is close to its asymptotics \smash{$(\rho_0,\rho_1)|_{\gamma_0=1,\gamma_1=\frac{1}{3}}$} with a~distance of order \smash{$O\big({\rm e}^{\frac{4}{3} s}\big)$}.
	
	\subsection{Case E1 \label{E1Case}}
	This subsection is devoted to the verification of the fine asymptotics of the E1 case.
	Note that the E1 case is parameterized by $-1<\gamma_0<3$ and $\gamma_1=1$.
	To fix the problem, we take $\gamma_0=1$ as an example to verify the E1 case.
	Substituting~${(\gamma_0,\gamma_1)=(1,1)}$
	to the connection formula \eqref{ConnectFormula},
	we immediately get $\big(s_1^\mathbb{R}, s_2^\mathbb{R}\big)=(2, -2)$.
	Similar to the general case of Section \ref{GeneralCase},
	the numerical integration is divided into two parts: for $r\in [1,45]$ and for $s\in [s_f,0]$.
	
	\subsubsection[Numerical integration from r=45 to r=1]{Numerical integration from $\boldsymbol{r=45}$ to $\boldsymbol{r=1}$}
	By the truncation of \eqref{iniApprox}, the initial values at $r=45$ are obtained (only the first $50$ digits are listed)
	\begin{gather}
			w_p(45)= -5.2843098725232974899221393911204991207504443469367\ldots \times 10^{-57},\nonumber \\
			w_p'(45)= 1.5004885502015739552694025310567337731833237644509\ldots \times 10^{-56},\nonumber\\
			w_m(45)=- 3.9902150828859022626192436154419666864562950795650\ldots \times 10^{-80},\nonumber\\
			w_m'(45)= 1.6005134816454403480052616718328015209213735935410\ldots \times 10^{-79}.
		\label{ini-E1}
	\end{gather}
	
	Comparing with the more accurate solution starting from $r=55$,
	the errors of the initial values \eqref{ini-E1} are obtained as shown by Table~\ref{tab5}.

	\begin{table}[ht]\renewcommand{\arraystretch}{1.2}\centering
		\caption{Errors of the initial values of case E1 with $\gamma_0=1$.}		\label{tab5}

\vspace{1mm}

		\begin{tabular}{ c|c c c c}
			\hline \hline
			$r=45$& $w_p$ & $w_p'$ & $w_m$ &$w_m'$\\
			\hline
		 Absolute error &$3.04 \times 10^{-170}$ &$2.59 \times 10^{-169}$ &$3.24 \times 10^{-193}$&$3.14\times 10^{-192}$ \\
	Relative error &$5.76\times 10^{-114}$ &$1.73 \times 10^{-113}$ &$8.12 \times 10^{-114}$&$1.96\times 10^{-113}$
		\end{tabular}
	\end{table}

	Integrating \eqref{wpwm} numerically from $r=45$ to $r=1$ by the Gauss--Legendre method with the same parameters as the ones in Section \ref{GeneralCase},
	the values of $w_p$, $w_p'$, $w_m$ and $w_m'$ at $r=1$ are obtained
	\begin{gather}
			w_p(1)=-3.8076020447615564848336037555396597913276640146800\ldots\times 10^{-2}, \nonumber\\
			w_p'(1)=1.2507257120725277318359466237894266588814464453818\ldots\times 10^{-1},\nonumber\\
			w_m(1)=-6.5181931373519405060356987540333399617643482502891\ldots\times 10^{-3},\nonumber\\
			w_m'(1)=2.8018632441288063804071518136255604932977444116709\ldots\times 10^{-2}.
		\label{values-E1-1}
	\end{gather}
	Comparing with the more accurate solution starting from $r=55$,
	the errors of \eqref{values-E1-1} are obtained as shown by Table \ref{tab6}.
	\begin{table}[ht]\renewcommand{\arraystretch}{1.2}\centering
		\caption{Errors of the numerical solution at $r=1$ of case E1 with $\gamma_0=1$.}
		\label{tab6}

\vspace{1mm}

		\begin{tabular}{ c|c c c c}
			\hline \hline
			$r=1$& $w_p$ & $w_p'$ & $w_m$ &$w_m'$\\
			\hline
			 Absolute error  &$4.38 \times 10^{-115}$ &$1.43 \times 10^{-114}$ &$8.52 \times 10^{-116}$&$3.55\times 10^{-115}$ \\
			 Relative error  &$1.15\times 10^{-113}$ &$1.15 \times 10^{-113}$ &$1.31 \times 10^{-113}$&$1.27\times 10^{-113}$
		\end{tabular}
	\end{table}

	\subsubsection[Near r=0]{Near $\boldsymbol{r=0}$}
	Let
	\begin{gather}
			\tilde w_0 =2 w_0-\gamma_0 s, \qquad
			\tilde w_1 =2 w_1 - s,
		 \label{transform-r0-E1}
	\end{gather}
	where $s=\ln(r)$ as defined by \eqref{s-DEF}.
	Then the differential equations for $\tilde w_0$ and $\tilde w_1$ are
	\begin{gather}
			\frac{1}{4} \frac{ {\rm d}^2\tilde w_0}{{\rm d}s^2}= {\rm e}^{2 \tilde w_0 +2 (\gamma_0+1) s}-{\rm e}^{\tilde w_1- \tilde w_0+(3-\gamma_0)s},\nonumber \\
			\frac{1}{4} \frac{ {\rm d}^2\tilde w_1}{{\rm d}s^2}={\rm e}^{\tilde w_1- \tilde w_0+(3-\gamma_0)s}-{\rm e}^{-2 \tilde w_1 }.
	\label{tu0tu1-E1}
	\end{gather}
	Note that \eqref{tu0tu1-E1} can also be obtained from \eqref{tu0tu1} by substituting $\gamma_1=1$ to it.
	
	We expect $\tilde w_0$ is of order $O(1)$ and that $\tilde w_1$ is of order $O(\ln(-s))$.
	Also considering ${-1<\!\gamma_0\!<3}$, we obtain the associated truncation of \eqref{tu0tu1-E1} near $s=-\infty$ for the fine asymptotic of the E1 case
	\begin{gather}
			\frac{1}{4} \frac{ {\rm d}^2\tilde w_0^{(0)}}{{\rm d}s^2}=0,\qquad
			\frac{1}{4} \frac{ {\rm d}^2\tilde w_1^{(0)}}{{\rm d}s^2}=-{\rm e}^{-2 \tilde w_1^{(0)} }.
	\label{tu0tu1-trunc-E1}
	\end{gather}
	The general solution of \eqref{tu0tu1-trunc-E1} is
	\[%\label{SOL-trunc-E1}
			\tilde w_0^{(0)}=k_{0E1}+k_{1E1} s,\qquad
			\tilde w_1^{(0)}=\ln \left( \pm \frac{2}{k_{2E1} } \sinh(k_{2E1} (s+k_{3E1})) \right).
	\]
	By \eqref{Asymp-E1} and \eqref{transform-r0-E1},
	we know the fine asymptotics of \eqref{TT-2} in the E1 case corresponds to
	${k_{0E1}=a_{E1}}$, $k_{1E1}=0$, $k_{2E1}\rightarrow 0$, $k_{3E1}=-\frac{1}{2}b_{E1}$ and the ``$\pm$" sign chosen to be minus.
	
	\begin{Remark}
		It is obvious that $k_{1E1}=0$ and $k_{2E1}\rightarrow 0$, or else $\tilde w_0$ and $\tilde w_1$ will have order $O(s)$ at $s=-\infty$,
		which is in contradiction with our assumption that $\tilde w_0$ and $\tilde w_1$ are of order $O(1)$ and $O(\ln(-s))$ respectively.
		Therefore, the consistent solution of \eqref{tu0tu1-trunc-E1} is
		\begin{gather}
				\tilde w_0^{(0)}=k_{0E1},\qquad
				\tilde w_1^{(0)}=\ln ( -2 (s+k_{3E1}) ).
		\label{SOL-trunc-E1-0}
		\end{gather}
		\eqref{SOL-trunc-E1-0} with \eqref{transform-r0-E1} gives a fine structure for solution of \eqref{TT-2} with $\gamma_0 \in (-1,3)$ and $\gamma_1=1$.
		In other words, any fixed set of $\{k_{0E1}, k_{3E1}\}$ for \eqref{SOL-trunc-E1-0} defines a well-posed initial value problem for \eqref{TT-2} from $r=0$.
		The E1 case has $k_{0E1}=a_{E1}$ and $k_{3E1}=-\frac{1}{2}b_{E1}$,
		which is distinguished by that $w_0(r)$ and $w_1(r)$ are smooth on $r\in (0, \infty)$ and that they have asymptotics \eqref{InfAsymp}.
	\end{Remark}
	
	In the truncation from \eqref{tu0tu1-E1} to \eqref{tu0tu1-trunc-E1},
	the ignored term for the differential equation of $\tilde w_1$ is~${\rm e}^{\tilde w_1- \tilde w_0+(3-\gamma_0)s}$, which is of order $O\big(s {\rm e}^{(3-\gamma_0) s}\big)$.
	Similarly, the ignored terms for the differential equation of $\tilde w_0$ are of order \smash{$O\big(s {\rm e}^{(3-\gamma_0) s}\big)$} and order \smash{$O\big({\rm e}^{2 (\gamma_0+1)s}\big)$}.
	In the current numerical experiment, $\gamma_0=1$.
	Therefore, the difference between the asymptotic solution and the exact solution is of order $O\big(s {\rm e}^{2 s}\big)$.
	So, we should do high-precision numerical integration from $s=0$ to about $s=s_f=-120$
	since \smash{$120 \times {\rm e}^{2 \times (-120)} \approx 7.055 \times 10^{-103}$}.
	Similar to the general case of Section~\ref{GeneralCase},
	the values of $\tilde w_0$, $\frac{ {\rm d}\tilde w_0}{{\rm d}s}$, $\tilde w_1$ and $\frac{ {\rm d}\tilde w_1}{{\rm d}s}$ at $s=0$
	are obtained by formula~\eqref{ini-s0-General}.
	Then, numerically integrating \eqref{tu0tu1-E1} by the Gauss--Legendre method, the high-precision numerical solution is obtained.
	Comparing it with the more accurate numerical solution starting from~${r=55}$,
	the errors of the numerical solution are obtained.
Table~\ref{tab7} shows that the numerical solution is as accurate as we expected.
	The large relative error of $\frac{{\rm d}\tilde w_0}{{\rm d}s}$ is nothing but the fact that~${\frac{{\rm d}\tilde w_0}{{\rm d}s}|_{s=-120} \approx -1.29\times 10^{-102}}$ is small.

\begin{table}[ht]\renewcommand{\arraystretch}{1.2}\centering
		\caption{Errors of the numerical solution at $s=-120$ for the E1 case with $\gamma_0=1$.}\label{tab7}

\vspace{1mm}

		\begin{tabular}{ c|c c c c}
			\hline \hline
			$s=-120$ & $\tilde w_0$ & $\frac{{\rm d}\tilde w_0}{{\rm d}s}$ & $\tilde w_1$ &$\frac{{\rm d} \tilde w_1}{{\rm d}s}$\\
			\hline
 Absolute error &$1.06 \times 10^{-111}$ &$8.84 \times 10^{-114}$ &$3.56 \times 10^{-110}$&$5.94\times 10^{-112}$ \\
 Relative error &$1.35 \times 10^{-111}$ &$6.84 \times 10^{-12}$ &$6.50 \times 10^{-111}$&$7.12\times 10^{-110}$\\
		\end{tabular}
	\end{table}
	
	Table \ref{tab8} shows how good the asymptotic solution \eqref{Asymp-E1} is.

	\begin{table}[ht]\renewcommand{\arraystretch}{1.2}\centering
		\caption{Approximate derivation from the asymptotic solution for the E1 case with $\gamma_0=1$.}\label{tab8}

\vspace{1mm}

			\begin{tabular}{@{\,}c|c c c c c c@{\,}}
				\hline \hline
				$s$ & $-20$ & $-40$ & $-60$ &$-80$ &$-100$& $-120$\\
				\hline
				$\ln\big(a_{E1}-\tilde w_0\big)$ &$-37.0566$ &$-76.3821$&$-115.983$&$-155.698$&$-195.477$&$-235.296$ \\
				$\ln\big(\tilde w_1-\ln(-2 s+b_{E1})\big)$&$-37.0553$&$-76.3818$ &$-115.983$&$-155.698$&$-195.477$&$-235.296$
			\end{tabular}
	\end{table}
	
	Table \ref{tab8} not only numerically verifies the asymptotics of the E1 case for $\gamma_0=1$,
	but also confirms our estimate that $\big(\tilde w_0, \tilde w_1\big)$ differs with its asymptotic solution
	by an order of $O\big(s {\rm e}^{2 s}\big)$.

	\subsection{Case E2}
	In this case, $\gamma_0=-1$ and $-3<\gamma_1<1$.
	As explained in the beginning of Section \ref{sec-verify}, the fine asymptotics of the E2 case can be obtained from the E1 case.
	So we omit the numerical verification for this case.
	
	\subsection{Case E3}
	This subsection will verify numerically the fine asymptotics of the E3 case.
	Note that in this case $\gamma_1=\gamma_0-2$ and $-1<\gamma_0<3$.
	Also note that
	\[
		a_{E3}
		=\lim_{\gamma_1 \rightarrow \gamma_0-2} (\rho_0(\gamma_0,\gamma_1)+\rho_1(\gamma_0,\gamma_1)),
	\]
	where $\rho_0$ and $\rho_1$ are defined by \eqref{rhos-DEF}.
	
	Let us take $\gamma_0=\frac{1}{3}$ as an example to verify \eqref{Asymp-E3} numerically.
	Then $\big(s_1^\mathbb{R},s_2^\mathbb{R}\big)=(-2,-3)$.
	Similar to the general case of Section \ref{GeneralCase},
	the numerical integration is divided into two parts: for $r\in [1,45]$ and for $s\in [s_f,0]$.
	
	\subsubsection[Numerical integration from r=45 to r=1]{Numerical integration from $\boldsymbol{r=45}$ to $\boldsymbol{r=1}$}
	By the truncation of \eqref{iniApprox}, the initial values at $r=45$ are obtained (only the first $50$ digits are listed)
	\begin{gather}
			w_p(45)= 5.2843098725232974899221393911204991207504443469367\ldots \times 10^{-57},\nonumber \\
			w_p'(45)=-1.5004885502015739552694025310567337731833237644509\ldots \times 10^{-56},\nonumber\\
			w_m(45)=- 5.9853226243288533939288654231629507224228092956986\ldots \times 10^{-80},\nonumber\\
			w_m'(45)= 2.4007702224681605220078925077492026747788333343193\ldots \times 10^{-79}.
	\label{ini-E3}
	\end{gather}
	It is not surprising that $w_p(45)$ and $w_p'(45)$ of \eqref{ini-E3} coincide with that of \eqref{ini-E1} with many digits
	since \smash{$s_1^\mathbb{R}=-2$} in the example for this case and \smash{$s_1^\mathbb{R}=2$} in the example for the E1 case.
	
	Comparing with the more accurate solution starting from $r=55$,
	the errors of the initial values \eqref{ini-E3} are obtained as shown by Table~\ref{tab9}.
	
\begin{table}[ht]\renewcommand{\arraystretch}{1.2}\centering
		\caption{Errors of the initial values of case E3 with $\gamma_0=\frac{1}{3}$.}\label{tab9}

\vspace{1mm}

		\begin{tabular}{ c|c c c c}
			\hline \hline
			$r=45$& $w_p$ & $w_p'$ & $w_m$ &$w_m'$\\
			\hline
Absolute error &$3.04 \times 10^{-170}$ &$2.59 \times 10^{-169}$ &$4.86 \times 10^{-193}$&$4.71\times 10^{-192}$ \\
Relative error &$5.76\times 10^{-114}$ &$1.73 \times 10^{-113}$ &$8.12 \times 10^{-114}$&$1.96\times 10^{-113}$
		\end{tabular}
	\end{table}
	
	Numerically integrating \eqref{wpwm} from $r=45$ to $r=1$ by the Gauss--Legendre method with the same parameters as the ones in Section \ref{GeneralCase},
	the values of $w_p$, $w_p'$, $w_m$ and $w_m'$ at $r=1$ are obtained
	\begin{gather}
			w_p(1)= 3.8027004168653915145363303284447255846983739527888\ldots\times 10^{-2}, \nonumber\\
			w_p'(1)=-1.2469806975938122928142121636698878096900701362539\ldots\times 10^{-1},\nonumber\\
			w_m(1)= -1.0071686775204061495316019356342162460012952192431\ldots\times 10^{-2},\nonumber\\
			w_m'(1)= 4.3926896299159549125370306923225572137558540540015\ldots\times 10^{-2}.
\label{values-E3-1}
	\end{gather}
	Comparing with the more accurate solution starting from $r=55$,
	the errors of \eqref{values-E3-1} are obtained as shown by Table \ref{tab10}.
	\begin{table}[ht]\renewcommand{\arraystretch}{1.2}\centering
	\caption{Errors of the numerical solution at $r=1$ of case E3 with $\gamma_0=\frac{1}{3}$.}	\label{tab10}

\vspace{1mm}

		\begin{tabular}{ c|c c c c}
			\hline \hline
			$r=1$& $w_p$ & $w_p'$ & $w_m$ &$w_m'$\\
			\hline
 Absolute error &$4.37 \times 10^{-115}$ &$1.42 \times 10^{-114}$ &$1.35 \times 10^{-115}$&$5.76\times 10^{-115}$ \\
Relative error &$1.15\times 10^{-113}$ &$1.14 \times 10^{-113}$ &$1.34 \times 10^{-113}$&$1.31\times 10^{-113}$
		\end{tabular}
	\end{table}
	
\subsubsection[Near r=0]{Near $\boldsymbol{r=0}$}
	Near $r=0$, we still use the transformation \eqref{transform-general}.
	So the differential equations for $\tilde w_0$ and $\tilde w_1$ are also~\eqref{tu0tu1}.
	
	We expect $\tilde w_0$ and $\tilde w_1$ are of order $o(s)$.
	Also considering $-1<\gamma_0<3$ and $\gamma_1=\gamma_0-2$, we get the associated truncation of \eqref{tu0tu1} near
	$s=-\infty$ for the E3 case:
	\begin{gather}
			\frac{1}{4} \frac{ {\rm d}^2\tilde w_0^{(0)}}{{\rm d}s^2}=-{\rm e}^{\tilde w_1^{(0)}-\tilde w_0^{(0)}},\qquad
			\frac{1}{4} \frac{ {\rm d}^2\tilde w_1^{(0)}}{{\rm d}s^2}={\rm e}^{\tilde w_1^{(0)}-\tilde w_0^{(0)}}.
		\label{tu0tu1-trunc-E3}
	\end{gather}
	The solution of \eqref{tu0tu1-trunc-E3} is
	\[%\label{SOL-trunc-E3}
			\tilde w_0^{(0)}+\tilde w_1^{(0)}=k_{0E3}+k_{1E3} s,\qquad
			\tilde w_1^{(0)}-\tilde w_0^{(0)}=\ln \left(- \frac{k_{2E3}^2}{8 \pm 8 \cosh(k_{2E3} (s+k_{3E3}))} \right).
	\]
	Because we expect $\tilde w_0$ and $\tilde w_1$ are of order $o(s)$,
	we should take $k_{1E3}=0$ and $k_{3E3} \rightarrow 0$ or else~$\tilde w_0$ and $\tilde w_1$ will be of order $O(s)$.
	So the consistent solution of \eqref{tu0tu1-trunc-E3} is
	\[%\label{SOL-trunc-E3-0}
			\tilde w_0^{(0)}+\tilde w_1^{(0)}=k_{0E3},\qquad
			\tilde w_1^{(0)}-\tilde w_0^{(0)}=-\ln \big( 4 (s+k_{3E3})^2 \big).
	\]
	By \eqref{Asymp-E3} and \eqref{transform-general},
	we know that the fine asymptotics of the E3 case is fixed by $k_{0E3}=a_{E3}$ and~${k_{3E2}=b_{E3}}$.
	
	In the truncation from \eqref{tu0tu1} to \eqref{tu0tu1-trunc-E3},
	the ignored terms for the differential equation of~${\tilde w_0\!+\!\tilde w_1}$
	are ${\rm e}^{2 \tilde w_0+2 (\gamma_0+1)s}$ and ${\rm e}^{-2 \tilde w_1+2 (1-\gamma_1)s}$,
	which are of order $O\big(s^2 {\rm e}^{2(\gamma_0+1) s}\big)$ and order $O\big(s^{-2} {\rm e}^{2(\gamma_0+1) s}\big)$.
	Similarly, the ignored terms for the differential equation of $\tilde w_1$ are also
	of order \smash{$O\big(s^{-2} {\rm e}^{2(3-\gamma_0) s}\big)$} and order \smash{$O\big(s^{-2} {\rm e}^{2(\gamma_0+1) s}\big)$}.
	In the current numerical experiment, $\gamma_0=\frac{1}{3}$.
	Therefore, the difference between the asymptotic solution and the exact solution is of order \smash{$O\big(s^2 {\rm e}^{\frac{8}{3} s}\big)$}.
	So, we should do high-precision numerical integration from $s=0$ to about $s=s_f=-90$
	since \smash{$90^2 \times {\rm e}^{\frac{8}{3} \times (-90)} \approx 4.76 \times 10^{-101}$}.
	Just as the general case,
	the values of $\tilde w_0$, $\frac{ {\rm d}\tilde w_0}{{\rm d}s}$, $\tilde w_1$ and $\frac{ {\rm d}\tilde w_1}{{\rm d}s}$ at $s=0$
	are obtained by formula \eqref{ini-s0-General}.
	Then, the high-precision numerical solution is obtained by numerically integrating \eqref{tu0tu1} by the Gauss--Legendre method.
	Comparing it with the more accurate numerical solution starting from $r=55$,
	the errors of the numerical solution are obtained.
	
	Table \ref{tab11} shows that the numerical solution is as accurate as we expected.
\begin{table}[ht]\renewcommand{\arraystretch}{1.2}\centering
		\caption{Errors of the numerical solution at $s=-90$ for the E3 case with $\gamma_0=\frac{1}{3}$.}\label{tab11}

\vspace{1mm}

		\begin{tabular}{ c|c c c c}
			\hline \hline
			$s=-90$ & $\tilde w_0$ & $\frac{{\rm d}\tilde w_0}{{\rm d}s}$ & $\tilde w_1$ &$\frac{{\rm d} \tilde w_1}{{\rm d}s}$\\
			\hline
 Absolute error &$1.30 \times 10^{-110}$ &$2.95 \times 10^{-112}$ &$1.41 \times 10^{-110}$&$3.08\times 10^{-112}$ \\
Relative error &$2.74 \times 10^{-111}$ &$2.66 \times 10^{-110}$ &$2.51 \times 10^{-111}$&$2.77\times 10^{-110}$
		\end{tabular}
	\end{table}
	
	Table \ref{tab12} shows how good the asymptotic solution \eqref{Asymp-E3} is.
	\begin{table}[ht]\renewcommand{\arraystretch}{1.2}\centering
		\caption{Approximate derivation from the asymptotic solution for the E3 case with $\gamma_0=\frac{1}{3}$.}\label{tab12}

\vspace{1mm}

\begin{tabular}{@{\,\,}c@{\,\,}|@{\,\,}c@{\,\,}c@{\,\,}c@{\,\,}c@{\,\,}c@{\,\,}c@{\,\,}}
				\hline \hline
				$s$ & $-15$ & $-30$ & $-45$ &$-60$ &$-75$& $-90$\\
				\hline
				$\ln\big(\tilde w_0+\tilde w_1-a_{E3}\big)$ &$-34.5568$ &$-73.2186$&$-112.424$&$-151.857$&$-191.415$&$-231.054$ \\
				$\ln\big(\tilde w_0-\tilde w_1-\ln\big(4 (s+b_{E3})^2\big) \big)$ &$-34.5556$ &$-73.2183$ &$-112.424$&$-151.857$&$-191.415$&$-231.054$
			\end{tabular}
	\end{table}
	
	Table \ref{tab12} not only numerically verifies the asymptotics of the E3 case for $\gamma_0=\frac{1}{3}$,
	but also confirms our estimate that $\tilde w_0+\tilde w_1$ and $\tilde w_1-\tilde w_0$ deviate from their asymptotics
	by an order of~$O\big(s^2 {\rm e}^{\frac{8}{3} s}\big)$.
	More detailed analysis shows that $\tilde w_0$ and $\tilde w_1$ deviate from their asymptotics
	by an order of \smash{$O\big(s^2 {\rm e}^{\frac{8}{3} s}\big)$} and an order of \smash{$O\big({\rm e}^{\frac{8}{3} s}\big)$}, respectively.

	\subsection{Case V1 \label{CaseV1}}
	This subsection is devoted to the verification of the fine asymptotics of the V1 case. Note that~${\gamma_0=3}$ and $\gamma_1=1$ in this case.
	
	$\big(s_1^\mathbb{R},s_2^\mathbb{R}\big) =(4,-6)$ by \eqref{ConnectFormula}.
	Similar to the general case of Section \ref{GeneralCase},
	the numerical integration is done on two intervals: $r\in [1,45]$ and $s\in [s_f,0]$.
	
	\subsubsection[Numerical integration from r=45 to r=1]{Numerical integration from $\boldsymbol{r=45}$ to $\boldsymbol{r=1}$}
	By the truncation of \eqref{iniApprox}, the initial values at $r=45$ are obtained (only the first $50$ digits are listed)
	\begin{gather}
			w_p(45)= -1.0568619745046594979844278782240998241500888693873\ldots \times 10^{-56},\nonumber\\
			w_p'(45)= 3.0009771004031479105388050621134675463666475289019\ldots \times 10^{-56},\nonumber\\
			w_m(45)= -1.1970645248657706787857730846325898673892151885992\ldots \times 10^{-79},\nonumber\\
			w_m'(45)= 4.8015404449363210440157850154984037759705748926074\ldots \times 10^{-79}.
\label{ini-V1}
	\end{gather}
	
	Comparing with the more accurate solution starting from $r=55$,
	the errors of the initial values \eqref{ini-V1} are obtained as shown by Table~\ref{tab13}.

	\begin{table}[ht]\renewcommand{\arraystretch}{1.2}\centering

		\caption{Errors of the initial values of case V1.}\label{tab13}

\vspace{1mm}

		\begin{tabular}{ c|c c c c}
			\hline \hline
			$r=45$& $w_p$ & $w_p'$ & $w_m$ &$w_m'$\\
			\hline
Absolute error &$2.43 \times 10^{-169}$ &$2.07 \times 10^{-168}$ &$3.89 \times 10^{-192}$&$3.77\times 10^{-191}$ \\
Relative error &$2.30\times 10^{-113}$ &$6.91 \times 10^{-113}$ &$3.25 \times 10^{-113}$&$7.85\times 10^{-113}$
		\end{tabular}
	\end{table}

	Numerically integrating \eqref{wpwm} from $r=45$ to $r=1$ by the Gauss--Legendre method with the same parameters as the ones in Section \ref{GeneralCase},
	the values of $w_p$, $w_p'$, $w_m$ and $w_m'$ at $r=1$ are obtained
	\begin{gather}
			w_p(1)= -7.5811708202722819337886291345224915096864160866088\ldots\times 10^{-2}, \nonumber\\
			w_p'(1)= 2.4764894905832982616275785124301997778251205645956\ldots\times 10^{-1},\nonumber\\
			w_m(1)= -1.8985818420083245736824441481547286887104902789335\ldots\times 10^{-2},\nonumber\\
			w_m'(1)= 8.0472024534463364925338502074404317836916130555680\ldots\times 10^{-2}.
	\label{values-V1-1}
	\end{gather}
	Comparing with the more accurate solution starting from $r=55$,
	the errors of \eqref{values-V1-1} are obtained as shown by Table~\ref{tab14}.

	\begin{table}[ht]\renewcommand{\arraystretch}{1.2}\centering
		\caption{Errors of the numerical solution at $r=1$ of case V1.}\label{tab14}

\vspace{1mm}

		\begin{tabular}{ c|c c c c}
			\hline \hline
			$r=1$& $w_p$ & $w_p'$ & $w_m$ &$w_m'$\\
			\hline
Absolute error &$3.47 \times 10^{-114}$ &$1.12 \times 10^{-113}$ &$9.71 \times 10^{-115}$&$3.94\times 10^{-114}$ \\
Relative error &$4.58\times 10^{-113}$ &$4.54 \times 10^{-113}$ &$5.11 \times 10^{-113}$&$4.89\times 10^{-113}$
		\end{tabular}
	\end{table}

	\subsubsection[Near r=0]{Near $\boldsymbol{r=0}$}
	Near $r=0$, the transformation is still \eqref{transform-general}.
	Hence, the differential equations for $\tilde w_0$ and $\tilde w_1$ are also \eqref{tu0tu1}.
	
	Now, $(\gamma_0,\gamma_1)=(3,1)$ and the expected $\tilde w_0$ and $\tilde w_1$ are of order $o(s)$.
	So the associated truncation of \eqref{tu0tu1} near $s=-\infty$ for the V1 case is
	\begin{gather}
			\frac{1}{4} \frac{ {\rm d}^2\tilde w_0^{(0)}}{{\rm d}s^2}=-{\rm e}^{\tilde w_1^{(0)}-\tilde w_0^{(0)}},\qquad
			\frac{1}{4} \frac{ {\rm d}^2\tilde w_1^{(0)}}{{\rm d}s^2}={\rm e}^{\tilde w_1^{(0)}-\tilde w_0^{(0)}}-{\rm e}^{-2 \tilde w_1^{(0)}}.
\label{tu0tu1-trunc-V1}
	\end{gather}
	Let $\tilde w_p^{(0)}= \tilde{w}_0^{(0)}+ \tilde{w}_1^{(0)}$.
	Then, we have
	\begin{gather}
			\frac{1}{4} \frac{{\rm d}^2\tilde w_0^{(0)}}{{\rm d}s^2}=-{\rm e}^{\tilde w_p^{(0)}-2 \tilde w_0^{(0)}},\qquad
			\frac{1}{4} \frac{{\rm d}^2\tilde w_p^{(0)}}{{\rm d}s^2}=-{\rm e}^{-2 \tilde w_p^{(0)}+2 \tilde w_0^{(0)}}.
\label{tu0tu1-trunc1-V1}
	\end{gather}
	Unlike the cases discussed before,
	we have not achieved the general solution of \eqref{tu0tu1-trunc1-V1}.
	Anyhow, equation \eqref{tu0tu1-trunc1-V1} itself deserves an independent investigation. Let us leave it as a future work.
	Surprisingly, a two parameter family of explicit solutions of \eqref{tu0tu1-trunc1-V1} can be constructed
	and the asymptotic solution near $r=0$ is just among them!
	By the hint of the asymptotic solution and for the convenience of comparison,
	we seek the solutions of \eqref{tu0tu1-trunc1-V1} of the form
	\begin{gather}
			\tilde w_0^{(0)}= \ln \big( \tilde{a}_3 (s-\ln 4)^3+\tilde{a}_2 (s-\ln 4)^2+\tilde{a}_1 (s-\ln 4)+\tilde{a}_0 \big),\nonumber\\
			\tilde w_p^{(0)}= \ln \big(\tilde{b}_4 (s-\ln 4)^4+ \tilde{b}_3 (s-\ln 4)^3+\tilde{b}_2 (s-\ln 4)^2+\tilde{b}_1 (s-\ln 4)+\tilde{b}_0 \big).
 \label{formAnsaz}
	\end{gather}
	
	There are only two sets of solutions that has form \eqref{formAnsaz}.
	
	Set A:
	\begin{gather*}
		\tilde{a}_3=\frac{4}{3}, \qquad \tilde{b}_4=\frac{4}{3},\qquad
		\tilde{a}_1=\frac{1}{4}\tilde{a}_2^2, \qquad \tilde{b}_3=\frac{4}{3} \tilde{a}_2, \qquad \tilde{b}_2=\frac{1}{2} \tilde{a}_2^2,\\
		\tilde{b}_1=\frac{1}{8}\big(\tilde{a}_2^3-16 \tilde{a}_0\big), \qquad \tilde{b}_0=\frac{1}{64} \big(\tilde{a}_2^4-32 \tilde{a}_0 \tilde{a}_2\big).
	\end{gather*}
	
	Set B:
	\begin{gather*}
		\tilde{a}_3=-\frac{4}{3}, \qquad \tilde{b}_4=\frac{4}{3},\qquad
		\tilde{a}_1=-\frac{1}{4}\tilde{a}_2^2, \qquad \tilde{b}_3=-\frac{4}{3} \tilde{a}_2, \qquad \tilde{b}_2=\frac{1}{2} \tilde{a}_2^2,\\
		\tilde{b}_1=\frac{1}{8}\big(16 \tilde{a}_0-\tilde{a}_2^3\big), \qquad \tilde{b}_0=\frac{1}{64} \big(\tilde{a}_2^4-32 \tilde{a}_0 \tilde{a}_2\big).
	\end{gather*}
	
	The fine asymptotic solution of the V1 case is in Set B with
	\begin{gather*}
		\tilde{a}_2=-4 \gamma_{{\rm Eu}}, \qquad
		\tilde{a}_0=-\frac{1}{24} \zeta(3)-\frac{4}{3} \gamma_{{\rm Eu}}^3.
	\end{gather*}
	
	The error of the truncation from \eqref{tu0tu1} to \eqref{tu0tu1-trunc-V1}
	is caused by the term ${\rm e}^{2 \tilde w_0 +8 s}$, which is of order $O\big(s^6 {\rm e}^{8s}\big)$.
	So we set $s_f=-32$ since $(-32)^6 {\rm e}^{8\times (-32)} \approx 7.1 \times 10^{-103}$ has been smaller than~$10^{-100}$.
	
	The high-precision numerical solution is obtained by numerically integrating~\eqref{tu0tu1} by the Gauss--Legendre method.
	Comparing it with the more accurate numerical solution starting from~${r=55}$,
	the errors of the numerical solution are obtained.
Table \ref{tab15} shows that our numerical solution is as accurate as we expected.

	\begin{table}[ht]\renewcommand{\arraystretch}{1.2}\centering
	\caption{Errors of the numerical solution at $s=-32$ for the V1 case.}	\label{tab15}

\vspace{1mm}

		\begin{tabular}{ c|c c c c}
			\hline \hline
			$s=-32$ & $\tilde w_0$ & $\frac{{\rm d}\tilde w_0}{{\rm d}s}$ & $\tilde w_1$ &$\frac{{\rm d} \tilde w_1}{{\rm d}s}$\\
			\hline
Absolute error &$1.14 \times 10^{-109}$ &$1.49 \times 10^{-110}$ &$3.94 \times 10^{-109}$&$4.77\times 10^{-110}$ \\
Relative error &$1.06 \times 10^{-110}$ &$1.62 \times 10^{-109}$ &$1.13 \times 10^{-109}$&$1.57\times 10^{-108}$
		\end{tabular}
	\end{table}

	Table \ref{tab16} shows how good the asymptotic solution \eqref{Asymp-V1} is.

	Table \ref{tab16} not only numerically verifies the asymptotics of the V1 case,
	but also confirms our estimate that $\tilde w_0$ and $\tilde w_1+\tilde w_0$ differ from their asymptotics
	by an order of $O\big(s^6 {\rm e}^{8 s}\big)$.

	\begin{table}[ht]\renewcommand{\arraystretch}{1.2}	\centering
\caption{Approximate derivation from the asymptotic solution for the V1 case.}		\label{tab16}

\vspace{1mm}

		\begin{tabular}{ c| c c c c c c}
			\hline \hline
			$s$ & $-7$ & $-12$ &$-17$ &$-22$& $-27$ & $-32$\\
			\hline
			$\ln\big(\tilde w_0-\ln(P_3)\big)$ &$-45.6682$&$-82.7772$&$-120.834$&$-159.368$&$-198.191$
			&$-237.207$ \\
			$\ln\big(\tilde w_0+\tilde w_1-\ln (P4) \big)$
			&$-45.6691$ &$-82.7775$&$-120.834$&$-159.368$&$-198.191$ & $-237.207$
		\end{tabular}
	\end{table}

	\subsection{Case V2}
	In this case $(\gamma_0,\gamma_1)=(-1,1)$.
	
	By the connection formula \eqref{ConnectFormula}, we have
	$\big(s_1^\mathbb{R}, s_2^\mathbb{R}\big)=(0,2)$.
	$s_1^\mathbb{R}=0$ means $w_1=-w_0$ at~${r=\infty}$.
	This leads to $w_1 \equiv -w_0$ for $r \in (0 ,\infty)$, considering that they satisfy \eqref{TT-2}.
	
	Let $w=w_0=-w_1$.
	Then, the differential equation for $w$ is
	\[
		\frac{1}{2} \left(\frac{{\rm d}^2}{{\rm d}r^2}+ \frac{1}{r}\frac{\rm d}{{\rm d}r}\right) w={\rm e}^{4 w}-{\rm e}^{-4 w},
	\]
	which is the radical reduction of the sinh-Gordon equation.
	Both the associated truncation
	\[%\label{w-trunc-V2}
		\frac{1}{4} \frac{ {\rm d}^2\tilde w^{(0)}}{{\rm d}s^2}= {\rm e}^{2 \tilde w^{(0)}}
	\]
	and the numerical experiments show that $2 w(r)$ differs from its asymptotics by an order of $O\big(s^2 {\rm e}^{4 s}\big)$ near $r=0$.
	
	\subsection{Case V3}
	In this case, $(\gamma_0,\gamma_1)=(-1,-3)$.
	Thus, $\big(s_1^\mathbb{R},s_2^\mathbb{R}\big)=(-4,-6)$ by \eqref{ConnectFormula}.
	As explained in the beginning of Section \ref{sec-verify},
	the fine asymptotics of the V3 case can be obtained from the V1 case.
	So we omit the verification.

	\section[Out of the curved triangle: generalizing the connection formula and the fine asymptotics]{Out of the curved triangle: generalizing\\ the connection formula and the fine asymptotics} \label{CONJ}
	First, let us divide the real plane of $\big(s_1^\mathbb{R}, s_2^\mathbb{R}\big)$ into $19$ parts:
	regions $\Omega_0$, $\Omega_1$, $\Omega_2$, $\Omega_3$, $\Omega_4$, $\Omega_5$,~$\Omega_6$;
	edges $E1$, $E2$, $E3$, $E_1^U$, $E_2^U$, $E_1^D$, $E_2^D$, $E_3^R$, $E_3^L$;
	and vertices V1, V2, V3. See Figure \ref{fig-3} for details.
	Note that the boundaries of $\Omega_i$ are
	line $s_2^\mathbb{R}=2 s_1^\mathbb{R}+2$, line $s_2^\mathbb{R}=-2 s_1^\mathbb{R}+2$
	and parabola~\smash{$s_2^\mathbb{R}=-\frac{1}{4} \big( s_1^\mathbb{R} \big)^2 -2$}.
	
	\input{Figure3.tikz}
	
	By the connection formula \eqref{ConnectFormula} (see also Figure \ref{fig-2}), on the Stokes data side,
	the solutions studied in Theorem \ref{thm-GIL-1} are those parameterized by the point in the region $\Omega_0$,
	on the edges E1, E2, E3, and the vertices V1, V2, V3.
	These solutions are all smooth for $r \in (0,\infty)$.
	Consider the case where $\big(s_1^\mathbb{R}, s_2^\mathbb{R}\big)$ lies outside the curved triangle.
	Then the corresponding $w_0(r)$, $w_1(r)$, or both must evolve to a singularity
	somewhere as $r$ decreases from $r=\infty$.
	Numerical experiments show that there is a cut around every singularity.
	But we have evidence that these singularities and cuts are artificial:
	they can be avoided by choosing appropriate variables.
	For example, if we use variables $v_0={\rm e}^{2 w_0}$ and $v_1={\rm e}^{2 w_1}$,
	then $v_0$ and $v_1$ will have no cuts for $r>0$.
	$v_0$ or $v_1$ may still have singularities,
	i.e., in general, $v_0$ and $v_1$ are not the final smooth variables.
	Fortunately, we were able to find two smooth variables for each part of Figure \ref{fig-3},
	see Conjecture~\ref{Conj}.
	From this point of view, Theorem \ref{thm-GIL-1} studies only those solutions
	that have ``positivity" property so that they are still real after taking logarithm.
	
	\subsection{The conjecture}
	
	The fine asymptotics for the cases of $\Omega_0$, E1, E2, E3, V1, V2 and V3
	have been rigorously proved in \cite{GIL-3} and numerically verified in Section \ref{sec-verify}.
	So the following conjecture only deals with the other remaining $12$ cases:
	$\Omega_1$, $\Omega_2$, $\Omega_3$, $\Omega_4$, $\Omega_5$, $\Omega_6$, $E_1^U$, $E_2^U$, $E_1^D$, $E_2^D$, $E_3^R$ and $E_3^L$.
	Similar to the explanation at the beginning of Section \ref{sec-verify}, the formulas of $\Omega_3$, $\Omega_4$, $E_1^U$,
	$E_2^D$ and $E_3^L$ are symmetrical to those of $\Omega_1$, $\Omega_6$, $E_2^U$,
	$E_1^D$ and $E_3^R$, respectively.
	But for convenience, we will list all formulas for the $12$ cases.
	
	\begin{Conjecture} \label{Conj}
		Let the inverse of connection formula \eqref{ConnectFormula} be
		\begin{gather}
				\gamma_0=\frac{4}{\pi}\arccos\left(-\frac{1}{4}s_1^\mathbb{R}+\frac{1}{4}\sqrt{8+\big(s_1^\mathbb{R}\big)^2+4 s_2^\mathbb{R}} \right)-1,\nonumber\\
				\gamma_1=\frac{4}{\pi}\arccos\left(-\frac{1}{4}s_1^\mathbb{R}-\frac{1}{4}\sqrt{8+\big(s_1^\mathbb{R}\big)^2+4 s_2^\mathbb{R}} \right)-3,
	\label{CF-Inverse}
		\end{gather}
		where the values of the $\arccos$ terms may be complex and, if multivalued, should be given by their principal values.
		Suppose that $w_0(r)$ and $w_1(r)$ are the solutions of system~\eqref{TT-2} with asymptotics~\eqref{InfAsymp} at $r=\infty$
		but may have singularities for $r \in (0, \infty)$.
		Given $\big(s_1^{\mathbb{R}}, s_2^{\mathbb{R}}\big)$, one can calculate $(\gamma_0, \gamma_1)$ from \eqref{CF-Inverse} and $(\rho_0, \rho_1)$ from \eqref{rhos-DEF}.
		We set $s=\ln (r)$,	 ${\gamma_i^\mathbb{R}=\operatorname{Re}(\gamma_i)}$, ${\gamma_i^\mathbb{I}=\operatorname{Im}(\gamma_i)}$, ${\rho_i^\mathbb{R}=\operatorname{Re}(\rho_i)}$,
		and $\rho_i^\mathbb{I}=\operatorname{Im}(\rho_i)$, where $i=0,1$.
		Then, the characteristics of the solution parameterized by a point in region $\Omega_i$, $i=1,\dots,6$, are the following.
		\begin{itemize}\itemsep=0pt
			\item[$\Omega_1\colon$] \smash{$\sqrt{8+\big(s_1^\mathbb{R}\big)^2+4 s_2^\mathbb{R}} \in \mathbb{R}$},
			$\gamma_0 \in \mathbb{R}$, $\gamma_1 \not\in \mathbb{R}$.
			${\rm e}^{2 w_0(r)}$ and ${\rm e}^{2 w_1(r)}$ are smooth for $r \in (0,\infty)$.
			Their asymptotics at $s=-\infty$ are
			\[
				{\rm e}^{2 w_0} \xlongrightarrow{ s \rightarrow -\infty} {\rm e}^{\gamma_0 s+\rho_0}, \qquad
				{\rm e}^{2 w_1} \xlongrightarrow{ s \rightarrow -\infty} 2 \operatorname{Re} \big( {\rm e}^{\gamma_1 s+\rho_1} \big).
			\]
			\item[$\Omega_2\colon$] \smash{$\sqrt{8+\big(s_1^\mathbb{R}\big)^2+4 s_2^\mathbb{R}} \in \mathbb{R}$},
			$\gamma_0 \not\in \mathbb{R}$, $\gamma_1 \not\in \mathbb{R}$.
			${\rm e}^{-2 w_0(r)}$ and ${\rm e}^{2 w_1(r)}$ are smooth for $r \in (0,\infty)$.
			Their asymptotics at $s=-\infty$ are
			\[
				{\rm e}^{-2 w_0} \xlongrightarrow{ s \rightarrow -\infty} 2 \operatorname{Re} \big( {\rm e}^{-\gamma_0 s-\rho_0} \big),\qquad
				{\rm e}^{2 w_1} \xlongrightarrow{ s \rightarrow -\infty} 2 \operatorname{Re} \big( {\rm e}^{\gamma_1 s+\rho_1} \big).
			\]
			\item[$\Omega_3\colon$] \smash{$\sqrt{8+\big(s_1^\mathbb{R}\big)^2+4 s_2^\mathbb{R}} \in \mathbb{R}$},
			$\gamma_0 \not\in \mathbb{R}$, $\gamma_1 \in \mathbb{R}$.
			${\rm e}^{-2 w_0(r)}$ and ${\rm e}^{-2 w_1(r)}$ are smooth for $r \in (0,\infty)$.
			Their asymptotics at $s=-\infty$ are
			\[
				{\rm e}^{-2 w_0} \xlongrightarrow{ s \rightarrow -\infty} 2 \operatorname{Re} \big( {\rm e}^{-\gamma_0 s-\rho_0} \big) ,\qquad
				{\rm e}^{-2 w_1} \xlongrightarrow{ s \rightarrow -\infty} {\rm e}^{-\gamma_1 s-\rho_1}.
			\]
			\item[$\Omega_4\colon$] \smash{$\sqrt{8+\big(s_1^\mathbb{R}\big)^2+4 s_2^\mathbb{R}} \in \mathbb{R}$},
			$\gamma_0 \not\in \mathbb{R}$, $\gamma_1 \not\in \mathbb{R}$.
			${\rm e}^{-2w_1(r)}$ and ${\rm e}^{-2w_0(r)-2w_1(r)}$ are smooth for $r \in (0,\infty)$.
			Their asymptotics at $s=-\infty$ are
			\begin{gather*}
				{\rm e}^{-2w_1} \xlongrightarrow{ s \rightarrow -\infty} {\rm e}^{-\gamma_1^{\mathbb{R}} s} \left( \frac{8 {\rm e}^{-\rho_0^{\mathbb{R}}}}{\big(\gamma_0^\mathbb{I}-\gamma_1^\mathbb{I}\big)^2}
				\cos \big(\gamma_0^{\mathbb{I} } s+\rho_0^{\mathbb{I}} \big)
				+2 {\rm e}^{-\rho_1^{\mathbb{R}}} \cos\big( \gamma_1^{\mathbb{I}}s+\rho_1^{\mathbb{I}}\big) \right),\\
				{\rm e}^{-2w_0-2w_1} \xlongrightarrow{ s \rightarrow -\infty} {\rm e}^{-(\gamma_0^\mathbb{R}+\gamma_1^\mathbb{R})s}
				\Biggl\{ 2{\rm e}^{-\rho_0^\mathbb{R}-\rho_1^\mathbb{R}} \frac{\big(\gamma_0^\mathbb{I}+\gamma_1^\mathbb{I}\big)^2}{\big(\gamma_0^\mathbb{I}-\gamma_1^\mathbb{I}\big)^2}
				\cos\big( \big(\gamma_0^\mathbb{I}-\gamma_1^\mathbb{I}\big)s+\rho_0^\mathbb{I}-\rho_1^\mathbb{I} \big)  \\
				 \phantom{{\rm e}^{-2w_0-2w_1} \xlongrightarrow{ s \rightarrow -\infty} }{} +\frac{16 {\rm e}^{-2 \rho_0^\mathbb{R}} \big(\gamma_0^\mathbb{I}\big)^2}{\big(\gamma_0^\mathbb{I}-\gamma_1^\mathbb{I}\big)^4}
				+{\rm e}^{-2 \rho_1^\mathbb{R}} \big(\gamma_1^\mathbb{I}\big)^2 \\
				 \phantom{{\rm e}^{-2w_0-2w_1} \xlongrightarrow{ s \rightarrow -\infty}}{}+2 {\rm e}^{-\rho_0^\mathbb{R}-\rho_1^\mathbb{R}} \cos\big( \big(\gamma_0^\mathbb{R}+\gamma_1^\mathbb{R}\big)s+\rho_0^\mathbb{I}+\rho_1^\mathbb{I}\big) \Biggr\}.
			\end{gather*}
			\item[$\Omega_5\colon $] \smash{$\sqrt{8+\big(s_1^\mathbb{R}\big)^2+4 s_2^\mathbb{R}} \not\in \mathbb{R}$},
			$\gamma_0 \not\in \mathbb{R}$, $\gamma_1 \not\in \mathbb{R}$.
			${\rm e}^{2 w_0(r)}$ and ${\rm e}^{-2 w_1(r)}$ are smooth for $r \in (0,\infty)$.
			Their asymptotics at $s=-\infty$ are
			\[
				{\rm e}^{2 w_0} \xlongrightarrow{ s \rightarrow -\infty}
				2 \operatorname{Re} \big( {\rm e}^{\gamma_0 s+\rho_0} \big), \qquad
				{\rm e}^{-2 w_1} \xlongrightarrow{ s \rightarrow -\infty} 2 \operatorname{Re} \big( {\rm e}^{-\gamma_1 s-\rho_1} \big).
			\]
			\item[$\Omega_6\colon$] \smash{$\sqrt{8+\big(s_1^\mathbb{R}\big)^2+4 s_2^\mathbb{R}} \in \mathbb{R}$},
			$\gamma_0 \not\in \mathbb{R}$, $\gamma_1 \not\in \mathbb{R}$.
			${\rm e}^{2w_0(r)}$ and ${\rm e}^{2w_0(r)+2w_1(r)}$ are smooth for $r \in (0,\infty)$.
			Their asymptotics at $s=-\infty$ are
			\begin{gather*}
				{\rm e}^{2w_0} \xlongrightarrow{ s \rightarrow -\infty} {\rm e}^{\gamma_0^{\mathbb{R}} s} \left( \frac{8 {\rm e}^{\rho_1^{\mathbb{R}}}}{\big(\gamma_0^\mathbb{I}-\gamma_1^\mathbb{I}\big)^2}
				\cos \big(\gamma_1^{\mathbb{I} } s+\rho_1^{\mathbb{I}} \big)
				+2 {\rm e}^{\rho_0^{\mathbb{R}}} \cos\big( \gamma_0^{\mathbb{I}}s+\rho_0^{\mathbb{I}}\big) \right) ,\\
				{\rm e}^{2w_0+2w_1} \xlongrightarrow{ s \rightarrow -\infty} {\rm e}^{(\gamma_0^\mathbb{R}+\gamma_1^\mathbb{R})s}
				\Biggl\{ 2{\rm e}^{\rho_0^\mathbb{R}+\rho_1^\mathbb{R}} \frac{\big(\gamma_0^\mathbb{I}+\gamma_1^\mathbb{I}\big)^2}{\big(\gamma_0^\mathbb{I}-\gamma_1^\mathbb{I}\big)^2}
				\cos\big( \big(\gamma_0^\mathbb{I}-\gamma_1^\mathbb{I}\big)s+\rho_0^\mathbb{I}-\rho_1^\mathbb{I} \big)  \\
			  \phantom{	{\rm e}^{2w_0+2w_1} \xlongrightarrow{ s \rightarrow -\infty}}{}+ \frac{16 {\rm e}^{2 \rho_1^\mathbb{R}} \big(\gamma_1^\mathbb{I}\big)^2}{\big(\gamma_0^\mathbb{I}-\gamma_1^\mathbb{I}\big)^4}
				+{\rm e}^{2 \rho_0^\mathbb{R}} \big(\gamma_0^\mathbb{I}\big)^2 \\
 \phantom{	{\rm e}^{2w_0+2w_1} \xlongrightarrow{ s \rightarrow -\infty}}{}
				+2{\rm e}^{\rho_0^\mathbb{R}+\rho_1^\mathbb{R}} \cos\big( \big(\gamma_0^\mathbb{R}+\gamma_1^\mathbb{R}\big)s+\rho_0^\mathbb{I}+\rho_1^\mathbb{I}\big) \Biggr\}.
			\end{gather*}
		\end{itemize}
		
		On the edges, $ 8+\big(s_1^\mathbb{R}\big)^2+4 s_2^\mathbb{R} $ is always non-negative.
		Define
		\begin{gather*}
			 b_1= \frac{1}{2} \psi\left(\frac{3-\gamma_0}{8}\right)+\frac{1}{2} \psi\left(\frac{5+\gamma_0}{8}\right) -\gamma_{{\rm Eu}} +4 \ln 2,\\
			b_2= \frac{1}{2} \psi\left(\frac{3+\gamma_1}{8}\right)+\frac{1}{2} \psi\left(\frac{5-\gamma_1}{8}\right) -\gamma_{{\rm Eu}} +4 \ln 2,\\
			 b_3=-\frac{1}{4} \psi\left(\frac{3-\gamma_0}{4}\right) -\frac{1}{4}\psi\left(\frac{\gamma_0-3}{4}\right)
			+\frac{1}{3-\gamma_0}-2 \ln 2+\frac{\gamma_{{\rm Eu}} }{2}.
		\end{gather*}
		Then, the characteristics of the solution parameterized by a point on an edge are the following.
		\begin{itemize}\itemsep=0pt
			\item[$E_1^U\colon $]$\gamma_0 \not\in \mathbb{R}$, $\gamma_1 =1$, $\gamma_0^\mathbb{R}=-1 $, $\rho_0 \not\in \mathbb{R}$
			and $\rho_1$ is not defined.
			${\rm e}^{-2w_0(r)}$ and ${\rm e}^{2w_1(r)}$ are smooth for $r \in (0,\infty)$.
			Their asymptotics at $s=-\infty$ are
			\[
				{\rm e}^{-2 w_0} \xlongrightarrow{ s \rightarrow -\infty} 2 \operatorname{Re} \big( {\rm e}^{-\gamma_0 s-\rho_0} \big), \qquad
				{\rm e}^{2w_1(r)} \xlongrightarrow{ s \rightarrow -\infty} -2 s+ b_1.
			\]
			\item[$E_2^U\colon$]$\gamma_0 =-1$, $\gamma_1 \not\in \mathbb{R}$, $\gamma_1^\mathbb{R}=1$, $\rho_1 \not\in \mathbb{R}$
			and $\rho_0$ is not defined.
			${\rm e}^{-2w_0(r)}$ and ${\rm e}^{2w_1(r)}$ are smooth for $r \in (0,\infty)$.
			Their asymptotics at $s=-\infty$ are
			\[
				{\rm e}^{-2 w_0} \xlongrightarrow{ s \rightarrow -\infty} -2 s+b_2,\qquad
				{\rm e}^{2w_1(r)} \xlongrightarrow{ s \rightarrow -\infty} 2 \operatorname{Re} \big( {\rm e}^{\gamma_1 s+\rho_1} \big).
			\]
			\item[$E_1^D\colon$]$\gamma_0 =3$, $\gamma_1 \not\in \mathbb{R}$, $\gamma_1^\mathbb{R}=1$, $\rho_1 \not\in \mathbb{R}$
			and $\rho_0$ is not defined.
			${\rm e}^{2w_1(r)}$ and ${\rm e}^{2w_0(r)}$ are smooth for~${r \in (0,\infty)}$.
			Their asymptotics at $s=-\infty$ are
			\begin{gather*}
				{\rm e}^{2w_0(r)} \xlongrightarrow{ s \rightarrow -\infty} {\rm e}^{\gamma_0 s} \left( -\frac{8}{\big(\gamma_1^\mathbb{I}\big)^2} s+d_0 -\frac{8}{\big(\gamma_1^\mathbb{I}\big)^3}\cos\big( \gamma_1^\mathbb{I} s+\rho_1^\mathbb{I} \big) \right), \\
				{\rm e}^{2 w_1} \xlongrightarrow{ s \rightarrow -\infty} 2 {\rm e}^{\gamma_1^\mathbb{R} s+\rho_1^\mathbb{R}}
				\left( \cos\big( \gamma_1^\mathbb{I} s+\rho_1^\mathbb{I} \big) + \frac{\big(1-\sin\big(\gamma_1^\mathbb{I}s+\rho_1^\mathbb{I}\big) \big)^2}{\gamma_1^\mathbb{I}s-\frac{(\gamma_1^\mathbb{I})^3}{8}d_0 +\cos\big(\gamma_1^\mathbb{I} s+\rho_1^\mathbb{I}\big)} \right),
			\end{gather*}
			where
\[d_0=\lim\limits_{s_1^\mathbb{R} \rightarrow 1-\frac{s_2^\mathbb{R}}{2}+0_-} 2 {\rm e}^{\rho_0^\mathbb{R}}
			\left(\rho_0^\mathbb{I}+\frac{\pi}{2}\right).
\]
			\item[$E_2^D\colon$]$\gamma_0 \not\in \mathbb{R}$, $\gamma_1=-3$, $\gamma_0^\mathbb{R}=-1$, $\rho_0 \not\in \mathbb{R}$
			and $\rho_1$ is not defined.
			${\rm e}^{-2w_0(r)}$ and ${\rm e}^{-2w_1(r)}$ are smooth for $r \in (0,\infty)$.
			Their asymptotics at $s=-\infty$ are
			\begin{gather*}
				{\rm e}^{-2 w_0} \xlongrightarrow{ s \rightarrow -\infty} 2\, {\rm e}^{-\gamma_0^\mathbb{R} s-\rho_0^\mathbb{R}}
				\left( \cos\big( \gamma_0^\mathbb{I} s+\rho_0^\mathbb{I} \big) + \frac{8\big(1+\sin\big(\gamma_0^\mathbb{I}s+\rho_0^\mathbb{I}\big) \big)^2}{-8\gamma_0^\mathbb{I}s+ \big(\gamma_0^\mathbb{I}\big)^3 \tilde{d}_0+8 \cos\big(\gamma_0^\mathbb{I} s+\rho_0^\mathbb{I}\big)} \right), \\
				{\rm e}^{-2w_1(r)} \xlongrightarrow{ s \rightarrow -\infty} {\rm e}^{-\gamma_1 s} \left(-\frac{8}{\big(\gamma_0^\mathbb{I}\big)^2} s+\tilde{d}_0+\frac{8}{\big(\gamma_0^\mathbb{I}\big)^3}\cos\big(\gamma_0^\mathbb{I} s+\rho_0^\mathbb{I}\big)\right),
			\end{gather*}
			where
\[\tilde{d}_0=\lim\limits_{s_1^\mathbb{R} \rightarrow \frac{s_2^\mathbb{R}}{2}-1+0_+} 2 {\rm e}^{-\rho_1^\mathbb{R}}
			\left( \frac{\pi}{2}-\rho_1^\mathbb{I} \right).
\]
			\item[ %verified again: no typo.
			$E_3^R\colon$]$\gamma_0 \not\in \mathbb{R}$, $\gamma_1 \not\in \mathbb{R}$,
			$\gamma_0^\mathbb{R}=3$, $\gamma_1^\mathbb{R}=1$ and~${\gamma_0^\mathbb{I}=\gamma_1^\mathbb{I}}$.
			Both $\rho_0$ and $\rho_1$ are not defined.
			${\rm e}^{2w_0(r)}$ and~${{\rm e}^{2w_0(r)+2w_1(r)}}$ are smooth for $r \in (0,\infty)$.
			Their asymptotics at $s=-\infty$ are
		\begin{gather*}
				{\rm e}^{2 w_0} \xlongrightarrow{ s \rightarrow -\infty}-{\rm e}^{\gamma_0^\mathbb{R} s} \left( \frac{4}{\big(\gamma_0^\mathbb{I}\big)^2} (s+ \operatorname{Re}(b_3) ) \sin\big(\gamma_0^\mathbb{I} s+\theta_0\big)+\frac{4}{\big(\gamma_0^\mathbb{I}\big)^3} \cos\big(\gamma_0^\mathbb{I} s+\theta_0\big) \right), \\
				{\rm e}^{2w_0+2w_1(r)} \xlongrightarrow{ s \rightarrow -\infty} {\rm e}^{\big(\gamma_0^\mathbb{R}+\gamma_1^\mathbb{R} \big)s} \left( \frac{4}{\big(\gamma_0^\mathbb{I}\big)^2} (s+\operatorname{Re}(b_3))^2
				-\frac{4}{\big(\gamma_0^\mathbb{I}\big)^4} \big( \cos\big(\gamma_0^\mathbb{I} s+\theta_0\big) \big)^2 \right),
			\end{gather*}
			where
\[\theta_0=\lim\limits_{s_1^\mathbb{R} \rightarrow 2 \sqrt{-2-s_2^\mathbb{R}} +0_+} \rho_0^\mathbb{I} .\]
			\item[$E_3^L\colon$]$\gamma_0 \not\in \mathbb{R}$, $\gamma_1 \not\in \mathbb{R}$,
			$\gamma_0^\mathbb{R}=-1$, $\gamma_1^\mathbb{R}=-3$ and $\gamma_0^\mathbb{I}=\gamma_1^\mathbb{I}$.
			Both $\rho_0$ and $\rho_1$ are not defined.
			${\rm e}^{-2w_1(r)}$ and ${\rm e}^{-2w_0(r)-2w_1(r)}$ are smooth for $r \in (0,\infty)$.
			Their asymptotics at $s=-\infty$ are
		 \begin{gather*}
				{\rm e}^{-2 w_1} \xlongrightarrow{ s \rightarrow -\infty} {\rm e}^{-\gamma_1^\mathbb{R} s}
				\left( \frac{4}{\big(\gamma_1^\mathbb{I}\big)^2} (s+ \operatorname{Re}(b_3) )
				\sin\big(\gamma_1^\mathbb{I} s-\tilde{\theta}_0\big)
				+\frac{4}{\big(\gamma_1^\mathbb{I}\big)^3} \cos\big(\gamma_1^\mathbb{I} s-\tilde{\theta}_0\big) \right), \\
				{\rm e}^{-2w_0-2w_1(r)} \xlongrightarrow{ s \rightarrow -\infty} {\rm e}^{-(\gamma_0^\mathbb{R}+\gamma_1^\mathbb{R} )s} \left( \frac{4}{\big(\gamma_1^\mathbb{I}\big)^2} (s+\operatorname{Re}(b_3))^2
				-\frac{4}{\big(\gamma_1^\mathbb{I}\big)^4} \big( \cos\big(\gamma_1^\mathbb{I} s-\tilde{\theta}_0\big) \big)^2 \right),
			\end{gather*}
			where
\[\tilde{\theta}_0=-\lim\limits_{s_1^\mathbb{R} \rightarrow -2 \sqrt{-2-s_2^\mathbb{R}} +0_-} \rho_1^\mathbb{I}.\]
		\end{itemize}	
	\end{Conjecture}
	
	\subsection[Numerically verify the conjecture: the Omega\_1 case as an example]{Numerically verify the conjecture: the $\boldsymbol{\Omega_1}$ case as an example}
	In this subsection, we will numerically verify Conjecture \ref{Conj} for the $\Omega_1$ case with $\big(s_1^\mathbb{R}, s_2^\mathbb{R}\big)=(2,1)$.
	Then by \eqref{CF-Inverse}, we get
	\begin{gather}
			\gamma_0|_{s_1^\mathbb{R}=2, s_2^\mathbb{R}=1}=\frac{1}{3},\qquad
			\gamma_1|_{s_1^\mathbb{R}=2, s_2^\mathbb{R}=1}=\frac{4}{\pi}\arccos\left(-\frac{3}{2}\right)-3
			=1+\frac{4 \mathrm{i}}{\pi}\ln\left(\frac{3-\sqrt{5}}{2}\right).
\label{gam12-Omega1}
	\end{gather}
	
	With $\big(s_1^\mathbb{R}, s_2^\mathbb{R}\big)=(2,1)$, $w_0$ and $w_1$ keep real as $r$ decreasing from $r=\infty$ to $r=1$.
	So we do not need adjust our numerical integration for $r>1$.
	
	By the truncation of \eqref{iniApprox}, the initial values at $r=45$ are obtained (only the first $50$ digits are listed)
	\begin{gather}
			w_p(45)= -5.2843098725232974899221393911204991207504443469367\ldots \times 10^{-57},\nonumber\\
			w_p'(45)= 1.5004885502015739552694025310567337731833237644509\ldots \times 10^{-56},\nonumber\\
			w_m(45)= 1.9951075414429511313096218077209854214432475688359\ldots \times 10^{-80},\nonumber\\
			w_m'(45)=-8.0025674082272017400263083591640194065100562879396\ldots \times 10^{-80}.\label{ini-Omega1}
	\end{gather}
	
	Comparing with the more accurate solution starting from $r=55$,
	the errors of the initial values \eqref{ini-Omega1} are obtained as shown by Table~\ref{tab17}.

	\begin{table}[ht]\renewcommand{\arraystretch}{1.2}\centering
		\caption{Errors of the initial values of case $\Omega_1$ with $\big(s_1^\mathbb{R}, s_2^\mathbb{R}\big)=(2,1)$.}\label{tab17}

\vspace{1mm}

		\begin{tabular}{ c|c c c c}
			\hline \hline
			$r=45$& $w_p$ & $w_p'$ & $w_m$ &$w_m'$\\
			\hline
Absolute error &$3.04 \times 10^{-170}$ &$2.59 \times 10^{-169}$ &$1.62 \times 10^{-193}$&$1.57\times 10^{-192}$ \\
Relative error &$5.76\times 10^{-114}$ &$1.73 \times 10^{-113}$ &$8.12 \times 10^{-114}$&$1.96\times 10^{-113}$
		\end{tabular}
	\end{table}

	Numerically integrating \eqref{wpwm} from $r=45$ to $r=1$
	by the Gauss--Legendre method with parameters as same as the ones used in Section~\ref{GeneralCase},
	the values of $w_p$, $w_p'$, $w_m$ and $w_m'$ at $r=1$ are obtained
		\begin{gather*}%\label{values-Omega1-1}
			w_p(1)= -3.8224055163443861381648888321249635590437848425393\ldots\times 10^{-2}, \\
			w_p'(1)= 1.2620798170393397054252193737795545512207073701669\ldots\times 10^{-1},\\
			w_m(1)= 4.1421810495867924927295926159960489963050832028643\ldots\times 10^{-3},\\
			w_m'(1)=-1.9704834137414281607395710259152505912708048802280\ldots\times 10^{-2}.
		\end{gather*}
	Comparing with the more accurate solution starting from $r=55$,
	the errors of \eqref{values-V1-1} are obtained as shown by Table~\ref{tab18}.

	\begin{table}[ht]\renewcommand{\arraystretch}{1.2}\centering

\caption{Errors of the numerical solution at $r=1$ of case $\Omega_1$ with $\big(s_1^\mathbb{R}, s_2^\mathbb{R}\big)=(2,1)$.} \label{tab18}

\vspace{1mm}

		\begin{tabular}{ c|c c c c}
			\hline \hline
			$r=1$& $w_p$ & $w_p'$ & $w_m$ &$w_m'$\\
			\hline
Absolute error &$4.42 \times 10^{-115}$ &$1.46 \times 10^{-114}$ &$6.30 \times 10^{-116}$&$3.09\times 10^{-115}$\\
Relative error &$1.16\times 10^{-113}$ &$1.16 \times 10^{-113}$ &$1.52 \times 10^{-113}$&$1.57\times 10^{-113}$
		\end{tabular}
	\end{table}

	When $r<1$, $w_0$ and $w_1$ may be complex.
	As Conjecture~\ref{Conj} suggests, we use $v_0$ and $v_1$
	\begin{gather}
			v_0={\rm e}^{2 w_0},\qquad
			v_1={\rm e}^{2 w_1}.\label{v0v1-DEF1}
	\end{gather}
	as dependent variables for the $\Omega_1$ case.
	Then, $v_0$ and $v_1$ will be real for $r>0$.
	
	To improve computation efficiency, we use $s=\ln(r)$ as independent variable.
	Then the equations for $v_0$ and $v_1$ are
	\begin{gather}
			\frac{{\rm d}^2 v_0}{{\rm d}s^2}=4 {\rm e}^{2 s} \big(v_0^3-v_1\big) +\frac{1}{v_0} \left(\frac{{\rm d} v_0}{{\rm d}s}\right)^2, \qquad
			\frac{{\rm d}^2 v_1}{{\rm d}s^2}=4 {\rm e}^{2 s} \left( \frac{v_1^2}{v_0}-\frac{1}{v_1} \right)
			+\frac{1}{v_1} \left(\frac{{\rm d} v_1}{{\rm d}s}\right)^2. \label{v0v1-DEQ1}
	\end{gather}
	
	The associated truncation of \eqref{v0v1-DEQ1} for the fine asymptotics of the $\Omega_1$ case should be
	\begin{gather}
			\frac{{\rm d}^2 v_0^{(0)}}{{\rm d}s^2}=\frac{1}{v_0^{(0)}} \left(\frac{{\rm d} v_0^{(0)}}{{\rm d}s} \right)^2, \qquad
			\frac{{\rm d}^2 v_1^{(0)}}{{\rm d}s^2}=-\frac{4 {\rm e}^{2 s}}{v_1^{(0)}}
			+\frac{1}{v_1^{(0)}} \left(\frac{{\rm d} v_1^{(0)}}{{\rm d}s} \right)^2.
		 \label{v0v1-trunc-DEQ1}
	\end{gather}
	In fact, after substituting \eqref{gam12-Omega1} to the $\Omega_1$ case of Conjecture \ref{Conj},
	it becomes obvious which terms of \eqref{v0v1-DEQ1} should be ignored.
	The solution of \eqref{v0v1-trunc-DEQ1} is known
	\begin{gather}
			v_0^{(0)}(s)= {\rm e}^{a_{1\Omega_1} s+b_{1\Omega_1}}, \qquad
			v_1^{(0)}(s)=- \frac{2}{a_{2\Omega_1}} {\rm e}^s \cos( a_{2\Omega_1} s+b_{2\Omega_1} ). \label{v0v1-trunc-DEQ1-Sol}
	\end{gather}
	Comparing \eqref{v0v1-trunc-DEQ1-Sol} with Conjecture \ref{Conj}, we know that $a_{1\Omega_1}=\gamma_0$,
	$b_{1\Omega_1}=\rho_0$, $a_{2\Omega_1}=\operatorname{Im} (\gamma_1)$ and~${b_{2\Omega_1}=\operatorname{Im} (\rho_1)}$.
	Note also that \smash{$-\frac{1}{\operatorname{Im} (\gamma_1)}={\rm e}^{\operatorname{Re} (\rho_1)}$} in the $\Omega_1$ case.
	The ignored terms of the truncation from~\eqref{v0v1-DEQ1} to~\eqref{v0v1-trunc-DEQ1}
	are $ 4 {\rm e}^{2 s} \big(v_0^3-v_1\big)$ and \smash{$4 {\rm e}^{2 s} \frac{v_1^2}{v_0} $},
	which, considering \eqref{gam12-Omega1}, are of order~$O\big({\rm e}^{3 s}\big)$ and \smash{$O\big({\rm e}^{\frac{11}{3}s}\big)$}, respectively.
	So the relative errors are both of order \smash{$O\big({\rm e}^{\frac{8}{3}s}\big)$} except near the zeros of $v_1(s)$.
	Since $v_0$ and $v_1$ are both small in this case, only the relative errors are relevant.
	To avoid the inconvenience brought by the relative error,
	we will take
		\begin{gather*}%\label{DelDel1-DEF}
			\Delta_0(s)=\big|{\rm e}^{2 w_0} {\rm e}^{-\gamma_0 s-\rho_0}-1\big|,\qquad
			\Delta_1(s)=\left|\frac{1}{2}{\rm e}^{2 w_1} {\rm e}^{-\operatorname{Re}(\gamma_1) s-\operatorname{Re}(\rho_1)} -\cos(\operatorname{Im}(\gamma_1)s+\operatorname{Im}(\rho_1))\right|
		\end{gather*}
	as the measurement of error.
	So $\Delta_0$ and $\Delta_1$ are both of order $O\big({\rm e}^{\frac{8}{3}s}\big)$.
	Solving ${\rm e}^{\frac{8}{3}s_f}=10^{-100}$, we get $s_f \approx -86.35$. For safety and convenience, we set $s_f=-87$.

	Numerical results show that $v_0(s)$ has no zero for $s \in (-\infty, 0]$ but $v_1(s)$ has, just as Conjecture~\ref{Conj} predicts.
	For the sake of numerical integration, it is better to integrate around the zeros of $v_1(s)$.
	In order to keep away from the zeros of $v_1(s)$, we first compute $v_1(s+\mathrm{i} \epsilon)$ with~${\epsilon=10^{-2}}$
	to determine the approximate zeros of $v_1(s)$ by solving $\mathrm{Re}(v_1(s+\mathrm{i} \epsilon))=0$.
	Then we get the approximate zeros $s_i$ of $v_1(s)$ within the range $-87 \le s \le 0$.
	Table~\ref{tab19} lists the first few of them.
	
	\begin{table}[ht]\renewcommand{\arraystretch}{1.2}\centering
		\caption{The first few approximate zeros $s_i$ of $v_1(s)$ for the $\Omega_1$ case with $\big(s_1^\mathbb{R}, s_2^\mathbb{R}\big)=(2,1)$.}\label{tab19}

\vspace{1mm}

		\begin{tabular}{c|c c c c c c c c}
			\hline \hline
			$s_i$ & $s_1$ & $s_2$ & $s_3$ &$s_4$&$s_5$&$s_6$&$s_7$&$s_8$\\
			\hline
			value& $-2.506$ &$-5.069$ &$-7.633$& $-10.197$ &$-12.760$ & $-15.324$&$-17.888 $&$-20.452 $
		\end{tabular}
	\end{table}

	Obviously, the distance between two adjacent zeros in Table~\ref{tab19} is about $2.5$.
	To avoid the numerical instabilities caused by those zeros, we use a contour in the complex plane of $s$,
	as shown in Figure \ref{fig-4}. The radii of the circles around the zeros are set to $\frac{1}{5}$.
	\input{Figure4.tikz}
	The values of $v_i(s)$ for $s$ on the contour can be obtained directly from the numerical integration.
	Then we should supplement the values of $v_i(s)$ in the circles in order to complete the numerical solution of $v_i(s)$.
	In principle, the values of $v_i(s)$ can be evaluated using the Cauchy integral formula
	\smash{$ v_i(s)=\frac{1}{2 \pi \mathrm{i}} \oint \frac{v_i(\xi)}{\xi-s}{\rm d}\xi$}.
	But here $v_i(\xi)$ is only a numerical solution, which has high-precision value only at some points on the circle.
	This restricts our choice of high-precision numerical integration method to calculate the Cauchy integral efficiently.
	Since $v_i$ are periodic functions on the circle, we use the trapezoidal rule to calculate them
	\begin{eqnarray}
		v_i(s)=\frac{1}{2  n} \sum_j \frac{\tilde v_i(\theta_j)}{R {\rm e}^{\mathrm{i} \theta_j}-s} R {\rm e}^{\mathrm{i} \theta_j},
		\qquad i=0,1, \label{Trapezoidal}
	\end{eqnarray}
	where $R=\frac{1}{5}$ denotes the radius of the circle,
	and $\tilde v_i(\theta_j)$ the value of $v_i$ at $\theta_j$ on the circle.
	The distance between the adjacent $\theta_j$ is $\frac{\pi}{n}$.
	Obviously, formula \eqref{Trapezoidal} is not appropriate for a point near the circle.
	Therefore, the contour has $2$ line segments in each circle. We use line segments of length $\frac{1}{10}$.
	Altogether, for $s \in \big(s_j-\frac{1}{10}, s_j+\frac{1}{10}\big)$ we obtain the numerical solution of $v_0(s)$ and~$v_1(s)$ by \eqref{Trapezoidal}
	rather than solving \eqref{v0v1-DEQ1} numerically.
	In our numerical experiments, $n$ is equal to $1000$, which is far more than enough to guarantee an accuracy better than $10^{-100}$.
	
	The plots of $v_0$ and $v_1$ are shown in Figure \ref{fig-5}.
	
	\input{Figure5.tikz}
	
	Table \ref{tab20} shows that the numerical solution is as accurate as we expected.
	
	\begin{table}[ht]\renewcommand{\arraystretch}{1.2}\centering
	\caption{Errors of the numerical solution at $s=-87$ for the $\Omega_1$ case with $\big(s_1^\mathbb{R},s_2^\mathbb{R} \big)=(2,1)$.}	\label{tab20}

\vspace{1mm}

		\begin{tabular}{ c|c c c c}
			\hline \hline
			$s=-87$ & $v_0$ & $\frac{{\rm d}v_0}{{\rm d}s}$ & $v_1$ &$\frac{{\rm d} v_1}{{\rm d}s}$\\
			\hline
Absolute error &$4.06 \times 10^{-125}$ &$1.31 \times 10^{-125}$ &$2.75 \times 10^{-149}$&$2.30\times 10^{-149}$ \\
Relative error &$1.48 \times 10^{-112}$ &$1.43 \times 10^{-112}$ &$7.69 \times 10^{-111}$&$6.35\times 10^{-112}$
		\end{tabular}
	\end{table}

	Table \ref{tab21} shows how good the asymptotic solution is.

	\begin{table}[h!]\renewcommand{\arraystretch}{1.2}\centering
	\caption{Approximate derivation from the asymptotic solution
			for the $\Omega_1$ case with $\big(s_1^\mathbb{R},s_2^\mathbb{R} \big)=(2,1)$.}	\label{tab21}

\vspace{1mm}

		\begin{tabular}{ c| c c c c c c c}
			\hline \hline
			$s$ &$-27$ & $-37$ & $-47$ &$-57$ &$-67$& $-77$ & $-87$\\
			\hline
			$\ln(\Delta_0(s))$&$-73.2379$ &$-100.773$&$-130.684$&$-155.003$&$-181.988$&$-208.457$	&$-233.699$ \\
			$\ln(\Delta_1(s))$
			&$-72.5076 $ &$-98.9279$ &$-125.524$&$-152.287$&$-179.229$&$-206.373$ & $-233.697$
		\end{tabular}
	\end{table}

	\section[Deviating from (2.2)]{Deviating from (\ref{rhos-DEF})}\label{DFG}
	This section is concerned with what the solution looks like when \eqref{rhos-DEF} is not satisfied,
	i.e., we drop the assumption of solutions of \eqref{TT-1} being smooth on $\mathbb{C}^*$.
	First, we derive a better asymptotics near $r=0$, which is suitable to give initial values for the numerical integration.
	Then, \eqref{TT-2} is integrated numerically from $r=0$ to $r=\infty$.
	The integration contour on the complex plane of $r$ is used to surround the singularities.
	We will find that the singularities are regularly distributed.
	But here we have not been able to derive precise formulas from the limited numerical results.
	This is very different from the situation in Section \ref{CONJ}, where we have formulated a conjecture with substantial formulas based on the numerical results.
	This is because the difficulties here are much greater than those encountered in Section \ref{CONJ}:
	here we have in fact four independent parameters $\gamma_0$, $\gamma_1$, $\rho_0$ and $\rho_1$,
	while in Section \ref{CONJ} we have essentially only two parameters~$s_1^\mathbb{R}$ and~$s_2^\mathbb{R}$.
	
	For convenience, in this section we will always use the dependent variables $v_0$ and $v_1$ as defined by \eqref{v0v1-DEF1}.
	As the independent variable, we use $s=\ln(r) $ for $r \le 1$ as before.
	So the equations for $v_0$ and $v_1$ are still \eqref{v0v1-DEQ1}.
	
	Let us take the following assumption first.

\begin{Assumption}\label{Assumption1} Both terms $4 {\rm e}^{2 s} \big(v_0^3-v_1\big)$ and $4 {\rm e}^{2 s} \big( \frac{v_1^2}{v_0}-\frac{1}{v_1} \big)$
			in \eqref{v0v1-DEQ1} are negligible near $s=-\infty$.
\end{Assumption}

	So \eqref{v0v1-DEQ1} becomes
	\begin{gather}
			\frac{ {\rm d}^2 v_0^{(0)} }{{\rm d}s^2} =\frac{1}{ v_0^{(0)}} \left( \frac{ {\rm d} v_0^{(0)}}{{\rm d}s} \right)^2, \qquad
			\frac{ {\rm d}^2 v_1^{(0)} }{{\rm d}s^2}=\frac{1}{v^{(0)}_1}\left( \frac{{\rm d} v_1^{(0)}} {{\rm d}s} \right)^2.
	 \label{wv0wv1-DEQ}
	\end{gather}
	
	The solution of \eqref{wv0wv1-DEQ} is
	\begin{eqnarray}
		v_0^{(0)}= c_0 {\rm e}^{\gamma_0 s}, \qquad v_1^{(0)}= c_1 {\rm e}^{\gamma_1 s} , \label{wv0wv1-SOLU}
	\end{eqnarray}
	where $c_0$, $c_1 $, $\gamma_0$ and $\gamma_1$ are constants,
	which should be real if we are only interested in the real solutions of \eqref{v0v1-DEQ1}.
	The immediate result of Assumption~\ref{Assumption1} is that $\gamma_0$ and $\gamma_1$ satisfy the constraints~${3 \gamma_0+2 >\gamma_0}$, $\gamma_1+2 >\gamma_0$, $2 \gamma_1-\gamma_0+2 >\gamma_1$
	and $2-\gamma_1>\gamma_1$, which is just the interior of the triangle in Figure \ref{fig-2}.
	So, if $(\gamma_0,\gamma_1)$ is a point inside the triangle in Figure \ref{fig-2},
	then \smash{$\big(v_0^{(0)}, v_1^{(0)}\big)$} of \eqref{wv0wv1-SOLU} is the primary approximate solution of $(v_0, v_1)$ near $s=-\infty$.
	If $c_0={\rm e}^{\rho_0}$ and $c_1={\rm e}^{\rho_1}$ with $\rho_0$ and $\rho_1$ defined by \eqref{rhos-DEF},
	then the solution is the one treated by Theorem~\ref{thm-GIL-1}.
	Here we are interested in the case where $c_0 \neq {\rm e}^{\rho_0}$ or $c_1 \neq {\rm e}^{\rho_1}$.
	
	Now, let us transform \eqref{v0v1-DEQ1} to its integral form
	\begin{gather}
			v_0(s) = c_0 {\rm e}^{\gamma_0 s} \exp\left\{ 4 \int_{-\infty}^s {\rm d}\xi \int_{-\infty}^\xi
				{\rm d}\zeta \left[ v_0(\zeta)^2- \frac{v_1(\zeta)}{v_0(\zeta)} \right]
				{\rm e}^{2 \zeta} \right\} ,\nonumber\\
			v_1(s) = c_1 {\rm e}^{\gamma_1 s} \exp\left\{ 4 \int_{-\infty}^s {\rm d}\xi \int_{-\infty}^\xi
				{\rm d}\zeta \left[ \frac{v_1(\zeta)}{v_0(\zeta)} -\frac{1}{v_1(\zeta)^2}\right]
				{\rm e}^{2 \zeta} \right\}.
		\label{v0v1-inteEQ}
	\end{gather}
	In principle, \eqref{v0v1-inteEQ} can be solved recursively near $s=-\infty$:
	\smash{$v_0^{(0)}$} and \smash{$v_1^{(0)}$} are given by \eqref{wv0wv1-SOLU};
	\smash{$v_0^{(1)}$} and \smash{$v_1^{(1)}$} are
	\begin{gather}
			v_0^{(1)}(s) =c_0 {\rm e}^{\gamma_0 s} \exp\left\{ \frac{c_0^2}{(1+\gamma_0)^2}{\rm e}^{2(1+\gamma_0)s}-
			\frac{4 c_1}{c_0 (2-\gamma_0+\gamma_1)^2} {\rm e}^{(2-\gamma_0+\gamma_1)s} \right\} ,\nonumber\\
			v_1^{(1)}(s)=c_1 {\rm e}^{\gamma_1 s} \exp\left\{ \frac{4 c_1}{c_0 (2-\gamma_0+\gamma_1)^2} {\rm e}^{(2-\gamma_0+\gamma_1)s} -\frac{1}{c_1^2 (1-\gamma_1)^2}{\rm e}^{2(1-\gamma_1)s} \right\},
\label{v0v1-SOLU-1}
	\end{gather}
	which are obtained by substituting \smash{$v_0=v_0^{(0)}$} and \smash{$v_1=v_1^{(0)}$} to the right of \eqref{v0v1-inteEQ};
	and so on and so forth.
	If $(\gamma_0,\gamma_1)$ is inside the triangle in Figure \ref{fig-2}, then \smash{$v_0^{(i)}$} and \smash{$v_1^{(i)}$} converge as $i$ increases.

	\subsection{Numerical solution}
	As in Section \ref{GeneralCase}, we still use $(\gamma_0,\gamma_1)=\big(1,\frac{1}{3}\big)$.
	To have some deviation from Section \ref{GeneralCase}, $c_0$~and~$c_1$ should be chosen as
	\[
		c_0={\rm e}^{\rho_0}+\delta c_0, \qquad c_1={\rm e}^{\rho_1}+\delta c_1,
	\]
	where $\delta c_0$ and $\delta c_1$ can not be $0$ simultaneously.
	In the following numerical experiment, we use
	\[
		\delta c_0=\frac{1}{2}, \qquad \delta c_1=\frac{1}{5} .
	\]
	
	To solve \eqref{v0v1-DEQ1} numerically, the initial values of $\big(v_0,\frac{{\rm d}v_0}{{\rm d}s},v_1,\frac{{\rm d}v_1}{{\rm d}s}\big)$ must be given.
	We start from~${s_1=-100}$ and give the initial values by~\eqref{v0v1-SOLU-1}.
	Since it is easy to compute the initial values by~\eqref{v0v1-SOLU-1}, the details of the initial values are omitted.
	We only list the errors of the initial value by Table \ref{tab22}.
	
	\begin{table}[ht]\renewcommand{\arraystretch}{1.2}\centering
		\caption{Errors of the numerical solution at $s=-100$ with $(\gamma_0,\gamma_1,c_0,c_1))=\big(1,\frac{1}{3},{\rm e}^{\rho_0}+\frac{1}{2},{\rm e}^{\rho_1}+\frac{1}{5}\big)$.}		\label{tab22}

\vspace{1mm}

		\begin{tabular}{ c|c c c c}
			\hline \hline
			$s=-100$ & $v_0$ & $\frac{{\rm d}v_0}{{\rm d}s}$ & $v_1$ &$\frac{{\rm d} v_1}{{\rm d}s}$\\
			\hline
Absolute error &$5.32 \times 10^{-160}$ &$1.95 \times 10^{-159}$ &$2.47 \times 10^{-131}$&$7.42\times 10^{-131}$ \\
Relative error &$4.86 \times 10^{-117}$ &$1.78 \times 10^{-116}$ &$5.12 \times 10^{-117}$&$4.61\times 10^{-116}$
		\end{tabular}
	\end{table}

	The errors of the values at $s=-100$ are obtained by comparing them with the numerical solution starting from $s=-140$,
	which is much more accurate.
	
	The numerical solution is smooth for $s \in [-100,0]$.
	
	As a comparison to \eqref{values-general-1}, the values of $v_0$ and others at $s=0$ are
	\begin{gather}
			v_0|_{s=0}= 1.3324864759152155716932764336782719490481063559703\ldots ,\nonumber\\
			\frac{{\rm d}v_0}{{\rm d}s}|_{s=0}= 0.49495834671586092263807187324781656576576424051419\ldots,\nonumber\\
			v_1|_{s=0}= 2.6783375094329925626474416219547736732331423595096\ldots,\nonumber\\
			\frac{{\rm d}v_1}{{\rm d}s}|_{s=0}= 6.2948008049596612397631881197126092308528410458148\ldots.
	\label{values-Deviation-0}
	\end{gather}
	
	Table \ref{tab23} gives the errors of \eqref{values-Deviation-0}.
	
	\begin{table}[ht]\renewcommand{\arraystretch}{1.2}\centering % Table 23
		\caption{Errors of the numerical solution at $s=0$.}\label{tab23}

\vspace{1mm}

		\begin{tabular}{ c|c c c c}
			\hline \hline
			$s=0$& $v_0$ & $\frac{{\rm d} v_0}{{\rm d}s}$ & $v_1$ &$\frac{{\rm d} v_1}{{\rm d}s}$\\
			\hline
Absolute error &$ 3.37 \times 10^{-113}$ &$ 1.69 \times 10^{-112}$ &$ 5.96 \times 10^{-113}$&$ 3.43\times 10^{-112}$ \\
Relative error &$2.53 \times 10^{-113}$ &$ 3.41 \times 10^{-112}$ &$ 2.23 \times 10^{-113}$&$5.44 \times 10^{-113}$
		\end{tabular}
	\end{table}
	
Again, the errors are evaluated by comparing the two numerical solutions starting from $s=-140$ and from $s=-100$, respectively.
	
For $s>0$, i.e., $r>1$, it is convenient to use the variable $r$ itself instead of $s$:
	the pattern of the singularities is more transparent with respect to $r$ than with respect to $s$.
	Then \eqref{v0v1-DEQ1} is converted to
	\begin{gather}
			\frac{{\rm d}v_0}{{\rm d}r}=\frac{1}{r} p_0,\qquad
			\frac{{\rm d}p_0}{{\rm d}r}=\frac{p_0^2}{r v_0}+4 r v_0^3-4 r v_1,\nonumber \\
			\frac{{\rm d}v_1}{{\rm d}r}=\frac{1}{r}p_1,\qquad
			\frac{{\rm d}p_1}{{\rm d}r}=\frac{p_1^2}{r v_1}-\frac{4 r}{v_1}+\frac{4 r v_1^2}{v_0}. \label{v0v1-DEQr}
	\end{gather}
	
	Then, we compute the numerical solution of \eqref{v0v1-DEQr},
	for which the initial values are given by~\eqref{values-Deviation-0}.
	Near $r \approx 1.539167317$, the numerical solution blows up.
	Figures \ref{fig-6} and \ref{fig-7} show the plots of $v_0$ and $v_1$ on the circle with a radius of about $0.239167317$ around the singular point.
	
	\input{Figure6.tikz}
	
	\input{Figure7.tikz}
	
	Obviously, $v_0$ and $v_1$ are smooth functions on the circle.
	Numerical results show that the singularity at $r \approx 1.539167317$ is a simple pole of $v_1$.
	By \eqref{v0v1-DEQr}, either $v_0=\infty$ or $v_0=0$ at the singularity of $v_1$.
	Numerical results indicate $v_0=0$ at this singularity of $v_1$.
	
	\input{Figure8.tikz}
	\input{Figure9.tikz}

	To show the pattern of the singularities of $v_0$ and $v_1$,
	we plot $v_i\big(r+10^{-2}\mathrm{i}\big) $, $i=0,1$ as Figures~\ref{fig-8} and~\ref{fig-9}.
	Although we cannot give a precise description of Figures \ref{fig-8} and~\ref{fig-9},
	we can still make several heuristic observations from the two figures.
	First, we can observe that both~$v_0(r)$ and $v_1(r)$ have infinitely many singularities since some adjacent singularities are almost equidistant.
	Second, $v_0(r)$ and $v_1(r)$ should be real since the imaginary parts of~${v_0\big(r+10^{-2}\mathrm{i}\big)}$ and~${v_1\big(r+10^{-2}\mathrm{i}\big)}$
	are small except near the singularities.
	Third, $v_1(r_{\rm singular}+0_-)>0$ and ${v_1(r_{\rm singular}+0_+)<0}$ and the imaginary part of ${v_1(r_{\rm singular}+0_+ \mathrm{i})}$ is always positive.
	Fourth, the singularities of $v_0(r)$ have two frequencies:
	the class of singularities with~${v_0(r_{\rm singular}+0_+ \mathrm{i})<0}$ have one frequency
	and the class of singularities with $v_0(r_{\rm singular}+0_+ \mathrm{i})>0$ have another frequency.
	The first two observations should be general for cases deviating from~\eqref{rhos-DEF}.
	It seems that there is no simple combination of $v_0$ and $v_1$ such that the composite variable is smooth for $r \in (0,\infty)$.

	\section{Conclusion and discussion}\label{sec6}
	This paper numerically studies equation \eqref{TT-2}, the case 4a of the tt*-Toda equation.
	The fine asymptotics of the solutions described by Theorem \ref{thm-GIL-1} are verified with an accuracy of order~$10^{-100}$.
	We enlarge the class of the solutions described by Theorem \ref{thm-GIL-1} from the Stokes data side
	by assuming that they have asymptotics \eqref{InfAsymp} for $\big(s_1^\mathbb{R}, s_2^\mathbb{R}\big) \in \mathbb{R}^2$ but may have singularities for $r \in (0, \infty)$.
	For the enlarged class of solutions, we construct the proper dependent variables (smooth for $r \in (0, \infty)$) for every case,
	and find all the fine asymptotic formulas for these proper dependent variables.
	The associated truncation equations of \eqref{TT-2} are crucial for the realization of the high-precision verifications and are indeed useful in the search for the new fine asymptotics.
	Some preliminary numerical studies are also made to investigate what happens when the fine asymptotics is broken at the $r=0$ side.
	However, the studies in Section~\ref{DFG} are far from complete in investigating the deviation from~\eqref{rhos-DEF}.
	The first problem is whether we can find two proper dependent variables that are smooth near $r=\infty$.
	It can be shown that the singularity of $\frac{1}{v_0(r)}$ coincides with $v_1(r)$, differing only in amplitude.
	But this does not help much in determining what are the proper variables.
	Without proper variables it will be almost impossible to talk about the asymptotics near $r=\infty$.
	The second problem is to find out the $r=\infty$ asymptotics of~\eqref{v0v1-DEQr}
	beyond $v_0(r) \xlongrightarrow{r \rightarrow \infty} \pm 1$, $v_1(r) \xlongrightarrow{r \rightarrow \infty} \pm 1$.\footnote{Obviously, $v_0(r) \xlongrightarrow{r \rightarrow \infty} 1$, $v_1(r) \xlongrightarrow{r \rightarrow \infty} 1$
		is equivalent to $w_0(r) \xlongrightarrow{r \rightarrow \infty} 0$, $w_1(r) \xlongrightarrow{r \rightarrow \infty} 0$.
		From the symmetry of \eqref{v0v1-DEQr}, the solutions of $v_0(r) \xlongrightarrow{r \rightarrow \infty} -1$, $v_1(r) \xlongrightarrow{r \rightarrow \infty} -1$
		can be obtained from the one of $v_0(r) \xlongrightarrow{r \rightarrow \infty} 1$, $v_1(r) \xlongrightarrow{r \rightarrow \infty} 1$
		by a substitution $v_0 \rightarrow -v_0$, $v_1 \rightarrow -v_1$.}
	Perhaps the best way to explain this is to look at the similar but opposite case.
	In Conjecture~\ref{Conj}, the problem starts from the~${r=\infty}$ side
	and we see no classification near~${r=\infty}$ until the solutions evolve to the~${r=0}$ side where different types of asymptotics near~${r=0}$ are observed.
	Section \ref{DFG} starts from the~${r=0}$ side and we see no classification near~${r=0}$ for the cases parameterized by points in the triangle.
	We expect the behavior of these solutions to separate near $r=\infty$ and provide a natural classification of the $r=\infty$ asymptotics of \eqref{v0v1-DEQr}.
	
\subsection*{Acknowledgements}

Part of this work was done while Y. Li was visiting the Department of Mathematical Sciences of IUPUI.
Y.~Li would like to thank A.~Its for his hospitality and the suggestion of verifying their formulas in~\cite[Corollary~8.3]{GIL-3}.
The work is partly supported by NSFC (12235007) and
Science and Technology Commission of Shanghai Municipality (No.\ 22DZ2229014).
The author would also like to thank the referees for their helpful suggestions and comments.

\pdfbookmark[1]{References}{ref}
\LastPageEnding

\end{document}

%% file: Figure1.tikz
\begin{figure}
%\begin{center}
%\begin{minipage}[c]{10.5cm}
\centering
\begin{tikzpicture}
\draw[fill=orange]  (-1,1)--(-1,-3)--(3,1)-- cycle;
\draw[->, line width=0.2 mm](-2,0)--(4,0); \draw[->,line width=0.2 mm](0,-3.5)--(0,2);
\draw[line width=0.5 mm,black](-0.8,1)--(2.8,1); 
\draw[line width=0.5 mm,black](-1,0.8)--(-1,-2.8); 
\draw[line width=0.5 mm,black](-0.85,-2.85)--(2.85,0.85);
\node at(4.2,0){$\gamma$}; \node at(0,2.2){$\delta$};
\filldraw (-1,1) circle (1 mm); \filldraw (-1,-3) circle (1 mm); \filldraw (3,1) circle (1 mm);
\node at(3.4,1.2){V1}; \node at(-1.4,1.2){V2}; \node at(-1.4,-3.2){V3};
\node at(1.3,1.2){E1};  \node at(-1.3,-1.4){E2};  \node at(1.25,-1.25){E3};
\node at(-1.3,-0.3){$-\frac{2}{a}$}; \node at(2.1,-0.2){$2$};
\node at(-0.2,1.3){$\frac{2}{b}$};  \node at(-0.3,-1.9){$-2$};
\end{tikzpicture}
%\end{minipage}
%\centerline{Figure 1. The triangular region for $(\gamma,\delta)$.}
%\end{center}
\caption{The triangular region for $(\gamma,\delta)$.}
\label{fig-1}
\end{figure}
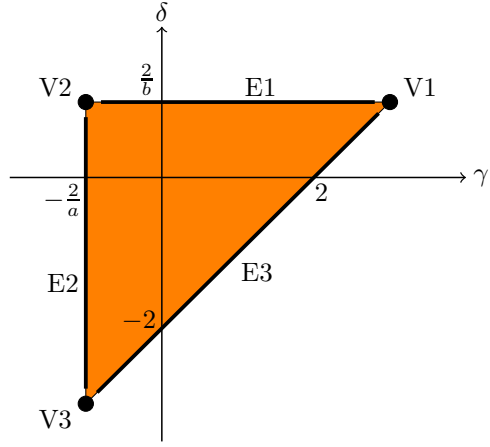

%% file: Figure2.tikz
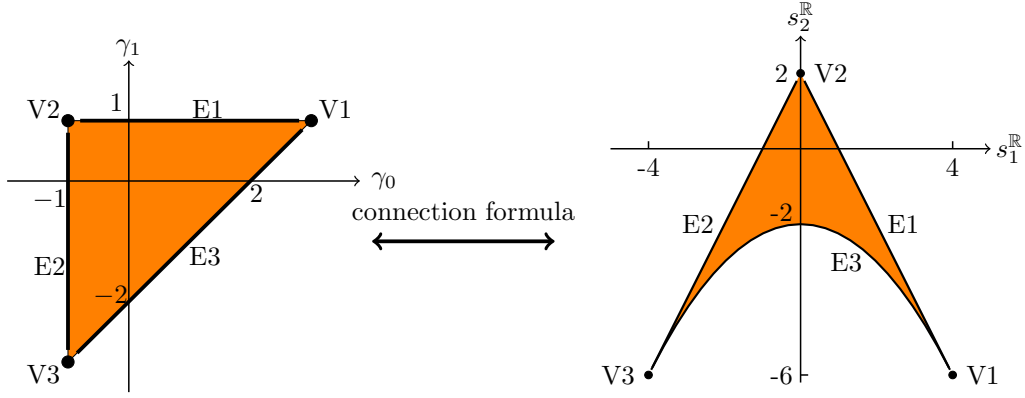
\begin{figure}
\centering
%\begin{center}
	%\begin{minipage}[c]{13cm}
		\begin{tikzpicture}[scale=0.8]
		\draw[fill=orange]  (-1,1)--(-1,-3)--(3,1)-- cycle;
		\draw[->, line width=0.2 mm](-2,0)--(3.8,0); \draw[->,line width=0.2 mm](0,-3.5)--(0,2);
		\draw[line width=0.5 mm,black](-0.8,1)--(2.8,1); 
		\draw[line width=0.5 mm,black](-1,0.8)--(-1,-2.8); 
		\draw[line width=0.5 mm,black](-0.85,-2.85)--(2.85,0.85);
		\node at(4.2,0){$\gamma_0$}; \node at(0,2.2){$\gamma_1$};
		\filldraw (-1,1) circle (1 mm); \filldraw (-1,-3) circle (1 mm); \filldraw (3,1) circle (1 mm);
		\node at(3.4,1.2){V1}; \node at(-1.4,1.2){V2}; \node at(-1.4,-3.2){V3};
		\node at(1.3,1.3){E1};  \node at(-1.3,-1.4){E2};  \node at(1.25,-1.25){E3};
		\node at(-1.5,-0.3){$-1$}; \node at(2.1,-0.2){$2$};
		\node at(-0.2,1.3){$1$};  \node at(-0.5,-1.9){$-2$};
		\draw[<->, line width=0.5 mm](4,-1)--(7,-1);
		\node at(5.5,-0.5){connection formula};
		\end{tikzpicture}
		\begin{tikzpicture}[scale=0.5]
		\fill[orange] (-4,-6)--(0,2)--(4,-6)--plot [domain=4:-4] (\x,{-((\x)^2+8)/4})--cycle;
		\draw[->, line width=0.2 mm](-5,0)--(5,0); \draw[->,line width=0.2 mm](0,-6.2)--(0,3);
		\draw[line width=0.2 mm](-4,0)--(-4,0.2); 	\draw[line width=0.2 mm](4,0)--(4,0.2); 	
		\draw[line width=0.2 mm](0,-6)--(0.2,-6); %\draw[line width=0.2 mm](0,2)--(0.2,2);
		\node at(5.5,0){$s_1^\mathbb{R}$}; \node at(0,3.5){$s_2^\mathbb{R}$};
		\filldraw (0,2) circle (1 mm); \filldraw (4,-6) circle (1 mm); \filldraw (-4,-6) circle (1 mm);
		\node at(0.8,2){V2}; \node at(4.8,-6){V1}; \node at(-4.8,-6){V3};
		\draw[line width=0.3 mm,black](-4+0.1,-6+0.2)--(0-0.1,2-0.2);
		\draw[line width=0.3 mm,black](4-0.1,-6+0.2)--(0+0.1,2-0.2);
		\draw[thick] plot[domain=-3.93:3.93] (\x, {-((\x)^2+8)/4});
		\node at(2.7,-2){E1}; \node at(-2.7,-2){E2}; \node at(1.2,-3){E3};
		\node at(-0.7,-1.7){$-2$}; \node at(-0.5,2){2}; \node at(-0.7,-6){$-6$}; \node at(4,-0.5){4}; \node at(-4,-0.5){$-4$};
		\end{tikzpicture}
	%\end{minipage}
	%\centerline{Figure 2. The region map of the connection formula (\ref{ConnectFormula}).}
%\end{center}
\caption{The region map of the connection formula (\ref{ConnectFormula}).}
\label{fig-2}
\end{figure}

%% file: Figure3.tikz
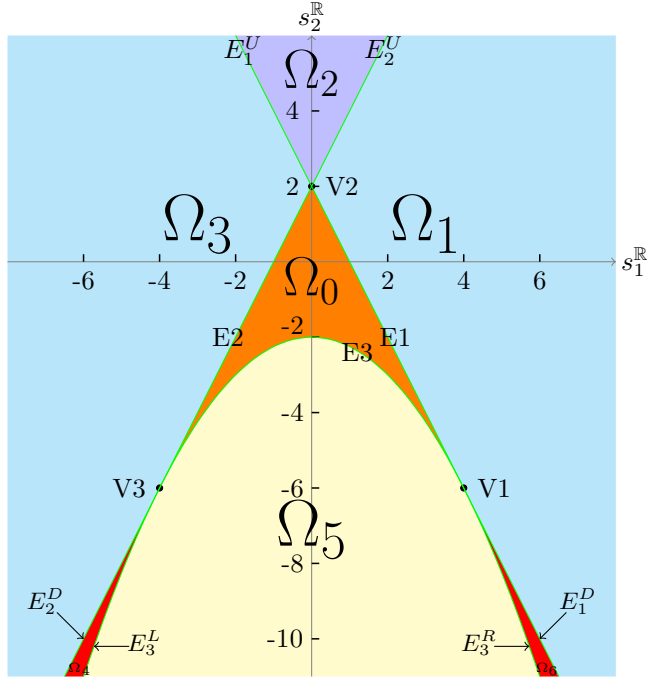
\begin{figure}
\centering
%\begin{center}
	%\begin{minipage}[c]{8cm}
		\begin{tikzpicture}[scale=0.5]
		\fill[orange] (-4,-6)--(0,2)--(4,-6)--plot [domain=4:-4] (\x,{-((\x)^2+8)/4})--cycle;
                \fill[yellow!25] (-6,-11)--(6,-11)--plot [domain=6:-6] (\x,{-((\x)^2+8)/4})--cycle;
                \fill[red] (6,-11)--(6.5,-11)--(4,-6)--plot [domain=4:6] (\x,{-((\x)^2+8)/4})--cycle;
                \fill[red] (-6,-11)--(-6.5,-11)--(-4,-6)--plot [domain=-4:-6] (\x,{-((\x)^2+8)/4})--cycle;
                \fill[cyan!25] (-8,-11)--(-6.5,-11)--(0,2)--(-2,6)--(-8,6)--cycle;
                \fill[cyan!25] (8,-11)--(6.5,-11)--(0,2)--(2,6)--(8,6)--cycle;
                \fill[blue!25] (-2,6)--(0,2)--(2,6)--cycle;
		\draw[->,gray, line width=0.1 mm](-8,0)--(8,0); \node at(8.5,0){$s_1^\mathbb{R}$}; 
                 \draw[->, gray, line width=0.1 mm](0,-11)--(0,6); \node at(0, 6.5){$s_2^\mathbb{R}$};
		\draw[line width=0.2 mm](-4,0)--(-4,0.2); \node at(-4,-0.5){$-4$};
                \draw[line width=0.2 mm](-2,0)--(-2,0.2); \node at(-2,-0.5){$-2$};
                \draw[line width=0.2 mm](2,0)--(2,0.2); \node at(2,-0.5){2};	
                \draw[line width=0.2 mm](4,0)--(4,0.2); \node at(4,-0.5){4};
                \draw[line width=0.2 mm](-6,0)--(-6,0.2); \node at(-6,-0.5){$-6$};
                \draw[line width=0.2 mm](6,0)--(6,0.2); \node at(6,-0.5){6};
                \draw[line width=0.2 mm](0,4)--(0.2,4);\node at(-0.5,4){4};
                \draw[line width=0.2 mm](0,2)--(0.2,2);\node at(-0.5,2){2};
                \draw[line width=0.2 mm](0,-2)--(0.2,-2);\node at(-0.6,-1.7){$-2$};
                \draw[line width=0.2 mm](0,-4)--(0.2,-4); \node at(-0.6,-4){$-4$};	
		\draw[line width=0.2 mm](0,-6)--(0.2,-6); \node at(-0.6,-6){$-6$};
                \draw[line width=0.2 mm](0,-8)--(0.2,-8); \node at(-0.6,-8){$-8$};
                \draw[line width=0.2 mm](0,-10)--(0.2,-10); \node at(-0.8,-10){$-10$};
		\filldraw (0,2) circle (0.8 mm); \filldraw (4,-6) circle (0.8 mm); \filldraw (-4,-6) circle (0.8 mm);
		\node at(0.8,2){V2}; \node at(4.8,-6){V1}; \node at(-4.8,-6){V3};
		\draw[thin,green](-6.5,-11)--(2,6);
		\draw[thin,green](6.5,-11)--(-2,6);
		\draw[thin,green] plot[smooth, domain=-6:6] (\x, {-((\x)^2+8)/4});
		\node at(2.2,-2){E1}; \node at(-2.2,-2){E2}; \node at(1.2,-2.4){E3}; 
                \node at(0,-0.5){\Large $\Omega_0$};  
                \node at(3,1){\Large $\Omega_1$};
                \node at(0,5){\Large $\Omega_2$};
                \node at(-3,1){\Large $\Omega_3$};
                \node at(-6.1,-10.8){\tiny $\Omega_4$};
                \node at(0,-7.1){\Large $\Omega_5$};
                \node at(6.2,-10.8){\tiny $\Omega_6$};
                \draw[->](6.7,-9.3)--(6, -10);  \node at(7,-9){\small $E_1^D$}; 
                \draw[->](4.8,-10.2)--(5.72713, -10.2);  \node at(4.4,-10.1){\small $E_3^R$};
                \draw[->](-4.8,-10.2)--(-5.72713, -10.2);  \node at(-4.4,-10.1){\small $E_3^L$};
                \draw[->](-6.7,-9.3)--(-6, -10);  \node at(-7,-9){\small $E_2^D$};
                \node at (1.9,5.6){$E_2^U$};  \node at (-1.8,5.6){$E_1^U$};
		\end{tikzpicture}

\caption{Regions of $\Omega_i$, $i=0,1,2,3,4,5,6$, edges of E1, E2, E3, $E_1^U$, $E_2^U$, $E_1^D$,
and vertex of~V1,~V2,~V3.}
\label{fig-3}
\end{figure}

%% file: Figure4.tikz
\begin{figure}
\centering
%\begin{center}
	%\begin{minipage}[c]{12 cm}
		%\begin{center}
		%\begin{minipage}[c]{10 cm}

\begin{tikzpicture}[scale=1]

\draw[->] (-8.5,0)--(1,0) node[right] {$\operatorname{Re}(s)$};
\draw[->] (0,-1)--(0,1) node[above] {$\operatorname{Im}(s)$};
\draw[-] (0,0.5)--(0.1,0.5); \node at (-0.37,0.5){$0.5$};
\draw[-] (0,-0.5)--(0.1,-0.5); \node at (-0.47,-0.5){$-0.5$};
\draw[-] (-1,0)--(-1,0.1);% node[below]{$-1$};
\draw[-] (-2,0)--(-2,0.1);\node at (-2,-0.2){$-2$};
\draw[-] (-3,0)--(-3,0.1);% node[below]{$-3$};
\draw[-] (-4,0)--(-4,0.1); \node at (-4,-0.2){$-4$};
\draw[-] (-5,0)--(-5,0.1);% node[below]{$-5$};
\draw[-] (-6,0)--(-6,0.1);\node at (-6,-0.2){$-6$}; 
\draw[-] (-7,0)--(-7,0.1);% node[below]{$-7$};
\draw[-] (-8,0)--(-8,0.1); \node at (-8.05,-0.2){$-8$};
\draw[-, thick, yellow] (0,0)--(-2.406,0);
\draw[thick, yellow] (-2.506,0) circle (0.2); \draw[thick] (-2.506,0) circle (0.02); \node at (-2.506,-0.35){$s_1$};
\draw[-, thick, yellow] (-2.606,0)--(-4.969,0);
\draw[thick, yellow] (-5.069,0) circle (0.2); \draw[thick] (-5.069,0) circle (0.02); \node at (-5.069,-0.35){$s_2$};
\draw[-, thick, yellow] (-5.169,0)--(-7.533,0);
\draw[thick, yellow] (-7.633,0) circle (0.2); \draw[thick] (-7.633,0) circle (0.02); \node at (-7.633,-0.35){$s_3$};
\draw[-, thick, yellow] (-7.733,0)--(-8.5,0);
%\node at(-0.1,0.15){$0$};
%\draw[-] (0,10*\yscal)--(0.1,10*\yscal); \node at(-0.2,10*\yscal){$10$};
%\draw[-] (0,-10*\yscal)--(0.1,-10*\yscal); \node at(-0.3,-10*\yscal){$-10$};
%\draw[-] (3.14159265,0)--(3.14159265,0.1); \node at(3.14159265,-0.2){$\pi$};
%\draw[-] (6.283185307,0)--(6.283185307,0.1); \node at(6.283185307,-0.2){$2 \pi$};
%\draw[color=red] plot[smooth] coordinates {\FigureAdataBR};
%\draw[color=green] plot[smooth] coordinates {\FigureAdataBI};
%\node at(3.5,9*\yscal){Re($v_1$)};
%\node at(1.5,11*\yscal){Im($v_1$)};
			
\end{tikzpicture}

\caption{Contour  in the complex plane of $s$ to compute $v_0(s)$ and $v_1(s)$.}
\label{fig-4}
\end{figure}
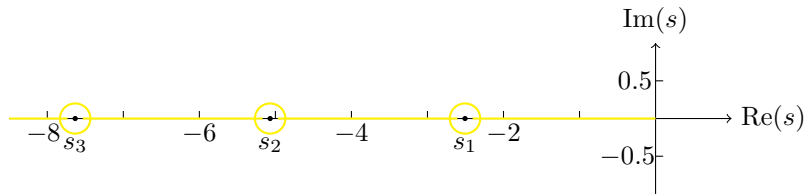

%% file: Figure5.tikz
\begin{figure}
\centering
%\begin{center}
%\begin{minipage}[c]{12cm}
\begin{tikzpicture}[scale=1.2]
%\filldraw[red](7.9,9.3) circle (2mm);

\draw[->,line width=0.2 mm](-8.7, 0)--(0.3,0); \node at(0.4,0){$s$}; 
\draw[->,line width=0.2 mm](0,-1.2)--(0,1.3);
\foreach \x in {-1,-2,-3,-4,-5,-6,-7,-8} \draw[thick](\x,0)--(\x,0.05);
\draw[thick](0,1)--(0.05,1); \node at(0.2,1){$1$};  \draw[thick](0,-1)--(0.05,-1); \node at(0.3,-1){$-1$};
\node at(-1,-0.2){$-10$}; \node at(-2,-0.2){$-20$}; \node at(-3,-0.2){$-30$}; \node at(-4,-0.2){$-40$};
\node at(-5,-0.2){$-50$};\node at(-6,-0.2){$-60$}; \node at(-7,-0.2){$-70$}; \node at(-8,-0.2){$-80$};
\draw[thick,red] plot[smooth] coordinates {(0,0.898019)(-0.01,0.917539)(-0.02,0.935197)  (-0.03,0.950812)(-0.04,0.964288)(-0.05,0.975617)  (-0.06,0.984876)(-0.07,0.992207)(-0.08,0.997806)  (-0.09,1.0019)(-0.1,1.00472)(-0.11,1.00651)  (-0.12,1.00749)(-0.13,1.00785)(-0.14,1.00777)  (-0.15,1.00738)(-0.16,1.00679)(-0.17,1.00611)  (-0.18,1.00538)(-0.19,1.00466)(-0.2,1.00398)  (-0.21,1.00335)(-0.22,1.00279)(-0.23,1.00229)  (-0.24,1.00187)(-0.25,1.00151)(-0.26,1.00121)  (-0.27,1.00096)(-0.28,1.00075)(-0.29,1.00059)  (-0.3,1.00045)(-0.31,1.00035)(-0.32,1.00027)  (-0.33,1.0002)(-0.34,1.00015)(-0.35,1.00011)  (-0.36,1.00008)(-0.37,1.00006)(-0.38,1.00004)  (-0.39,1.00003)(-0.4,1.00002)(-0.41,1.00002)  (-0.42,1.00001)(-0.43,1.00001)(-0.44,1.)(-0.45,1.)  (-0.46,1.)(-0.47,1.)(-0.48,1.)(-0.49,1.)(-0.5,1.)  (-0.51,1.)(-0.52,1.)(-0.53,1.)(-0.54,1.)(-0.55,1.)  (-0.56,1.)(-0.57,1.)(-0.58,1.)(-0.59,1.)(-0.6,1.)  (-0.61,1.)(-0.62,1.)(-0.63,1.)(-0.64,1.)(-0.65,1.)  (-0.66,1.)(-0.67,1.)(-0.68,1.)(-0.69,1.)(-0.7,1.)  (-0.71,1.)(-0.72,1.)(-0.73,1.)(-0.74,1.)(-0.75,1.)  (-0.76,1.)(-0.77,1.)(-0.78,1.)(-0.79,1.)(-0.8,1.)  (-0.81,1.)(-0.82,1.)(-0.83,1.)(-0.84,1.)(-0.85,1.)  (-0.86,1.)(-0.87,1.)(-0.88,1.)(-0.89,1.)(-0.9,1.)  (-0.91,1.)(-0.92,1.)(-0.93,1.)(-0.94,1.)(-0.95,1.)  (-0.96,1.)(-0.97,1.)(-0.98,1.)(-0.99,1.)(-1.,1.)  (-1.01,1.)(-1.02,1.)(-1.03,1.)(-1.04,1.)(-1.05,1.)  (-1.06,1.)(-1.07,1.)(-1.08,1.)(-1.09,1.)(-1.1,1.)  (-1.11,1.)(-1.12,1.)(-1.13,1.)(-1.14,1.)(-1.15,1.)  (-1.16,1.)(-1.17,1.)(-1.18,1.)(-1.19,1.)(-1.2,1.)  (-1.21,1.)(-1.22,1.)(-1.23,1.)(-1.24,1.)(-1.25,1.)  (-1.26,1.)(-1.27,1.)(-1.28,1.)(-1.29,1.)(-1.3,1.)  (-1.31,1.)(-1.32,1.)(-1.33,1.)(-1.34,1.)(-1.35,1.)  (-1.36,1.)(-1.37,1.)(-1.38,1.)(-1.39,1.)(-1.4,1.)  (-1.41,1.)(-1.42,1.)(-1.43,1.)(-1.44,1.)(-1.45,1.)  (-1.46,1.)(-1.47,1.)(-1.48,1.)(-1.49,1.)(-1.5,1.)  (-1.51,1.)(-1.52,1.)(-1.53,1.)(-1.54,1.)(-1.55,1.)  (-1.56,1.)(-1.57,1.)(-1.58,1.)(-1.59,1.)(-1.6,1.)  (-1.61,1.)(-1.62,1.)(-1.63,1.)(-1.64,1.)(-1.65,1.)  (-1.66,1.)(-1.67,1.)(-1.68,1.)(-1.69,1.)(-1.7,1.)  (-1.71,1.)(-1.72,1.)(-1.73,1.)(-1.74,1.)(-1.75,1.)  (-1.76,1.)(-1.77,1.)(-1.78,1.)(-1.79,1.)(-1.8,1.)  (-1.81,1.)(-1.82,1.)(-1.83,1.)(-1.84,1.)(-1.85,1.)  (-1.86,1.)(-1.87,1.)(-1.88,1.)(-1.89,1.)(-1.9,1.)  (-1.91,1.)(-1.92,1.)(-1.93,1.)(-1.94,1.)(-1.95,1.)  (-1.96,1.)(-1.97,1.)(-1.98,1.)(-1.99,1.)(-2.,1.)  (-2.01,1.)(-2.02,1.)(-2.03,1.)(-2.04,1.)(-2.05,1.)  (-2.06,1.)(-2.07,1.)(-2.08,1.)(-2.09,1.)(-2.1,1.)  (-2.11,1.)(-2.12,1.)(-2.13,1.)(-2.14,1.)(-2.15,1.)  (-2.16,1.)(-2.17,1.)(-2.18,1.)(-2.19,1.)(-2.2,1.)  (-2.21,1.)(-2.22,1.)(-2.23,1.)(-2.24,1.)(-2.25,1.)  (-2.26,1.)(-2.27,1.)(-2.28,1.)(-2.29,1.)(-2.3,1.)  (-2.31,1.)(-2.32,1.)(-2.33,1.)(-2.34,1.)(-2.35,1.)  (-2.36,1.)(-2.37,1.)(-2.38,1.)(-2.39,1.)(-2.4,1.)  (-2.41,1.)(-2.42,1.)(-2.43,1.)(-2.44,1.)(-2.45,1.)  (-2.46,1.)(-2.47,1.)(-2.48,1.)(-2.49,1.)(-2.5,1.)  (-2.51,1.)(-2.52,1.)(-2.53,1.)(-2.54,1.)(-2.55,1.)  (-2.56,1.)(-2.57,1.)(-2.58,1.)(-2.59,1.)(-2.6,1.)  (-2.61,1.)(-2.62,1.)(-2.63,1.)(-2.64,1.)(-2.65,1.)  (-2.66,1.)(-2.67,1.)(-2.68,1.)(-2.69,1.)(-2.7,1.)  (-2.71,1.)(-2.72,1.)(-2.73,1.)(-2.74,1.)(-2.75,1.)  (-2.76,1.)(-2.77,1.)(-2.78,1.)(-2.79,1.)(-2.8,1.)  (-2.81,1.)(-2.82,1.)(-2.83,1.)(-2.84,1.)(-2.85,1.)  (-2.86,1.)(-2.87,1.)(-2.88,1.)(-2.89,1.)(-2.9,1.)  (-2.91,1.)(-2.92,1.)(-2.93,1.)(-2.94,1.)(-2.95,1.)  (-2.96,1.)(-2.97,1.)(-2.98,1.)(-2.99,1.)(-3.,1.)  (-3.01,1.)(-3.02,1.)(-3.03,1.)(-3.04,1.)(-3.05,1.)  (-3.06,1.)(-3.07,1.)(-3.08,1.)(-3.09,1.)(-3.1,1.)  (-3.11,1.)(-3.12,1.)(-3.13,1.)(-3.14,1.)(-3.15,1.)  (-3.16,1.)(-3.17,1.)(-3.18,1.)(-3.19,1.)(-3.2,1.)  (-3.21,1.)(-3.22,1.)(-3.23,1.)(-3.24,1.)(-3.25,1.)  (-3.26,1.)(-3.27,1.)(-3.28,1.)(-3.29,1.)(-3.3,1.)  (-3.31,1.)(-3.32,1.)(-3.33,1.)(-3.34,1.)(-3.35,1.)  (-3.36,1.)(-3.37,1.)(-3.38,1.)(-3.39,1.)(-3.4,1.)  (-3.41,1.)(-3.42,1.)(-3.43,1.)(-3.44,1.)(-3.45,1.)  (-3.46,1.)(-3.47,1.)(-3.48,1.)(-3.49,1.)(-3.5,1.)  (-3.51,1.)(-3.52,1.)(-3.53,1.)(-3.54,1.)(-3.55,1.)  (-3.56,1.)(-3.57,1.)(-3.58,1.)(-3.59,1.)(-3.6,1.)  (-3.61,1.)(-3.62,1.)(-3.63,1.)(-3.64,1.)(-3.65,1.)  (-3.66,1.)(-3.67,1.)(-3.68,1.)(-3.69,1.)(-3.7,1.)  (-3.71,1.)(-3.72,1.)(-3.73,1.)(-3.74,1.)(-3.75,1.)  (-3.76,1.)(-3.77,1.)(-3.78,1.)(-3.79,1.)(-3.8,1.)  (-3.81,1.)(-3.82,1.)(-3.83,1.)(-3.84,1.)(-3.85,1.)  (-3.86,1.)(-3.87,1.)(-3.88,1.)(-3.89,1.)(-3.9,1.)  (-3.91,1.)(-3.92,1.)(-3.93,1.)(-3.94,1.)(-3.95,1.)  (-3.96,1.)(-3.97,1.)(-3.98,1.)(-3.99,1.)(-4.,1.)  (-4.01,1.)(-4.02,1.)(-4.03,1.)(-4.04,1.)(-4.05,1.)  (-4.06,1.)(-4.07,1.)(-4.08,1.)(-4.09,1.)(-4.1,1.)  (-4.11,1.)(-4.12,1.)(-4.13,1.)(-4.14,1.)(-4.15,1.)  (-4.16,1.)(-4.17,1.)(-4.18,1.)(-4.19,1.)(-4.2,1.)  (-4.21,1.)(-4.22,1.)(-4.23,1.)(-4.24,1.)(-4.25,1.)  (-4.26,1.)(-4.27,1.)(-4.28,1.)(-4.29,1.)(-4.3,1.)  (-4.31,1.)(-4.32,1.)(-4.33,1.)(-4.34,1.)(-4.35,1.)  (-4.36,1.)(-4.37,1.)(-4.38,1.)(-4.39,1.)(-4.4,1.)  (-4.41,1.)(-4.42,1.)(-4.43,1.)(-4.44,1.)(-4.45,1.)  (-4.46,1.)(-4.47,1.)(-4.48,1.)(-4.49,1.)(-4.5,1.)  (-4.51,1.)(-4.52,1.)(-4.53,1.)(-4.54,1.)(-4.55,1.)  (-4.56,1.)(-4.57,1.)(-4.58,1.)(-4.59,1.)(-4.6,1.)  (-4.61,1.)(-4.62,1.)(-4.63,1.)(-4.64,1.)(-4.65,1.)  (-4.66,1.)(-4.67,1.)(-4.68,1.)(-4.69,1.)(-4.7,1.)  (-4.71,1.)(-4.72,1.)(-4.73,1.)(-4.74,1.)(-4.75,1.)  (-4.76,1.)(-4.77,1.)(-4.78,1.)(-4.79,1.)(-4.8,1.)  (-4.81,1.)(-4.82,1.)(-4.83,1.)(-4.84,1.)(-4.85,1.)  (-4.86,1.)(-4.87,1.)(-4.88,1.)(-4.89,1.)(-4.9,1.)  (-4.91,1.)(-4.92,1.)(-4.93,1.)(-4.94,1.)(-4.95,1.)  (-4.96,1.)(-4.97,1.)(-4.98,1.)(-4.99,1.)(-5.,1.)  (-5.01,1.)(-5.02,1.)(-5.03,1.)(-5.04,1.)(-5.05,1.)  (-5.06,1.)(-5.07,1.)(-5.08,1.)(-5.09,1.)(-5.1,1.)  (-5.11,1.)(-5.12,1.)(-5.13,1.)(-5.14,1.)(-5.15,1.)  (-5.16,1.)(-5.17,1.)(-5.18,1.)(-5.19,1.)(-5.2,1.)  (-5.21,1.)(-5.22,1.)(-5.23,1.)(-5.24,1.)(-5.25,1.)  (-5.26,1.)(-5.27,1.)(-5.28,1.)(-5.29,1.)(-5.3,1.)  (-5.31,1.)(-5.32,1.)(-5.33,1.)(-5.34,1.)(-5.35,1.)  (-5.36,1.)(-5.37,1.)(-5.38,1.)(-5.39,1.)(-5.4,1.)  (-5.41,1.)(-5.42,1.)(-5.43,1.)(-5.44,1.)(-5.45,1.)  (-5.46,1.)(-5.47,1.)(-5.48,1.)(-5.49,1.)(-5.5,1.)  (-5.51,1.)(-5.52,1.)(-5.53,1.)(-5.54,1.)(-5.55,1.)  (-5.56,1.)(-5.57,1.)(-5.58,1.)(-5.59,1.)(-5.6,1.)  (-5.61,1.)(-5.62,1.)(-5.63,1.)(-5.64,1.)(-5.65,1.)  (-5.66,1.)(-5.67,1.)(-5.68,1.)(-5.69,1.)(-5.7,1.)  (-5.71,1.)(-5.72,1.)(-5.73,1.)(-5.74,1.)(-5.75,1.)  (-5.76,1.)(-5.77,1.)(-5.78,1.)(-5.79,1.)(-5.8,1.)  (-5.81,1.)(-5.82,1.)(-5.83,1.)(-5.84,1.)(-5.85,1.)  (-5.86,1.)(-5.87,1.)(-5.88,1.)(-5.89,1.)(-5.9,1.)  (-5.91,1.)(-5.92,1.)(-5.93,1.)(-5.94,1.)(-5.95,1.)  (-5.96,1.)(-5.97,1.)(-5.98,1.)(-5.99,1.)(-6.,1.)  (-6.01,1.)(-6.02,1.)(-6.03,1.)(-6.04,1.)(-6.05,1.)  (-6.06,1.)(-6.07,1.)(-6.08,1.)(-6.09,1.)(-6.1,1.)  (-6.11,1.)(-6.12,1.)(-6.13,1.)(-6.14,1.)(-6.15,1.)  (-6.16,1.)(-6.17,1.)(-6.18,1.)(-6.19,1.)(-6.2,1.)  (-6.21,1.)(-6.22,1.)(-6.23,1.)(-6.24,1.)(-6.25,1.)  (-6.26,1.)(-6.27,1.)(-6.28,1.)(-6.29,1.)(-6.3,1.)  (-6.31,1.)(-6.32,1.)(-6.33,1.)(-6.34,1.)(-6.35,1.)  (-6.36,1.)(-6.37,1.)(-6.38,1.)(-6.39,1.)(-6.4,1.)  (-6.41,1.)(-6.42,1.)(-6.43,1.)(-6.44,1.)(-6.45,1.)  (-6.46,1.)(-6.47,1.)(-6.48,1.)(-6.49,1.)(-6.5,1.)  (-6.51,1.)(-6.52,1.)(-6.53,1.)(-6.54,1.)(-6.55,1.)  (-6.56,1.)(-6.57,1.)(-6.58,1.)(-6.59,1.)(-6.6,1.)  (-6.61,1.)(-6.62,1.)(-6.63,1.)(-6.64,1.)(-6.65,1.)  (-6.66,1.)(-6.67,1.)(-6.68,1.)(-6.69,1.)(-6.7,1.)  (-6.71,1.)(-6.72,1.)(-6.73,1.)(-6.74,1.)(-6.75,1.)  (-6.76,1.)(-6.77,1.)(-6.78,1.)(-6.79,1.)(-6.8,1.)  (-6.81,1.)(-6.82,1.)(-6.83,1.)(-6.84,1.)(-6.85,1.)  (-6.86,1.)(-6.87,1.)(-6.88,1.)(-6.89,1.)(-6.9,1.)  (-6.91,1.)(-6.92,1.)(-6.93,1.)(-6.94,1.)(-6.95,1.)  (-6.96,1.)(-6.97,1.)(-6.98,1.)(-6.99,1.)(-7.,1.)  (-7.01,1.)(-7.02,1.)(-7.03,1.)(-7.04,1.)(-7.05,1.)  (-7.06,1.)(-7.07,1.)(-7.08,1.)(-7.09,1.)(-7.1,1.)  (-7.11,1.)(-7.12,1.)(-7.13,1.)(-7.14,1.)(-7.15,1.)  (-7.16,1.)(-7.17,1.)(-7.18,1.)(-7.19,1.)(-7.2,1.)  (-7.21,1.)(-7.22,1.)(-7.23,1.)(-7.24,1.)(-7.25,1.)  (-7.26,1.)(-7.27,1.)(-7.28,1.)(-7.29,1.)(-7.3,1.)  (-7.31,1.)(-7.32,1.)(-7.33,1.)(-7.34,1.)(-7.35,1.)  (-7.36,1.)(-7.37,1.)(-7.38,1.)(-7.39,1.)(-7.4,1.)  (-7.41,1.)(-7.42,1.)(-7.43,1.)(-7.44,1.)(-7.45,1.)  (-7.46,1.)(-7.47,1.)(-7.48,1.)(-7.49,1.)(-7.5,1.)  (-7.51,1.)(-7.52,1.)(-7.53,1.)(-7.54,1.)(-7.55,1.)  (-7.56,1.)(-7.57,1.)(-7.58,1.)(-7.59,1.)(-7.6,1.)  (-7.61,1.)(-7.62,1.)(-7.63,1.)(-7.64,1.)(-7.65,1.)  (-7.66,1.)(-7.67,1.)(-7.68,1.)(-7.69,1.)(-7.7,1.)  (-7.71,1.)(-7.72,1.)(-7.73,1.)(-7.74,1.)(-7.75,1.)  (-7.76,1.)(-7.77,1.)(-7.78,1.)(-7.79,1.)(-7.8,1.)  (-7.81,1.)(-7.82,1.)(-7.83,1.)(-7.84,1.)(-7.85,1.)  (-7.86,1.)(-7.87,1.)(-7.88,1.)(-7.89,1.)(-7.9,1.)  (-7.91,1.)(-7.92,1.)(-7.93,1.)(-7.94,1.)(-7.95,1.)  (-7.96,1.)(-7.97,1.)(-7.98,1.)(-7.99,1.)(-8.,1.)  (-8.01,1.)(-8.02,1.)(-8.03,1.)(-8.04,1.)(-8.05,1.)  (-8.06,1.)(-8.07,1.)(-8.08,1.)(-8.09,1.)(-8.1,1.)  (-8.11,1.)(-8.12,1.)(-8.13,1.)(-8.14,1.)(-8.15,1.)  (-8.16,1.)(-8.17,1.)(-8.18,1.)(-8.19,1.)(-8.2,1.)  (-8.21,1.)(-8.22,1.)(-8.23,1.)(-8.24,1.)(-8.25,1.)  (-8.26,1.)(-8.27,1.)(-8.28,1.)(-8.29,1.)(-8.3,1.)  (-8.31,1.)(-8.32,1.)(-8.33,1.)(-8.34,1.)(-8.35,1.)  (-8.36,1.)(-8.37,1.)(-8.38,1.)(-8.39,1.)(-8.4,1.)  (-8.41,1.)(-8.42,1.)(-8.43,1.)(-8.44,1.)(-8.45,1.)  (-8.46,1.)(-8.47,1.)(-8.48,1.)(-8.49,1.)(-8.5,1.)  (-8.51,1.)(-8.52,1.)(-8.53,1.)(-8.54,1.)(-8.55,1.)  (-8.56,1.)(-8.57,1.)(-8.58,1.)(-8.59,1.)(-8.6,1.)  (-8.61,1.)(-8.62,1.)(-8.63,1.)(-8.64,1.)(-8.65,1.)  (-8.66,1.)(-8.67,1.)(-8.68,1.)(-8.69,1.)(-8.7,1.)};
\draw[thick,green] plot[smooth] coordinates {(0,0.587282)(-0.01,0.63834)(-0.02,0.690702)  (-0.03,0.743489)(-0.04,0.795642)(-0.05,0.845942)  (-0.06,0.893044)(-0.07,0.935516)(-0.08,0.971896)  (-0.09,1.00074)(-0.1,1.02071)(-0.11,1.03056)  (-0.12,1.0293)(-0.13,1.01612)(-0.14,0.990488)  (-0.15,0.952167)(-0.16,0.901196)(-0.17,0.837904)  (-0.18,0.762894)(-0.19,0.677025)(-0.2,0.581388)  (-0.21,0.477272)(-0.22,0.366141)(-0.23,0.249597)  (-0.24,0.129346)(-0.25,0.00716752)(-0.26,-0.115117)  (-0.27,-0.235679)(-0.28,-0.35271)(-0.29,-0.464455)  (-0.3,-0.569235)(-0.31,-0.665476)(-0.32,-0.751734)  (-0.33,-0.826713)(-0.34,-0.889289)(-0.35,-0.938523)  (-0.36,-0.973676)(-0.37,-0.994222)(-0.38,-0.999854)  (-0.39,-0.990488)(-0.4,-0.966265)(-0.41,-0.927548)  (-0.42,-0.874919)(-0.43,-0.809168)(-0.44,-0.731281)  (-0.45,-0.642426)(-0.46,-0.543936)(-0.47,-0.437288)  (-0.48,-0.324082)(-0.49,-0.206016)(-0.5,-0.0848604)  (-0.51,0.0375682)(-0.52,0.159433)(-0.53,0.278907)  (-0.54,0.394199)(-0.55,0.503578)(-0.56,0.605405)  (-0.57,0.698153)(-0.58,0.78043)(-0.59,0.851004)  (-0.6,0.908814)(-0.61,0.952995)(-0.62,0.982884)  (-0.63,0.998032)(-0.64,0.998213)(-0.65,0.983423)  (-0.66,0.953885)(-0.67,0.910041)(-0.68,0.852549)  (-0.69,0.782271)(-0.7,0.700262)(-0.71,0.60775)  (-0.72,0.506124)(-0.73,0.396907)(-0.74,0.281738)  (-0.75,0.162344)(-0.76,0.0405152)(-0.77,-0.0819213)  (-0.78,-0.203129)(-0.79,-0.321291)(-0.8,-0.434634)  (-0.81,-0.541459)(-0.82,-0.640164)(-0.83,-0.729267)  (-0.84,-0.807434)(-0.85,-0.873492)(-0.86,-0.92645)  (-0.87,-0.965514)(-0.88,-0.990097)(-0.89,-0.999832)  (-0.9,-0.994573)(-0.91,-0.974397)(-0.92,-0.939609)  (-0.93,-0.890729)(-0.94,-0.82849)(-0.95,-0.753827)  (-0.96,-0.667858)(-0.97,-0.571873)(-0.98,-0.467312)  (-0.99,-0.355743)(-1.,-0.238838)(-1.01,-0.118351)  (-1.02,0.00391022)(-1.03,0.126113)(-1.04,0.246425)  (-1.05,0.363041)(-1.06,0.474212)(-1.07,0.578271)  (-1.08,0.673658)(-1.09,0.758942)(-1.1,0.832844)  (-1.11,0.894256)(-1.12,0.942257)(-1.13,0.976126)  (-1.14,0.995356)(-1.15,0.999659)(-1.16,0.988969)  (-1.17,0.963448)(-1.18,0.923478)(-1.19,0.869658)  (-1.2,0.802796)(-1.21,0.723894)(-1.22,0.634136)  (-1.23,0.534868)(-1.24,0.427578)(-1.25,0.313875)  (-1.26,0.195466)(-1.27,0.0741248)(-1.28,-0.048328)  (-1.29,-0.170056)(-1.3,-0.289233)(-1.31,-0.404073)  (-1.32,-0.512853)(-1.33,-0.613942)(-1.34,-0.705823)  (-1.35,-0.787119)(-1.36,-0.85661)(-1.37,-0.913255)  (-1.38,-0.956203)(-1.39,-0.984811)(-1.4,-0.99865)  (-1.41,-0.997511)(-1.42,-0.981413)(-1.43,-0.950597)  (-1.44,-0.905524)(-1.45,-0.846871)(-1.46,-0.775517)  (-1.47,-0.692532)(-1.48,-0.599162)(-1.49,-0.496806)  (-1.5,-0.386999)(-1.51,-0.271389)(-1.52,-0.151708)  (-1.53,-0.0297519)(-1.54,0.0926503)(-1.55,0.213663)  (-1.56,0.331471)(-1.57,0.444308)(-1.58,0.550482)  (-1.59,0.6484)(-1.6,0.736594)(-1.61,0.813741)  (-1.62,0.878685)(-1.63,0.93045)(-1.64,0.968262)  (-1.65,0.991552)(-1.66,0.999972)(-1.67,0.993395)  (-1.68,0.97192)(-1.69,0.935868)(-1.7,0.885782)  (-1.71,0.822411)(-1.72,0.746707)(-1.73,0.659803)  (-1.74,0.563005)(-1.75,0.457764)(-1.76,0.345657)  (-1.77,0.228366)(-1.78,0.10765)(-1.79,-0.0146798)  (-1.8,-0.13679)(-1.81,-0.256848)(-1.82,-0.373055)  (-1.83,-0.483666)(-1.84,-0.587024)(-1.85,-0.681579)  (-1.86,-0.765911)(-1.87,-0.838757)(-1.88,-0.899024)  (-1.89,-0.945809)(-1.9,-0.978409)(-1.91,-0.996335)  (-1.92,-0.999319)(-1.93,-0.987317)(-1.94,-0.960507)  (-1.95,-0.919292)(-1.96,-0.864291)(-1.97,-0.796328)  (-1.98,-0.716422)(-1.99,-0.625772)(-2.,-0.525737)  (-2.01,-0.417817)(-2.02,-0.303632)(-2.03,-0.184892)  (-2.04,-0.0633802)(-2.05,0.0590824)(-2.06,0.180659)  (-2.07,0.299526)(-2.08,0.413901)(-2.09,0.522069)  (-2.1,0.622407)(-2.11,0.713411)(-2.12,0.793716)  (-2.13,0.862117)(-2.14,0.917589)(-2.15,0.9593)  (-2.16,0.986624)(-2.17,0.999151)(-2.18,0.996694)  (-2.19,0.979289)(-2.2,0.947198)(-2.21,0.900902)  (-2.22,0.841094)(-2.23,0.768673)(-2.24,0.684723)  (-2.25,0.590505)(-2.26,0.48743)(-2.27,0.377046)  (-2.28,0.261007)(-2.29,0.141054)(-2.3,0.018985)  (-2.31,-0.103368)(-2.32,-0.224172)(-2.33,-0.341613)  (-2.34,-0.453931)(-2.35,-0.559441)(-2.36,-0.656562)  (-2.37,-0.743836)(-2.38,-0.819954)(-2.39,-0.883775)  (-2.4,-0.934343)(-2.41,-0.970897)(-2.42,-0.992891)  (-2.43,-0.999995)(-2.44,-0.992101)(-2.45,-0.969329)  (-2.46,-0.932019)(-2.47,-0.880732)(-2.48,-0.816236)  (-2.49,-0.7395)(-2.5,-0.651672)(-2.51,-0.554072)  (-2.52,-0.448162)(-2.53,-0.335531)(-2.54,-0.217867)  (-2.55,-0.0969367)(-2.56,0.0254477)(-2.57,0.14745)  (-2.58,0.267242)(-2.59,0.383025)(-2.6,0.493065)  (-2.61,0.595709)(-2.62,0.68942)(-2.63,0.772791)  (-2.64,0.844573)(-2.65,0.903688)(-2.66,0.949251)  (-2.67,0.980578)(-2.68,0.997198)(-2.69,0.998864)  (-2.7,0.98555)(-2.71,0.957455)(-2.72,0.915)  (-2.73,0.858824)(-2.74,0.789767)(-2.75,0.708867)  (-2.76,0.617335)(-2.77,0.516545)(-2.78,0.408008)  (-2.79,0.293353)(-2.8,0.174297)(-2.81,0.0526283)  (-2.82,-0.0698301)(-2.83,-0.191241)(-2.84,-0.309784)  (-2.85,-0.423681)(-2.86,-0.531225)(-2.87,-0.630801)  (-2.88,-0.720917)(-2.89,-0.800221)(-2.9,-0.867525)  (-2.91,-0.921817)(-2.92,-0.962286)(-2.93,-0.988322)  (-2.94,-0.999537)(-2.95,-0.995761)(-2.96,-0.977052)  (-2.97,-0.94369)(-2.98,-0.896175)(-2.99,-0.83522)  (-3.,-0.761739)(-3.01,-0.676834)(-3.02,-0.581779)  (-3.03,-0.477998)(-3.04,-0.367049)(-3.05,-0.250595)  (-3.06,-0.130383)(-3.07,-0.00821598)(-3.08,0.114075)  (-3.09,0.234654)(-3.1,0.351715)(-3.11,0.463501)  (-3.12,0.568336)(-3.13,0.664647)(-3.14,0.750991)  (-3.15,0.826071)(-3.16,0.888763)(-3.17,0.938126)  (-3.18,0.97342)(-3.19,0.994116)(-3.2,0.999902)  (-3.21,0.990693)(-3.22,0.966626)(-3.23,0.928062)  (-3.24,0.87558)(-3.25,0.809967)(-3.26,0.732207)  (-3.27,0.643466)(-3.28,0.545074)(-3.29,0.438508)  (-3.3,0.325366)(-3.31,0.207344)(-3.32,0.0862119)  (-3.33,-0.0362126)(-3.34,-0.158094)(-3.35,-0.277604)  (-3.36,-0.392952)(-3.37,-0.502406)(-3.38,-0.604325)  (-3.39,-0.697181)(-3.4,-0.779582)(-3.41,-0.850291)  (-3.42,-0.908248)(-3.43,-0.952583)(-3.44,-0.982633)  (-3.45,-0.997946)(-3.46,-0.998293)(-3.47,-0.983668)  (-3.48,-0.954291)(-3.49,-0.910602)(-3.5,-0.853257)  (-3.51,-0.783115)(-3.52,-0.701229)(-3.53,-0.608826)  (-3.54,-0.507293)(-3.55,-0.398152)(-3.56,-0.283039)  (-3.57,-0.163682)(-3.58,-0.0418703)(-3.59,0.0805696)  (-3.6,0.201801)(-3.61,0.320006)(-3.62,0.433412)  (-3.63,0.540318)(-3.64,0.639121)(-3.65,0.728339)  (-3.66,0.806634)(-3.67,0.872831)(-3.68,0.925939)  (-3.69,0.96516)(-3.7,0.989906)(-3.71,0.999807)  (-3.72,0.994713)(-3.73,0.974701)(-3.74,0.940072)  (-3.75,0.891344)(-3.76,0.829249)(-3.77,0.754717)  (-3.78,0.668867)(-3.79,0.572985)(-3.8,0.468511)  (-3.81,0.35701)(-3.82,0.240155)(-3.83,0.119698)  (-3.84,-0.00255401)(-3.85,-0.124768)(-3.86,-0.24511)  (-3.87,-0.361777)(-3.88,-0.473017)(-3.89,-0.577164)  (-3.9,-0.672655)(-3.91,-0.758059)(-3.92,-0.832093)  (-3.93,-0.893648)(-3.94,-0.941802)(-3.95,-0.97583)  (-3.96,-0.995225)(-3.97,-0.999693)(-3.98,-0.989169)  (-3.99,-0.963811)(-4.,-0.923997)(-4.01,-0.870327)  (-4.02,-0.803604)(-4.03,-0.724829)(-4.04,-0.635184)  (-4.05,-0.536013)(-4.06,-0.428803)(-4.07,-0.315163)  (-4.08,-0.196796)(-4.09,-0.0754772)(-4.1,0.0469733)  (-4.11,0.168719)(-4.12,0.287935)(-4.13,0.402832)  (-4.14,0.511688)(-4.15,0.612871)(-4.16,0.704862)  (-4.17,0.786282)(-4.18,0.85591)(-4.19,0.912701)  (-4.2,0.955805)(-4.21,0.984575)(-4.22,0.998578)  (-4.23,0.997606)(-4.24,0.981673)(-4.25,0.951017)  (-4.26,0.906098)(-4.27,0.847591)(-4.28,0.776372)  (-4.29,0.69351)(-4.3,0.600247)(-4.31,0.497983)  (-4.32,0.388249)(-4.33,0.272694)(-4.34,0.153048)  (-4.35,0.0311074)(-4.36,-0.0912998)(-4.37,-0.212338)  (-4.38,-0.330191)(-4.39,-0.443093)(-4.4,-0.549349)  (-4.41,-0.647367)(-4.42,-0.735676)(-4.43,-0.812952)  (-4.44,-0.878036)(-4.45,-0.929952)(-4.46,-0.967922)  (-4.47,-0.991375)(-4.48,-0.99996)(-4.49,-0.993549)  (-4.5,-0.972238)(-4.51,-0.936345)(-4.52,-0.88641)  (-4.53,-0.823182)(-4.54,-0.747608)(-4.55,-0.660822)  (-4.56,-0.564126)(-4.57,-0.458969)(-4.58,-0.346929)  (-4.59,-0.229686)(-4.6,-0.108998)(-4.61,0.0133237)  (-4.62,0.135446)(-4.63,0.255537)(-4.64,0.371796)  (-4.65,0.482479)(-4.66,0.585926)(-4.67,0.680586)  (-4.68,0.765039)(-4.69,0.838018)(-4.7,0.89843)  (-4.71,0.945367)(-4.72,0.978127)(-4.73,0.996218)  (-4.74,0.999368)(-4.75,0.987531)(-4.76,0.960883)  (-4.77,0.919825)(-4.78,0.864973)(-4.79,0.797148)  (-4.8,0.717368)(-4.81,0.626829)(-4.82,0.52689)  (-4.83,0.419049)(-4.84,0.304923)(-4.85,0.186225)  (-4.86,0.0647336)(-4.87,-0.0577285)(-4.88,-0.179325)  (-4.89,-0.298232)(-4.9,-0.412666)(-4.91,-0.520912)  (-4.92,-0.621345)(-4.93,-0.71246)(-4.94,-0.79289)  (-4.95,-0.861429)(-4.96,-0.917049)(-4.97,-0.958916)  (-4.98,-0.986402)(-4.99,-0.999094)(-5.,-0.996803)  (-5.01,-0.979563)(-5.02,-0.947632)(-5.03,-0.901489)  (-5.04,-0.841827)(-5.05,-0.769539)(-5.06,-0.685711)  (-5.07,-0.591599)(-5.08,-0.488614)(-5.09,-0.378302)  (-5.1,-0.262316)(-5.11,-0.142396)(-5.12,-0.020341)  (-5.13,0.102019)(-5.14,0.22285)(-5.15,0.340338)  (-5.16,0.452722)(-5.17,0.558317)(-5.18,0.655538)  (-5.19,0.742928)(-5.2,0.819177)(-5.21,0.88314)  (-5.22,0.933858)(-5.23,0.970572)(-5.24,0.992729)  (-5.25,0.999998)(-5.26,0.99227)(-5.27,0.969661)  (-5.28,0.93251)(-5.29,0.881374)(-5.3,0.817019)  (-5.31,0.740412)(-5.32,0.6527)(-5.33,0.5552)  (-5.34,0.449374)(-5.35,0.336808)(-5.36,0.219191)  (-5.37,0.0982865)(-5.38,-0.0240919)(-5.39,-0.146109)  (-5.4,-0.265935)(-5.41,-0.381772)(-5.42,-0.491884)  (-5.43,-0.594619)(-5.44,-0.688437)(-5.45,-0.77193)  (-5.46,-0.843846)(-5.47,-0.903107)(-5.48,-0.948824)  (-5.49,-0.980311)(-5.5,-0.997096)(-5.51,-0.998928)  (-5.52,-0.985778)(-5.53,-0.957845)(-5.54,-0.915547)  (-5.55,-0.859518)(-5.56,-0.790599)(-5.57,-0.709823)  (-5.58,-0.618401)(-5.59,-0.517706)(-5.6,-0.409246)  (-5.61,-0.294649)(-5.62,-0.175633)(-5.63,-0.0539826)  (-5.64,0.0684771)(-5.65,0.18991)(-5.66,0.308494)  (-5.67,0.422453)(-5.68,0.530075)(-5.69,0.629748)  (-5.7,0.719976)(-5.71,0.799407)(-5.72,0.866849)  (-5.73,0.921291)(-5.74,0.961916)(-5.75,0.988115)  (-5.76,0.999495)(-5.77,0.995885)(-5.78,0.97734)  (-5.79,0.944138)(-5.8,0.896776)(-5.81,0.835965)  (-5.82,0.762617)(-5.83,0.677832)(-5.84,0.582881)  (-5.85,0.479189)(-5.86,0.36831)(-5.87,0.251908)  (-5.88,0.131728)(-5.89,0.00957214)(-5.9,-0.112727)  (-5.91,-0.233336)(-5.92,-0.350445)(-5.93,-0.462299)  (-5.94,-0.567219)(-5.95,-0.663633)(-5.96,-0.750094)  (-5.97,-0.825306)(-5.98,-0.888141)(-5.99,-0.937656)  (-6.,-0.973109)(-6.01,-0.993968)(-6.02,-0.99992)  (-6.03,-0.990876)(-6.04,-0.966972)(-6.05,-0.928566)  (-6.06,-0.876235)(-6.07,-0.810762)(-6.08,-0.73313)  (-6.09,-0.644503)(-6.1,-0.546211)(-6.11,-0.439726)  (-6.12,-0.326648)(-6.13,-0.20867)(-6.14,-0.087563)  (-6.15,0.0348572)(-6.16,0.156755)(-6.17,0.276301)  (-6.18,0.391704)(-6.19,0.501233)(-6.2,0.603244)  (-6.21,0.696208)(-6.22,0.778732)(-6.23,0.849576)  (-6.24,0.907679)(-6.25,0.95217)(-6.26,0.982381)  (-6.27,0.997858)(-6.28,0.998371)(-6.29,0.983911)  (-6.3,0.954695)(-6.31,0.911162)(-6.32,0.853963)  (-6.33,0.783958)(-6.34,0.702195)(-6.35,0.609902)  (-6.36,0.508461)(-6.37,0.399396)(-6.38,0.28434)  (-6.39,0.16502)(-6.4,0.0432253)(-6.41,-0.0792177)  (-6.42,-0.200473)(-6.43,-0.318721)(-6.44,-0.43219)  (-6.45,-0.539177)(-6.46,-0.638077)(-6.47,-0.727409)  (-6.48,-0.805831)(-6.49,-0.872168)(-6.5,-0.925426)  (-6.51,-0.964804)(-6.52,-0.989713)(-6.53,-0.999779)  (-6.54,-0.994851)(-6.55,-0.975004)(-6.56,-0.940534)  (-6.57,-0.891958)(-6.58,-0.830006)(-6.59,-0.755606)  (-6.6,-0.669874)(-6.61,-0.574096)(-6.62,-0.469708)  (-6.63,-0.358276)(-6.64,-0.241471)(-6.65,-0.121044)  (-6.66,0.00119779)(-6.67,0.123422)(-6.68,0.243795)  (-6.69,0.360512)(-6.7,0.471822)(-6.71,0.576056)  (-6.72,0.671651)(-6.73,0.757173)(-6.74,0.83134)  (-6.75,0.893039)(-6.76,0.941345)(-6.77,0.975533)  (-6.78,0.995091)(-6.79,0.999726)(-6.8,0.989367)  (-6.81,0.964171)(-6.82,0.924515)(-6.83,0.870994)  (-6.84,0.80441)(-6.85,0.725763)(-6.86,0.636231)  (-6.87,0.537158)(-6.88,0.430028)(-6.89,0.31645)  (-6.9,0.198125)(-6.91,0.0768294)(-6.92,-0.0456185)  (-6.93,-0.167382)(-6.94,-0.286636)(-6.95,-0.401591)  (-6.96,-0.510523)(-6.97,-0.611799)(-6.98,-0.703899)  (-6.99,-0.785443)(-7.,-0.855207)(-7.01,-0.912146)  (-7.02,-0.955406)(-7.03,-0.984336)(-7.04,-0.998505)  (-7.05,-0.997699)(-7.06,-0.98193)(-7.07,-0.951435)  (-7.08,-0.906671)(-7.09,-0.84831)(-7.1,-0.777226)  (-7.11,-0.694487)(-7.12,-0.601332)(-7.13,-0.499158)  (-7.14,-0.389499)(-7.15,-0.273998)(-7.16,-0.154388)  (-7.17,-0.032463)(-7.18,0.0899492)(-7.19,0.211012)  (-7.2,0.328911)(-7.21,0.441877)(-7.22,0.548216)  (-7.23,0.646333)(-7.24,0.734757)(-7.25,0.812162)  (-7.26,0.877387)(-7.27,0.929453)(-7.28,0.96758)  (-7.29,0.991196)(-7.3,0.999947)(-7.31,0.993702)  (-7.32,0.972554)(-7.33,0.936821)(-7.34,0.887037)  (-7.35,0.823951)(-7.36,0.748508)(-7.37,0.661839)  (-7.38,0.565245)(-7.39,0.460173)(-7.4,0.348201)  (-7.41,0.231006)(-7.42,0.110347)(-7.43,-0.0119676)  (-7.44,-0.134102)(-7.45,-0.254226)(-7.46,-0.370537)  (-7.47,-0.48129)(-7.48,-0.584826)(-7.49,-0.679591)  (-7.5,-0.764164)(-7.51,-0.837277)(-7.52,-0.897833)  (-7.53,-0.944924)(-7.54,-0.977844)(-7.55,-0.996099)  (-7.56,-0.999416)(-7.57,-0.987744)(-7.58,-0.961258)  (-7.59,-0.920357)(-7.6,-0.865652)(-7.61,-0.797966)  (-7.62,-0.718312)(-7.63,-0.627885)(-7.64,-0.528042)  (-7.65,-0.42028)(-7.66,-0.306215)(-7.67,-0.187557)  (-7.68,-0.0660869)(-7.69,0.0563745)(-7.7,0.177991)  (-7.71,0.296937)(-7.72,0.411431)(-7.73,0.519754)  (-7.74,0.620282)(-7.75,0.711508)(-7.76,0.792063)  (-7.77,0.86074)(-7.78,0.916508)(-7.79,0.95853)  (-7.8,0.986178)(-7.81,0.999036)(-7.82,0.996911)  (-7.83,0.979835)(-7.84,0.948064)(-7.85,0.902076)  (-7.86,0.842558)(-7.87,0.770405)(-7.88,0.686697)  (-7.89,0.592692)(-7.9,0.489797)(-7.91,0.379557)  (-7.92,0.263625)(-7.93,0.143739)(-7.94,0.0216969)  (-7.95,-0.10067)(-7.96,-0.221527)(-7.97,-0.339062)  (-7.98,-0.451513)(-7.99,-0.557191)(-8.,-0.654513)  (-8.01,-0.74202)(-8.02,-0.818398)(-8.03,-0.882503)  (-8.04,-0.933372)(-8.05,-0.970244)(-8.06,-0.992565)  (-8.07,-1.)(-8.08,-0.992438)(-8.09,-0.969992)  (-8.1,-0.932999)(-8.11,-0.882014)(-8.12,-0.817801)  (-8.13,-0.741323)(-8.14,-0.653727)(-8.15,-0.556328)  (-8.16,-0.450585)(-8.17,-0.338084)(-8.18,-0.220514)  (-8.19,-0.099636)(-8.2,0.022736)(-8.21,0.144767)  (-8.22,0.264627)(-8.23,0.380518)(-8.24,0.490703)  (-8.25,0.593528)(-8.26,0.687453)(-8.27,0.771067)  (-8.28,0.843117)(-8.29,0.902524)(-8.3,0.948395)  (-8.31,0.980042)(-8.32,0.996992)(-8.33,0.99899)  (-8.34,0.986005)(-8.35,0.958234)(-8.36,0.916091)  (-8.37,0.86021)(-8.38,0.791428)(-8.39,0.710777)  (-8.4,0.619467)(-8.41,0.518866)(-8.42,0.410483)  (-8.43,0.295945)(-8.44,0.176968)(-8.45,0.0553368)  (-8.46,-0.067124)(-8.47,-0.188578)(-8.48,-0.307204)  (-8.49,-0.421223)(-8.5,-0.528925)(-8.51,-0.628694)  (-8.52,-0.719034)(-8.53,-0.798592)(-8.54,-0.866172)  (-8.55,-0.920763)(-8.56,-0.961544)(-8.57,-0.987905)  (-8.58,-0.999451)(-8.59,-0.996007)(-8.6,-0.977626)  (-8.61,-0.944584)(-8.62,-0.897375)(-8.63,-0.836709)  (-8.64,-0.763494)(-8.65,-0.678828)(-8.66,-0.583983)  (-8.67,-0.480379)(-8.68,-0.369571)(-8.69,-0.25322)  (-8.7,-0.133072)}; 
\end{tikzpicture}

\caption{${\rm e}^{-\gamma_0 s-\rho_0} v_0(s) $ (red); $\frac{1}{2}   {\rm e}^{-\operatorname{Re}(\gamma_1)s-\operatorname{Re}(\rho_1)} v_1(s)$  (green).}
\label{fig-5}
\end{figure}